\documentclass[twocolumn]{aastex7}

\usepackage{amsmath}
\usepackage{xspace}
\usepackage{graphicx}
\usepackage{adjustbox}
\usepackage{natbib}
\usepackage{soul}
\usepackage{multirow}
\usepackage{rotating}
\usepackage{tabularx}

\newcommand{\Msol}{\ensuremath{\mathrm{M}_{\odot}}\xspace}

\newcommand{\kms}{km~s\ensuremath{^{-1}}\xspace}

\newcommand{\Ha}{H\ensuremath{\alpha}\xspace}
\newcommand{\Hb}{H\ensuremath{\beta}\xspace}
\newcommand{\Hc}{H\ensuremath{\gamma}\xspace}

\newcommand{\Nifs}{\ensuremath{^{56}}Ni\xspace}

\newcommand{\mic}{\ensuremath{\mu}m\xspace}

\newcommand{\ixf}{SN~2023ixf\xspace}

\newcommand{\jwst}{{\it JWST}\xspace}

\newcommand{\HeI}{He~{\sc i}}

\newcommand{\OI}{O~{\sc i}}

\newcommand{\CI}{C~{\sc i}}

\newcommand{\NeII}{Ne~{\sc ii}}
\newcommand{\ArII}{Ar~{\sc ii}}
\newcommand{\ArIII}{Ar~{\sc iii}}
\newcommand{\BaII}{Ba~{\sc ii}}
\newcommand{\NaI}{Na~{\sc i}}

\newcommand{\MgI}{Mg~{\sc i}}
\newcommand{\SiI}{Si~{\sc i}}

\newcommand{\SI}{S~{\sc i}}

\newcommand{\CaII}{Ca~{\sc ii}}

\newcommand{\FeI}{Fe~{\sc i}}
\newcommand{\FeII}{Fe~{\sc ii}}

\newcommand{\CoI}{Co~{\sc i}}
\newcommand{\CoII}{Co~{\sc ii}}
\newcommand{\CoIII}{Co~{\sc iii}}
\newcommand{\NiII}{Ni~{\sc ii}}

\newcommand{\Nineb}{[Ni~{\sc i}]}
\newcommand{\SrII}{Sr~{\sc ii}}
\newcommand{\Pone}{Paper~{I}}

\newcommand{\E}{$E\left(B-V\right)$}

\newcommand{\R}{$f_\mathrm{em}/f_\mathrm{abs}$}

\newcommand{\mzams}{$M_\mathrm{ZAMS}$}

\renewcommand{\micron}{\ensuremath{\mu}m\xspace}
\defcitealias{DerKacy_2025}{Paper~{I}}

\DeclareRobustCommand{\okina}{%
  \raisebox{\dimexpr\fontcharht\font`A-\height}{%
    \scalebox{0.8}{`}%
  }%
}

\begin{document}

\title{{\it JWST} Observations of SN~2023ixf II: The Panchromatic Evolution Between 250 and 720 Days After the Explosion}

\DeclareRobustCommand{\okina}{%
  \raisebox{\dimexpr\fontcharht\font`A-\height}{%
    \scalebox{0.8}{`}%
  }%
}

\newcommand{\PSI}{\affiliation{Planetary Science Institute, 1700 East Fort
  Lowell Road, Suite 106,Tucson, AZ 85719-2395 USA}}
\newcommand{\HS}{\affiliation{Hamburger Sternwarte, Gojenbergsweg 112, 21029 Hamburg, Germany}}
\newcommand{\IFA}{\affiliation{Institute for Astronomy, University of Hawai’i at Manoa, 2680 Woodlawn Dr., Hawai\okina i, HI 96822, USA}}
\newcommand{\VT}{\affiliation{Department of Physics, Virginia Tech,
    850 West Campus  Drive, Blacksburg VA, 24061, USA}}
\newcommand{\GRFP}{\altaffiliation{National Science Foundation Graduate Research Fellow}}
\newcommand{\FINESST}{\altaffiliation{NASA FINESST Future Investigator}}
\newcommand{\STSci}{\affiliation{Space Telescope Science Institute, 3700 San Martin Drive, Baltimore, MD 21218-2410, USA}}
\newcommand{\FSU}{\affiliation{Department of Physics, Florida State
    University, Tallahassee, FL 32306, USA}}
\newcommand{\Carnegie}{\affiliation{Observatories of the Carnegie
    Institution for Science, 813 Santa Barbara St., Pasadena, CA 91101, USA}}
\newcommand{\MSU}{\affiliation{Department of Physics \& Astronomy,
    Michigan State University, East Lansing, MI, USA}}
\newcommand{\TAMU}{\affiliation{George P. and Cynthia Woods Mitchell
    Institute for Fundamental Physics and Astronomy,
    Department of Physics and Astronomy, Texas 
             A\&M University, College Station, TX 77843, USA}}
\newcommand{\IALP}{\affiliation{Instituto de Astrof\'isica de La Plata
    (IALP), CONICET, Paseo del Bosque S/N, B1900FWA La Plata, Argentina}}
\newcommand{\LaPlata}{\affiliation{Facultad de Ciencias Astron\'omicas
    y Geof\'isicas Universidad Nacional de La Plata, Paseo del Bosque,
    B1900FWA, La Plata, Argentina}}
\newcommand{\WPI}{\affiliation{Kavli Institute for the Physics and
    Mathematics of the Universe (WPI), The University of Tokyo,
    Kashiwa, 277-8583 Chiba, Japan}} 

\newcommand{\ICE}{\affiliation{Institute of Space Sciences (ICE,
    CSIC), Campus UAB, Carrer de Can Magrans, s/n, E-08193 Barcelona, Spain}}

\newcommand{\IEEC}{\affiliation{Institut d’Estudis Espacials de
    Catalunya (IEEC), E-08034  Barcelona, Spain}} 

\newcommand{\LCO}{\affiliation{Las Campanas Observatory, Carnegie
    Observatories, Casilla 601, La Serena, Chile}} 

\newcommand{\Aarhus}{\affiliation{Department of Physics and Astronomy,
    Aarhus University, Ny  Munkegade 120, DK-8000 Aarhus C, Denmark.}} 

\newcommand{\OU}{\affiliation{Homer L.~Dodge Department of Physics and
  Astronomy, University of Oklahoma, 440 W. Brooks, Rm 100, Norman, OK
  73019-2061}}  

\newcommand{\UCSC}{\affiliation{Department of Astronomy and Astrophysics,
  University of California, Santa Cruz, CA 95064, USA}} 
\newcommand{\Melbourne}{\affiliation{School of Physics, The University of
  Melbourne, VIC 3010, Australia}}

\newcommand{\LPNHE}{\affiliation{LPNHE, (CNRS/IN2P3, Sorbonne
  Universit\'e, Universit\'e Paris Cit\'e), Laboratoire de Physique
  Nucl\'eaire et de Hautes \'Energies, 75005, Paris, France}}

\newcommand{\SETI}{\affiliation{SETI Institute, 339 Bernardo Ave., Ste. 200, Mountain View, CA 94043, USA}} 

\newcommand{\Princeton}{\affiliation{Princeton University, 4 Ivy Lane,
    Princeton, NJ 08544, USA}} 

\newcommand{\Berkeley}{\affiliation{Department of Astronomy,
    University of California, Berkeley, CA 94720-3411, USA}}

\newcommand{\Tsinghua}{\affiliation{Physics Department, Tsinghua
    University, Beijing, 100084, China}}

\newcommand{\Thailand}{\affiliation{National Astronomical Research
    Institute of Thailand, 260 Moo 4, Donkaew, Maerim, Chiang Mai,
    50180, Thailand}}

\newcommand{\UVA}{\affiliation{Department of Astronomy, University of
    Virginia, 530 McCormick Rd, Charlottesville, VA 22904, USA}}

\newcommand{\LJMU}{\affiliation{Astrophysics Research Institute,
    Liverpool John Moores University, 146 Brownlow Hill, Liverpool L3
    5RF, UK}}

\newcommand{\MPIA}{\affiliation{Max-Planck-Institut f\"ur Astrophysik,
    Karl-Schwarzschild Stra{\ss}e 1, 85748 Garching, Germany}}

\newcommand{\JHU}{\affiliation{Physics and Astronomy Department,
    Johns Hopkins University, Baltimore, MD 21218, USA}}

\newcommand{\OSU}{\affiliation{Department of Astronomy, The Ohio State
    University, Columbus, OH, USA}}

\newcommand{\CCAP}{\affiliation{Center for Cosmology and Astroparticle
    Physics, The Ohio State University, Columbus, OH, USA}}

\newcommand{\MIT}{\affiliation{Department of Physics and Kavli Institute for Astrophysics and Space Research, Massachusetts Institute of Technology, 77 Massachusetts Avenue, Cambridge, MA 02139, USA}}

\newcommand{\USzeged}{\affiliation{Department of Experimental Physics, Institute of Physics, University of Szeged, D{\'o}m t{\'e}r 9, 6720 Szeged, Hungary}}

\newcommand{\Stockholm}{\affiliation{Department of Physics, The Oskar Klein Center, Stockholm University, AlbaNova, 10691 Stockholm, Sweden}}

\newcommand{\Cardiff}{\affiliation{Cardiff Hub for Astrophysical Research and Technology (CHART), School of Physics and Astronomy, Cardiff University, The Parade, Cardiff CF24 3AA, UK}}

\newcommand{\Perdue}{\affiliation{Purdue University, Department of Physics and Astronomy, 525 Northwestern Ave, West Lafayette, IN 4790720, USA}}
\newcommand{\Harvard}{\affiliation{Harvard}}
\newcommand{\IPAC}{\affiliation{Caltech/IPAC, Mailcode 100-22, Pasadena, CA 91125, USA}}
\newcommand{\Arizona}{\affiliation{Steward Observatory, University of Arizona, 933 N. Cherry St, Tucson, AZ 85721, USA}}

\newcommand{\nextinstitute}{\affiliation{Put the institute of the new author here}}

\author[0000-0001-7186-105X]{K. Medler}
\email{kmedler@hawaii.edu}
\IFA

\author[0000-0002-5221-7557]{C. Ashall}
\email{cashall@hawaii.edu}
\IFA

\author[0000-0002-4338-6586]{P.~Hoeflich}
\email{phoeflich77@gmail.com}
\FSU

\author[0000-0001-5393-1608]{E.~Baron}
\email{ebaron@psi.edu}
\affiliation{Planetary Science Institute, 1700 East Fort Lowell Road, Suite 106, Tucson, AZ 85719-2395, USA}
\affiliation{Hamburger Sternwarte, Gojenbergsweg 112, D-21029 Hamburg, Germany}
 
\author[0000-0002-4338-6586]{J.~M.~DerKacy}
\email{jmderkacy@gmail.com}
\STSci

\author[0000-0002-9301-5302]{M.~Shahbandeh}
\email{mshahbandeh@stsci.edu}
\STSci

\author[0000-0001-5888-2542]{T.~Mera}
\email{tycomera@gmail.com}
\FSU

\author[0000-0002-7305-8321]{C.~M.~Pfeffer}
\email{cmpfeffer@hawaii.edu}
\GRFP
\IFA

\author[0000-0003-3953-9532]{W.~B.~Hoogendam}
\email{willemh@hawaii.edu} 
\GRFP
\IFA

\author[0000-0002-6230-0151]{D.~O.~Jones}
\affiliation{Institute for Astronomy, University of Hawai'i, 640 N. A'ohoku Pl., Hilo, HI 96720, USA}
\email{dojones@hawaii.edu}

\author[0000-0001-6107-0887]{S. Shiber}
\FSU
\email{sshiber1@lsu.edu} 

\author[0009-0001-9148-8421]{E.~Fereidouni}
\email{ef22g@fsu.edu}
\FSU

\author[0000-0003-2238-1572]{O.~D.~Fox}
\email{ofox@stsci.edu}
\STSci

\author[0000-0003-2238-1572]{J.~Jencson}
\email{jjencson@ipac.caltech.edu}
\IPAC

\author[0000-0002-1296-6887]{L.~Galbany}
\email{lluisgalbany@gmail.com}
\ICE
\IEEC

\author[0000-0001-9668-2920]{J.~T.~Hinkle}
\FINESST
\IFA
\email{jhinkle6@hawaii.edu}

\author[0000-0002-2471-8442]{M.~A.~Tucker}
\email{tuckerma95@gmail.com}
\CCAP
\OSU

\author[0000-0003-4631-1149]{B.~J.~Shappee}
\email{shappee@hawaii.edu}
\IFA

\author[0000-0003-1059-9603]{M.~E.~Huber}
\IFA
\email{mehuber7@hawaii.edu}

\author[0000-0002-5063-0751]{K.~Auchettl}
\affiliation{Ozgrav, School of Physics, The University of Melbourne, Parkville, VIC 3010, Australia}
\affiliation{Department of Astronomy and Astrophysics, University of California, Santa Cruz, CA 95064, USA}
\email{katie.auchettl@unimelb.edu.au}

\author[0000-0002-4269-7999]{C.~R.~Angus}
\affiliation{Astrophysics Research Centre, School of Mathematics and Physics, Queen’s University Belfast, Belfast BT7 1NN, UK}
\affiliation{DARK, Niels Bohr Institute, University of Copenhagen, Jagtvej 128, DK-2200 Copenhagen {\O} Denmark}
\email{c.angus@qub.ac.uk} 

\author[0000-0002-2164-859X]{D.~D.~Desai}
\email{dddesai@hawaii.edu}
\IFA

\author[0000-0003-3429-7845]{A.~Do}
\affiliation{Institute of Astronomy and Kavli Institute for Cosmology, Madingley Road, Cambridge CB3 0HA, UK}
\email{ajmd6@cam.ac.uk}

\author[0000-0003-3490-3243]{A.~V.~Payne}
\email{apayne@stsci.edu}
\STSci

\author[0009-0008-3724-1824]{J.~Shi}
\affiliation{Ozgrav, School of Physics, The University of Melbourne, Parkville, VIC 3010, Australia}
\email{jennifer.shi@student.unimelb.edu.au}

\author[0009-0005-5121-2884]{M.~Y.~Kong}
\IFA
\email{ykong2@hawaii.edu}

\author[0009-0003-8153-9576]{S.~Romagnoli}
\affiliation{Ozgrav, School of Physics, The University of Melbourne, Parkville, VIC 3010, Australia}
\email{romagnolis@student.unimelb.edu.au}

\author[0009-0000-6821-9285]{A.~Syncatto}
\affiliation{Department of Physics and Astronomy, University of Hawai`i at Hilo, Hilo, HI, USA}
\email{jaq45@hawaii.edu}

\author[0000-0002-0141-7436]{G.~Clayton}
\email{gclayton@SpaceScience.org}
\affiliation{Department of Physics \& Astronomy, Louisiana State University, Baton Rouge, LA 70803, USA}
\affiliation{Space Science Institute, 4765 Walnut St, Suite B Boulder, CO 80301, USA
}

\author{M.~Dulude}
\email{dulude@stsci.edu}
\STSci

\author{M.~Engesser}
\email{mengesser@stsci.edu}
\STSci

\author[0000-0003-3460-0103]{A.~V.~Filippenko} 
\Berkeley
\email{afilippenko@berkeley.edu}

\author[0000-0001-6395-6702]{S.~Gomez}
\email{sebastian.gomez@austin.utexas.edu}
\affiliation{Department of Astronomy, The University of Texas at Austin, 2515 Speedway, Stop C1400, Austin, TX 78712, USA}

\author[0000-0003-1039-2928]{E.~Y.~Hsiao}
\email{yichi.hsiao@gmail.com}
\FSU

\author[0000-0001-6069-1139]{T.~de~Jaeger}
\email{dejaeger.thomas@gmail.com}
\LPNHE

\author{J.~Johansson}
\email{joeljo@fysik.su.se}
\Stockholm

\author[0000-0002-6650-694X]{K.~Krisciunas}
\email{krisciunas@physics.tamu.edu}
\TAMU

\author[0000-0001-8367-7591]{S.~Kumar}
\email{sahanak@gmail.com}
\UVA

\author[0000-0002-3900-1452]{J.~Lu}
\email{lujingeve158@gmail.com}
\MSU

\author[0000-0002-5529-5593]{M.~Matsuura}
\email{MatsuuraM@cardiff.ac.uk}
\Cardiff

\author[0000-0001-6876-8284]{P.~A.~Mazzali}
\email{P.Mazzali@ljmu.ac.uk}
\LJMU
\MPIA

\author[0000-0002-0763-3885]{D.~Milisavljevic}
\email{dmilisav@purdue.edu}
\Perdue

\author[0000-0003-2535-3091]{N.~Morrell}
\email{nmorrell@carnegiescience.edu}
\LCO

\author[0000-0002-2432-8946]{R.~O’Steen}
\email{rosteen@stsci.edu}
\STSci

\author[0000-0001-7488-4337]{S.~Park}
\email{rogersh0125@snu.ac.kr}
\affiliation{Department of Physics and Astronomy, Seoul National University, Gwanak-ro 1, Gwanak-gu, Seoul, 08826, South Korea}

\author[0000-0003-2734-0796]{M.~M.~Phillips}
\email{mmp@lco.cl}
\LCO

\author[0000-0002-7352-7845]{A.~P.~Ravi}
\email{apazhayathravi@ucdavis.edu}
\affiliation{Department of Physics and Astronomy, University of California, 1 Shields Avenue, Davis, CA 95616-5270, USA}

\author[[0000-0002-4410-5387]{A.~Rest}
\email{arest@stsci.edu}
\STSci
\JHU

\author[0000-0003-3643-839X]{J.~Rho}
\email{jrho@seti.org}
\affiliation{SETI Institute, 339 Bernardo Ave., Ste. 200, Mountain View, CA 94043, USA}

\author[0000-0002-8102-181X]{N.~B.~Suntzeff}
\email{nsuntzeff@tamu.edu}
\TAMU

\author[0000-0002-9820-679X]{A.~Sarangi}
\email{sarangi@NBI.KU.DK}
\SETI

\author[0000-0001-5510-2424]{N.~Smith}
\email{nathansmith@arizona.edu}
\Arizona

\author[0000-0002-5571-1833]{M.~D.~Stritzinger}
\email{max@phys.au.dk}
\Aarhus

\author[0000-0002-7756-4440]{L.~Strolger}
\email{strolger@stsci.edu}
\STSci

\author[0000-0003-4610-1117]{T.~Szalai}
\affiliation{Department of Experimental Physics, Institute of Physics, University of Szeged, H-6720 Szeged, D{\'o}m t{\'e}r 9, Hungary}
\affiliation{MTA-ELTE Lend\"ulet "Momentum" Milky Way Research Group,
Szent Imre H. st. 112, 9700 Szombathely, Hungary}
\email{szaszi@titan.physx.u-szeged.hu}

\author[0000-0001-7380-3144]{T.~Temim}
\email{temim@astro.princeton.edu}
\Princeton

\author[0000-0002-1481-4676]{S.~Tinyanont}
\email{samaporn@narit.or.th}
\Thailand

\author[0000-0001-9038-9950]{S.~D.~Van Dyk}
\email{vandyk@ipac.caltech.edu}
\IPAC

\author[0000-0001-7092-9374]{L.~Wang}
\email{lifan@tamu.edu}
\TAMU

\author[0000-0001-5233-6989]{Q.~Wang}
\email{qnwang@mit.edu}
\MIT

\author[0000-0002-4000-4394]{R.~Wesson}
\email{rw@nebulousresearch.org}
\Cardiff

\author[0000-0002-6535-8500]{Y.~Yang}
\email{yiyangtamu@gmail.com}
\Tsinghua
\Berkeley

\author[0000-0001-7473-4208]{S.~Zs{\'i}ros}
\affiliation{Department of Experimental Physics, Institute of Physics,
University of Szeged, D{\'o}m t{\'e}r 9, 6720 Szeged, Hungary}
\affiliation{HUN-REN CSFK Konkoly Observatory, Konkoly Thege M. ut
15-17, Budapest, 1121, Hungary}
\email{szannazsiros@titan.physx.u-szeged.hu}

\submitjournal{ApJ}

\received{\today}
\revised{}
\accepted{}

\begin{abstract}
 We present the nebular phase spectroscopic and photometric observations of the nearby hydrogen-rich core-collapse supernova (CC-SN) 2023ixf, obtained through our \jwst\ programs. These observations, combined with ground-based optical and near-infrared spectra, cover $+252.67-719.96$~d, creating a comprehensive, panchromatic time-series dataset spanning $0.32 -  30$~\micron. In this second paper of the series, we focus on identifying key spectral emission features and tracking their evolution through the nebular phase. The \jwst\ data reveal hydrogen emission from the Balmer to Humphreys series, as well as prominent forbidden lines from Ne, Ar, Fe, Co, and Ni. NIRSpec observations display strong emission from the first overtone and fundamental bands of carbon monoxide, which weaken with time as the ejecta cools and dust emission dominates.
The spectral energy distribution shows a clear infrared excess emerging by $+252.67$~d peaking around $10.0$~\micron, with a secondary bump at $18.0$~\micron\ developing by $+719.96$~d. We suggest that this evolution could arises from multiple warm dust components.
In upcoming papers in this series, we will present detailed modeling of the molecular and dust properties. Overall, this dataset significantly advances our understanding of the mid-infrared properties of CC-SNe, providing an unprecedented view of their late-time line, molecule, and dust emission.
\end{abstract}

\section{Introduction}\label{sec:intro}

Core-collapse supernovae (CC-SNe), which arise from the collapse of massive stars with zero-age main sequence mass $\mathrm{M_{ZAMS}} \gtrsim 8$~\Msol, provide crucial insights into the final stages of stellar evolution \citep{Smartt_2009}. These explosions enrich galaxies with iron-group and $\alpha$ elements \citep{Woosley_1995, Janka_2012, Wang_2024}, and contribute to the formation of galactic dust \citep{Dwek_2007, Gall_2011}. While public, optical, all-sky \citep[e.g.,][]{Shappee14_ASASSN, Tonry18} and targeted \citep[e.g.,][]{Huber15, Groot22_BlackGEM, Dyer24_GOTO} surveys now discover thousands of CC-SNe annually, the majority of these events are primarily studied in the optical or near-infrared (NIR), with far fewer observations at the ultraviolet (UV) or mid-infrared (MIR) wavelengths. This observational bias, confined typically to $\sim 0.3 - 2.5$~\micron, limits our ability to comprehensively explore the parameter space of the physical conditions, and explosion physics of these objects, as well as their role in galactic chemical and dust enrichment.

The MIR regime, spanning $5 - 30$~\micron, remains relatively under-explored in the study of CC-SNe, yet it contains crucial features that are otherwise inaccessible.  While strong emission lines are prevalent at all wavelengths, the emission lines located within the optical and NIR can be heavily blended making species identification and modeling more difficult. In contrast, lines at longer wavelengths ($\geq 3$~\micron) tend to be significantly less blended, due to reduced overlap of the Doppler broadened lines, allowing for clearer identification and modeling \citep[e.g.,][]{Jerkstrand_2012, Ashall_2024}. Additionally, the IR flux is significantly less affected by extinction from dust along the line of sight, allowing for more accurate spectroscopic modeling. Several strong C, O, Ne, and Na lines are located in the IR along with many Fe, Ni, and Co lines, which dominate at late times \citep[e.g.,][]{Kotak_2005, Kotak_2006, Jerkstrand_2012}. Modeling these IR emission lines provides deeper insight into the progenitor properties than optical modeling alone \citep[e.g.,][]{Woosley_1995, Jerkstrand_2012}. 

MIR observations of SNe offer more than access to unique emission lines; they are essential for tracing dust formation.
CC-SNe are predicted to contribute significantly to the large dust masses observed in early universe galaxies \citep{Dwek_1980, Wooden_1993, Hughes_1997, Fan_2003, Maiolino_2004, Schneider_2004, Dwek_2006, Gall_2011, Watson_2015}, as they form dust considerably faster than Asymptotic Giant Branch Stars, which dominate dust production in the local universe \citep{Ferrarotti_2006, Valiante_2009, Williams_2014, Maund_2017}. Both pre-existing circumstellar dust heated by the explosion and newly formed dust within the SN ejecta emit strongly at NIR and MIR wavelengths, making these regimes crucial for identifying and characterizing dust production in SNe \citep{Ercolano_2007, Kotak_2009, Wesson_2015, Szalai_2019, Sarangi22, Shahbandeh_2023}. 

Before the onset of dust formation, CC-SNe are expected to synthesize molecules of Carbon Monoxide (CO) and Silicon Monoxide (SiO) \citep{Liu_1992, Liu_1994, Clayton_2001, Biscaro_2014, Liljegren_2020}. Molecules enhance dust formation by acting as nucleation points onto which dust grains can condense \citep{Sluder_2018}. Additionally, molecules rapidly cool the ejecta to below the dust condensation point, enabling dust formation \citep{Liu_1995, Liljegren_2020}. This cooling happens through re-emission via the molecular ro-vibrational bands whose spectral signatures are only located at IR wavelengths, with the first overtone of CO ($\Delta v = 2$) located at $\sim 2.3 - 2.5\,$~\micron, and the CO and SiO fundamentals ($\Delta v = 1$) located at $\sim 4.3 - 5.2\,$~\micron\ and $\sim 7.5 - 9.3$~\micron, respectively \citep{Snow_1929, Singh_1975}. The location, formation rate, and total mass of these molecules are highly dependent on the initial mass of the progenitor and offer a unique probe to constrain the progenitor properties and explosion physics \citep{Woosley_2002, Muller_2016}. 

Several hydrogen-rich Type-II CC-SNe (SNe~II) have displayed molecular features. The CO first overtone was first detected in SN~1987A \citep{Catchpole_1987, Spyromilio_1988, Liu_1992, Wooden_1993} and later in other events via ground-based (GB) NIR spectroscopy \citep[e.g.,][]{Spyromilio_1996, Gerardy_2000, Pozzo_2007, Kotak_2005, Meikle_2011}. Limited MIR coverage has resulted in, with the exception of SN~1987A \citep{Wooden_1993}, only partial observations of the CO fundamental band \citep{Wooden_1993, Kotak_2005, Szalai_2011}. Finally, the SiO fundamental band, typically emerging several months after the CO \citep{Roche_1991}, has been identified in several late-time MIR spectra \citep{Wooden_1993, Kotak_2006, Kotak_2009, Szalai_2011, Szalai_2013}.

Once the ejecta is sufficiently cool, $T \leq 2000$~K, the formation of fresh dust begins \citep{Clayton_2001, Sarangi_2015, Sluder_2018}. Signatures of dust typically emerge as an IR excess observed one hundred days to a few years after the explosion \citep{Roche_1989}. A time series of NIR and MIR spectroscopy enables detailed tracking of the evolving IR excess and places constraints on key dust properties \citep[e.g., location, composition, and quantity;][]{Meikle_2007, Kotak_2009, Szalai_2013, Shahbandeh_2023}. While hot dust ($\mathrm{T_{dust}} \geq 800~\mathrm{K}$) can be observed with ground based NIR (GB-NIR) instruments \citep{Pozzo_2004, Tinyanont_2019b, Rho_2021, Ravi_2023}, cooler dust requires MIR observations \citep{Rho_2021, Shahbandeh_2023, Ravi_2023}. 

The extraordinary capabilities of \jwst\ have dramatically improved our ability to study molecule and dust formation in CC-SNe. \jwst\ can simultaneously cover the NIR and MIR regions with the Near-Infrared Spectrograph (NIRSpec, $1.6 - 5.2$~\micron) and Mid-Infrared Instrument (MIRI, $4.9 - 30.0$~\micron), allowing for comprehensive, simultaneous observations of the CO and SiO emission features, as well as detection of both hot and cool dust in nearby CC-SNe \citep[e.g.,][]{Shahbandeh_2023, Shahbandeh_2024, Zsiros_2024, Clayton_2025, Shahbandeh_2025}.

In this paper, we present \jwst\ spectroscopic and photometric observations of \ixf\ spanning approximately $+250$ to $+720$~d past explosion. The four epochs of spectroscopic data were obtained as part of our programs JWST-DD-4575 and JWST-GO-5290 \citep{Ashall2023_cycle2_23ixf, Ashall2024_cycle3_23ixf}, while the photometric data was acquired through our programs JWST-GO-3921 and JWST-GO-5290 \citep{Fox_2024, Ashall2024_cycle3_23ixf}. This paper is the second in a series investigating the evolution of \ixf's panchromatic observations. Here, we provide a detailed identification and analysis of emission lines associated with \ixf\ and compare the IR data of \ixf\ with those of previous SNe~II observed during the Spitzer Space Telescope (\textit{SST}) era and prior.

First, in Section ~\ref{sec:prev} we present a brief overview of previous work on \ixf. Followed by Section~\ref{sec:Obs_sec}, where we describe the panchromatic $0.32 -  30$~\micron\ photometric and spectroscopic observations and data reductions.
Then in Section~\ref{sec:SED}, we discuss the evolving panchromatic data of \ixf\  and in Section~\ref{sec:Line_spec} we examine the key spectral features in the full optical to MIR spectrum of \ixf. In Section~\ref{sec:Spec_comp_sec}, we compare the NIR and MIR spectra of \ixf\ to those of other SNe~II, enabling us to identify spectral differences. The evolving line velocities of several prominent species are discussed in Section~\ref{sec:Line_vels}. Followed by a discussion on the location of the dust in the ejecta in Section~\ref{sec:dust}. Finally, we present a summary of the results derived from the panchromatic observations of \ixf\ in Section~\ref{sec:Conc_sec}.

\section{Previous Studies on SN~2023ixf} \label{sec:prev}

\ixf, discovered on 2023 May 19.72 (MJD=60083.73; \citealt{Itagaki_2023}), was rapidly classified as a SN~II displaying prominent flash ionization features \citep{Perley_2023, Zimmerman_2023}. As one of the closest CC-SNe in the last decade, the discovery of \ixf\ triggered an unprecedented follow-up campaign across the electromagnetic spectrum ranging from $\gamma$-ray to radio, along with multiple multi-messenger searches \citep[e.g.,][]{Berger_2023, Bostroem_2023, Chandra_2023, Grefenstette_2023, Hiramatsu_2023, Jacobson-Gal_2023, Sarmah_2023, Singh_2024, Smith_2023, Stritzinger_2023, Teja_2023a, Yamanaka_2023, Zimmerman_2023, Guetta_2023,  Thwaites_2023, Marti-Devesa_2024, Panjkov_2024, Ravensburg_2024, Vandyk_2024, Iwata_2025, Abac_2025, Nayana_2025}, including \jwst\ observations.

Early-time spectra of \ixf\ revealed a blue continuum with prominent narrow ($v \lesssim 300$~\kms) emission lines from H and ionized He, C, and N, which persisted for several days \citep{Perley_2023, Smith_2023, Stritzinger_2023, Vasylyev_2023, Yamanaka_2023, Zimmerman_2023}. These features arise from the ejecta interacting with a dense, asymmetric CSM located $\sim (0.5 - 1) \times 10^{15}$~cm from the progenitor \citep{Bostroem_2023, Jacobson-Gal_2023, Vasylyev_2023, Nayana_2025}. The spectra of \ixf\ evolved over several months, exhibiting typical SNe~II characteristics with a $\sim 70$~d long hydrogen recombination plateau phase \citetext{\citealp{Zimmerman_2023, Singh_2024, Zheng_2025}}.

Pre-explosion multi-wavelength imaging at the site of \ixf\ indicates a red supergiant progenitor \citep[e.g.][]{Kilpatrick_2023, Jencson_2023, Niu_2023, Pledger_2023, Soraisam_2023, Ransome_2024}. Estimates of the progenitor's mass are bi-modal, with models suggesting either a low-mass progenitor (\mzams~$= 8 - 12$~\Msol; \citealp{Kilpatrick_2023, Pledger_2023, Vandyk_2024a, Bersten_2024, Neustadt_2024, Singh_2024, Xiang_2024}) or a more massive one (\mzams~$ = 17 - 22$~\Msol; \citealp{Jencson_2023, Liu_2023, Niu_2023, Soraisam_2023, Qin_2024}). Analysis of nebular-phase spectra did not decidedly confirm a low or high mass progenitor, with mass estimates between \mzams$ = 12 - 15$~\Msol\ \citep{Ferrari_2024, Folatelli_2025, Kumar_2025, Michel_2025}.

Pre-explosion imaging of \ixf\ reveals clear evidence of enhanced mass loss in the final years before explosion \citep{Jencson_2023, Soraisam_2023, Flinner_2023, Zhang_2023, Panjkov_2024, Qin_2024, Xiang_2024}. While mass-loss rates inferred from imaging were modest ($10^{-6} - 10^{-4}$ \Msol$\mathrm{yr}^{-1}$), early-time light curve and spectral modeling suggests a significantly higher rate ($10^{-3} - 10^{-2}$ \Msol $\mathrm{yr^{-1}}$), implying a major mass-loss episode shortly before explosion \citep{Bostroem_2023, Jacobson-Gal_2023, Teja_2023a}. Pre-explosion observations from the \textit{SST} show a strong IR signal from the progenitor, consistent with heavy obscuration by dust surrounding the progenitor \citep{Szalai_2023, Soraisam_2023, Vandyk_2024a}.

The proximity of \ixf\, at $\sim 6.9 \pm 0.1$ Mpc \citep{Riess_2022}, makes it an ideal candidate for \jwst\ follow-up. An initial \jwst\ observation of \ixf\ was obtained during the plateau phase at $+33.6$~d post explosion under our project JWST-DD-4722 \citep{Ashall2023_cycle1_23ixf}, and is presented in the first paper of this series \citep[][hereafter, \Pone]{DerKacy_2025}. These observations showed no signs of molecules or dust. 

\begin{figure*}
    \centering
    \includegraphics[width=\linewidth, trim={0 0cm 0 0}, clip=True]{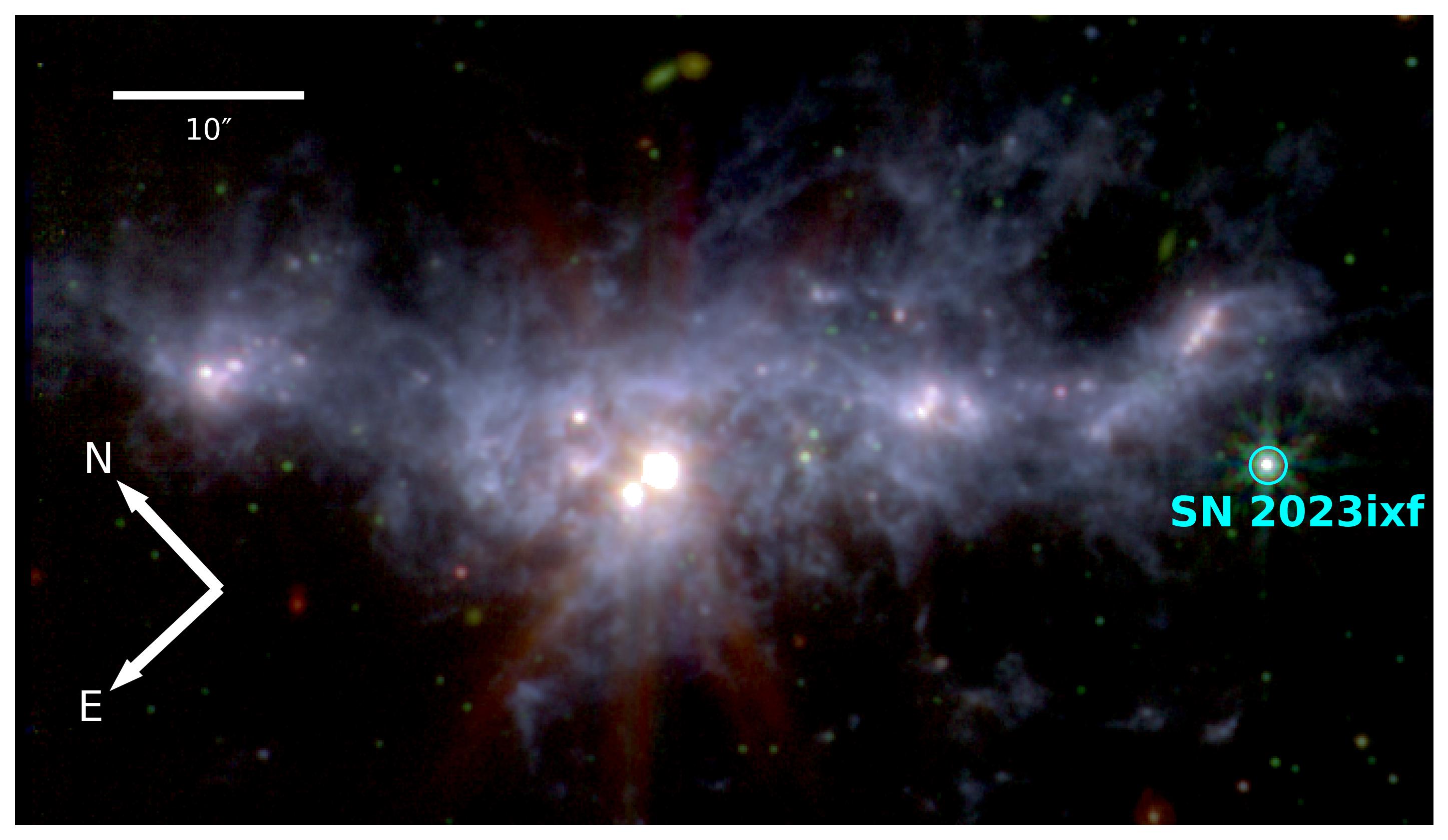}
    \caption{A stacked three-color image of the host-galaxy spiral arm using the \emph{F770W}, \emph{F1000W}, and \emph{F1500W} MIRI filters from epoch 301.72~d. The location of \ixf\ is highlighted by the light blue circle.} \label{fig:JWST_Stacked}
\end{figure*}

\begin{table}
    \centering
    \caption{Basic observational parameters of \ixf
    \label{ixf_details} }
    \begin{tabular}{ccc}
    \hline
    \hline
    Parameter & Value & Source \\
    \hline
    Right Ascension & 14$^{h}$03$^{m}$38$^{s}$.56 & (1)  \\
    Declination & $+$54\arcdeg18\arcmin41\arcsec.94 & (1) \\
    $T_{\rm exp}$ (MJD) & $60082.75$ & (2) \\
    $V_{\rm max}$ (mag) & $\sim-18.1$ & (3, 4, 5) \\
    $z$ & 0.0008 & (6) \\
    Distance [Mpc] & $6.9 \pm 0.1$ & (6) \\
    $\mu$ [Mpc] & $ 29.18 \pm 0.04 $ & (6) \\ 
    $E$(B-V)$_{MW}$ [mag] & $0.0077 \pm 0.0002$ & (7) \\
    $E$(B-V)$_{Host}$ [mag]& $0.031 \pm 0.012$ & (8) \\
    \hline
    \end{tabular}
    \tablerefs{(1) TNS, (2) \cite{Hosseinzadeh_2023}, (3) \cite{Teja_2023a}, (4) \cite{Jacobson-Gal_2023}, (5) \cite{Zimmerman_2023}, (6) \cite{Riess_2022}, (7) \cite{Schlafly_2011}, (8) \cite{Smith_2023}}
\end{table}

\section{Observations and data reductions} \label{sec:Obs_sec}

In this section, we describe the \jwst\ and ground-based observations, and the data reduction methods. The key properties of \ixf, such as distance, extinction, and explosion date, adopted throughout this work are listed in Table~\ref{ixf_details}. All phases discussed here after are reported in rest-frame days relative to the date of explosion provided in Table~\ref{ixf_details}.  

\subsection{JWST MIRI Imaging} \label{sec:imaging}

We obtained three epochs of \jwst\ MIRI imaging of \ixf\ through two of our programs: JWST-GO-3921 \citep{Fox_2024} and JWST-GO-5290 \citep{Ashall2024_cycle3_23ixf}. The first epoch, acquired by JWST-GO-3921 at $+301.72$~d post-explosion, includes imaging in the \emph{F770W}, \emph{F1000W}, \emph{F1500W}, and \emph{F2100W} filters. The second and third epochs, obtained by JWST-GO-5290 at $+600.21$ and $+719.96$~d, use the same instrument with the \emph{F1500W}, \emph{F1800W}, \emph{F2100W}, and \emph{F2500W} filters. A stacked color image from the first epoch is presented in Fig.~\ref{fig:JWST_Stacked}.
\setcounter{footnote}{0} 
We perform point spread function (PSF) photometry using a custom Python notebook built on the \textit{WebbPSF}-based \texttt{space\_phot} package \citep{Perrin14}\footnote{\url{https://github.com/orifox/psf_phot/blob/main/space_phot/MIRI/miri_1028.ipynb}}. We select the PSF width to minimize residuals in the fit, accounting for the expected wavelength-dependent broadening. For each dither in which \ixf\ is detected, we average the individual flux measurements and estimate the photometric uncertainty from the standard deviation across the dithers. The AB magnitudes and corresponding flux densities (in mJy) for all filters and epochs are summarized in Table~\ref{tab:JWST_phot}.

\renewcommand{\tabcolsep}{4pt}
\begin{table*}[]
    \centering
    \caption{Observation log of \jwst\ photometric imaging taken through our program JWST-GO-3921 \citep{Fox_2024} at $+~301.72$~d and our program JWST-GO-5290 \citep{Ashall2024_cycle3_23ixf} at $+~600.21$ and $+~719.96$~d. The epochs are given in days from explosion. Magnitudes and fluxes are not corrected for extinction}\label{tab:JWST_phot}
    \begin{tabularx}{\textwidth}{cccccccc}
    \hline
    \hline
    & Epoch  & \textit{F770W} & \textit{F1000W} & \textit{F1500W} & \textit{F1800W} & \textit{F2100W} & \textit{F2500W} \\
    \hline
    \multirow{3}{*}{AB Mag. [mag]} & 301.72 & 14.803 $ \pm $ 0.013 & 14.51 $ \pm $ 0.008 & 15.023 $ \pm $ 0.005 & - & 15.346 $ \pm $ 0.002 & - \\
    & 600.21 & - & - & 15.214 $ \pm $ 0.001 & 15.180 $ \pm $ 0.002 & 15.274 $ \pm $ 0.002 & 15.351 $ \pm $ 0.004 \\
    & 719.96 & - & - & 15.269 $ \pm $ 0.001 & 14.601 $ \pm $ 0.001 & 15.283 $ \pm $ 0.003 & 15.348 $ \pm $ 0.004 \\
    \hline
    \multirow{3}{*}{Flux Density [mJy] } & 301.72 & 4.353  $\pm$  0.052 & 5.702  $\pm$  0.042 & 3.555  $\pm$  0.016 & - & 2.64  $\pm$  0.005 & - \\
    & 600.21 & - & - & 2.983  $\pm$  0.003 & 3.076  $\pm$  0.005 & 2.822  $\pm$  0.004 & 2.627  $\pm$  0.01 \\
    & 719.96 & - & - & 2.834  $\pm$  0.003 & 5.222  $\pm$  0.002 & 2.798  $\pm$  0.007 & 2.635  $\pm$  0.01 \\ 
    \hline
    \end{tabularx}%
\end{table*}

\subsection{JWST Spectroscopy} \label{sec:JWST_obs}
We obtained \jwst\ spectra of \ixf\ on 2024-01-26, 2024-05-26, 2025-01-08, and 2025-05-02 using NIRSpec \citep{Jakobsen_2022, Boker_2023} and MIRI in Low Resolution Spectroscopy mode \citep[MIRI/LRS;][]{Kendrew_2015}. We reduced all spectra using the standard NIRSpec fixed-slit and MIRI LRS-slit Jupyter notebooks \citep{law_2025}, with version 1.18.0 of the \jwst\ calibration pipeline \citep{Bushouse_2022_JWST_reduc}, \jwst\ Build version 11.3, and Calibration Reference Data System file \texttt{jwst\_1364.pmap}.

\subsubsection{NIRSpec} \label{sec:NIRSpec_redu}
The NIRSpec observations used the F170LP/G235M and F290LP/G395M filter/grating combinations, covering the $1.66$--$3.07$~\micron\ and $2.87$--$5.10$~\micron\ ranges, respectively. The observations employed the S400A1 subarray and a 3-Point-Nod dither pattern. Exposure times were increased over time to account for the fading SN flux, ensuring a high enough signal-to-noise ratio (SNR) to trace the evolution of the CO first overtone, fundamental band, and other spectral features. Full details of the NIRSpec configuration are provided in Appendix~\ref{app:A}.

We modified the standard reduction parameters to include a scaling factor in the \texttt{outlier\_detection} step during stage 3 spectroscopic processing. This scale value varies between $0$ and $1.2$ to minimize noise spikes, caused by bad pixels when combining the three dithers, while preserving the source flux. All reduction stages; stage 1 detector processing, stage 2 image processing, stage 2 spectroscopic processing, and stage 3 spectroscopic processing, were run on uncalibrated NIRSpec images downloaded from the Barbara A. Mikulski Archive for Space Telescopes (MAST\footnote{\url{https://mast.stsci.edu/portal/Mashup/Clients/Mast/Portal.html}}).

\subsubsection{ MIRI-LRS } \label{sec:MIRI_redu}
The MIRI observations utilized the LRS mode, covering a wavelength range of $\sim5$--$14$~\micron\ \citep{Kendrew_2015}. All epochs were taken using the full subarray, the FASTR1 readout pattern, and the Along-Slit-Nod dither pattern. Exposure times were increased to account for the declining flux of \ixf. Details on the exposure times and readout settings for each epoch are provided in Appendix~\ref{app:A}. We use the standard reduction parameters to run all processing stages on the uncalibrated MIRI-LRS images downloaded from MAST.

\begin{figure*}
    \centering
    \begin{tabular}{c}
        \includegraphics[width=\linewidth]{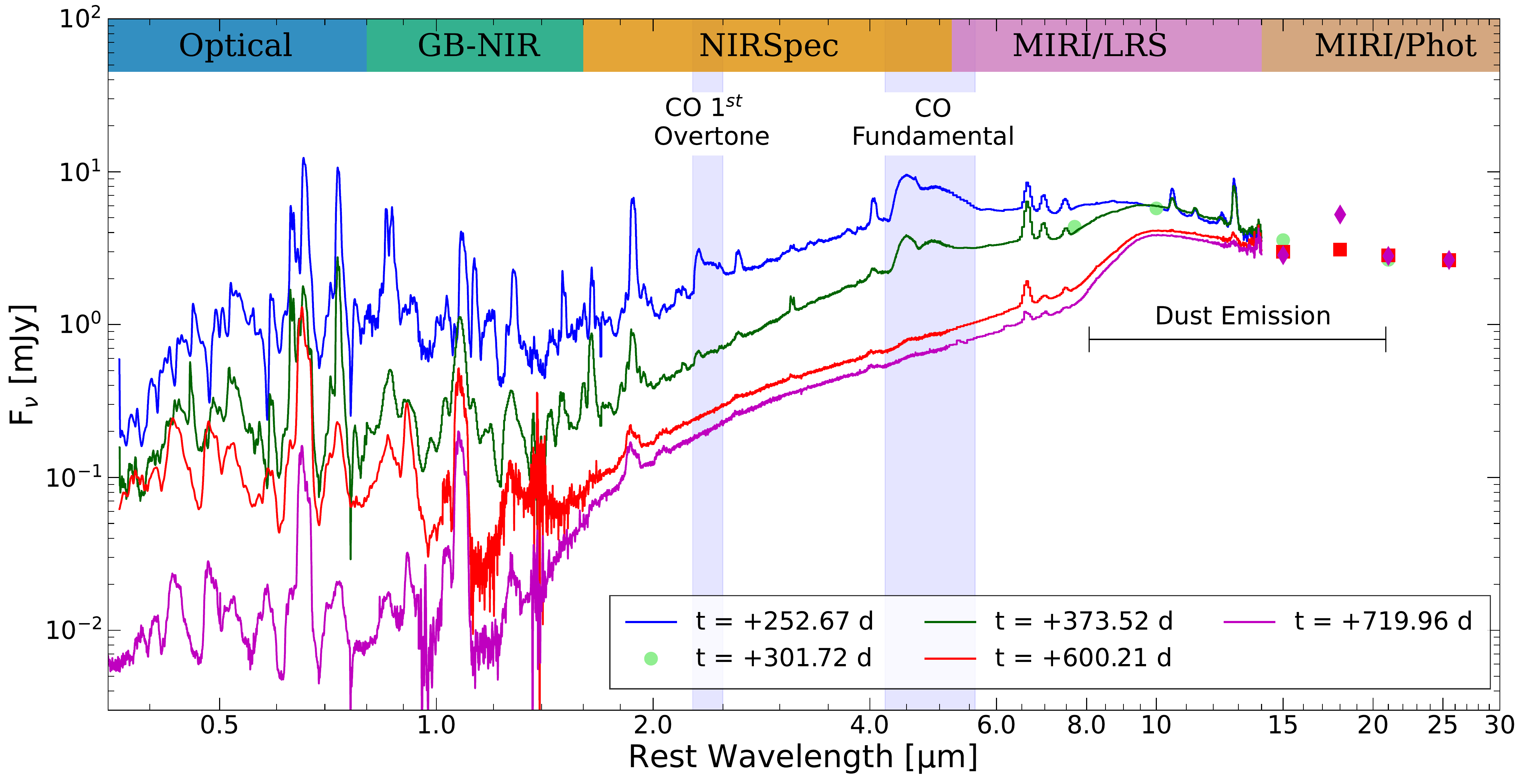}
    \end{tabular}
    \caption{The evolution of the SED covering $0.32 -  14.0$~\micron\ of \ixf\ from $+252.67$~d to $+719.96$~d post explosion. Photometry is shown by the colored symbols, at $+301.72$~d, $+600.21$~d, and $+719.96$~d. The data have been corrected for extinction and are presented in the rest frame. The SED's are dominated by a large number of emission lines that decline in strength as \ixf\ evolves. CO molecule and dust features are highlighted. In addition, there is a strong IR excess that grows in strength relative to the SN flux at late times. }
    \label{fig:SED_plots}
\end{figure*}

\begin{figure*}
    \centering
    \includegraphics[width=0.95\linewidth]{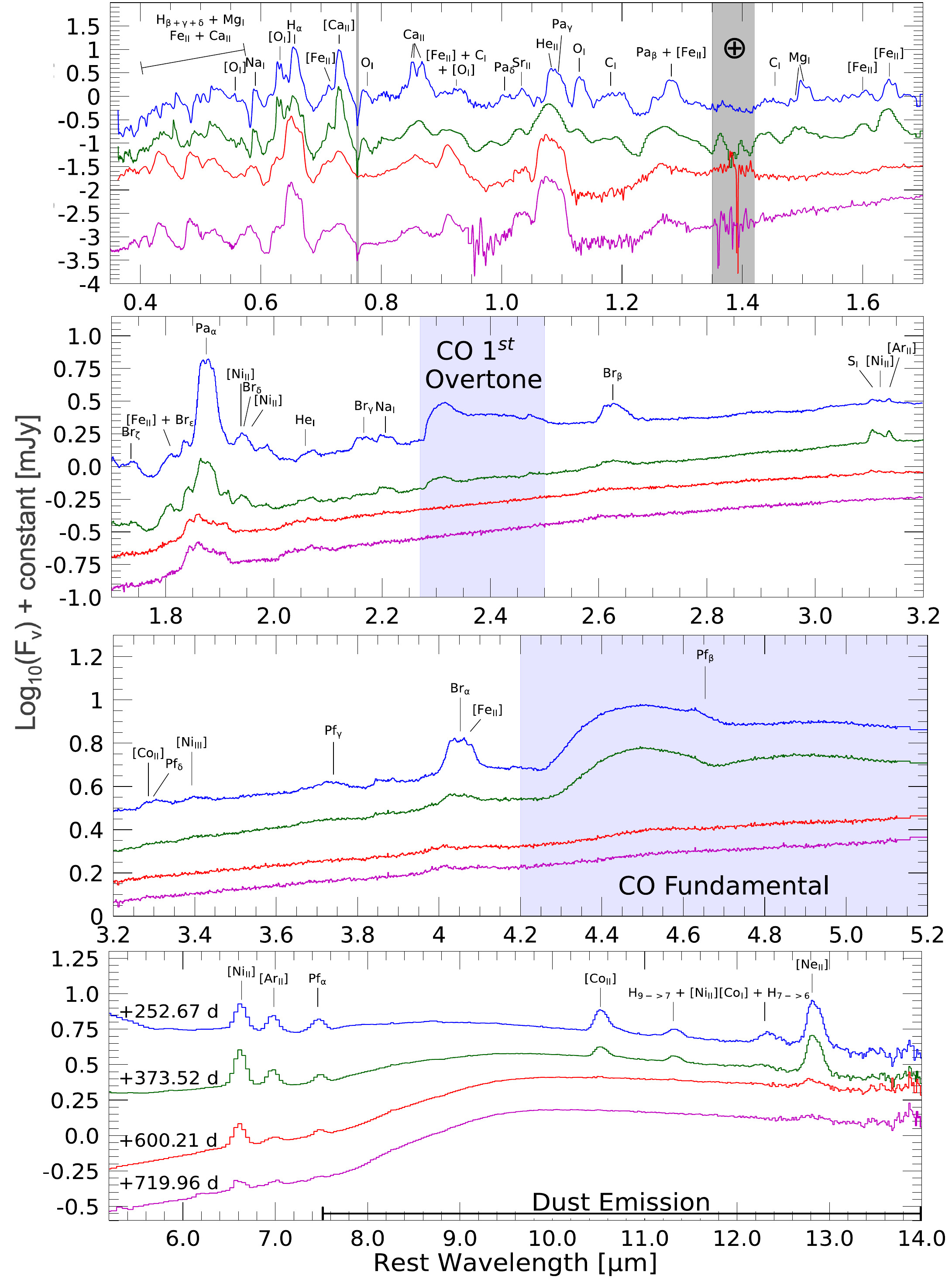}
    \caption{Lines identified within the ground based optical and NIR (top row), \jwst\ NIRSpec G235M/F170LP (second row), G395/F290LP (third row) and MIRI (fourth row) spectra of \ixf\ at $\sim 250$~d (blue),  $\sim 373$~d (green), $\sim 600$~d (red), and $\sim 720$~d (purple) post explosion. The spectra have been vertically shifted for clarity. The CO $\mathrm{1^{st}}$ overtone, and CO fundamental band are shown by the shaded blue regions. The full list of identified lines is given in Tables \ref{tab:GB_line_ids} and \ref{tab:JWST_line_ids}. }
    \label{fig:JWST_IDs}
\end{figure*}

\subsection{Ground-Based Spectroscopy} \label{sec:groundbasespec}

The ground based optical and NIR observations were obtained as close to contemporaneous with the \jwst\ observations as possible to ensure no significant evolution of the emission features between the observations. Details on the optical and ground-based NIR observations, and the full spectra obtained are presented in Appendix~\ref{app:A}, and Figures \ref{app:All_opt_spec} and \ref{app:All_NIR_spec} respectively.

\subsubsection{Optical} \label{sec:opt_redu}
Optical spectra were obtained at similar epochs to the \jwst\ spectral observations using the Optical System for Imaging and low-Intermediate-Resolution Integrated Spectroscopy (OSIRIS+) on the 10.4-m Gran Telescopio Canarias \citep[GTC;][]{Alverez_1998}. 
We reduce the GTC data using standard procedures, including bias subtraction, flat-fielding of the two-dimensional images, wavelength calibration of the extracted spectrum from arc lamp exposures, and the removal and correction of telluric features and cosmic rays. Multiple exposures were obtained in the R1000B and R1000R filters and median-combined to increase the signal-to-noise of the optical spectra. An additional optical spectrum was obtained using the $10$-m Keck-I telescope with the Low-Resolution Imaging Spectrometer \citep[LRIS;][]{Oke_1995, McCarthy_1998, Rockosi_2010} to complement the $+713$-d \jwst\ spectra. The LRIS observations used the 600/4000 grism on the blue side and the 400/8500 grating on the red side. Multiple images were obtained with exposure times of $1800$~s and $900$~s for the blue and red arms. We reduce the individual arms separately using the automated spectroscopic reduction code \texttt{Pypeit}\footnote{\url{https://pypeit.readthedocs.io/en/1.17.3/index.html}}, version 1.17.3.  

Optical spectra, covering $72 - 279$~d, were taken on the University of Hawai\okina i 88-inch telescope (UH88) located on Maunakea using the SuperNova Integral Field Spectrograph \citep[SNIFS:][]{Aldering_2002, Lantz_2004} through the Spectroscopic Classification of Astronomical Transients \citep[SCAT;][]{Tucker_2022} survey, as well as on the Intermediate Dispersion Spectrograph (IDS\footnote{\url{https://www.ing.iac.es/Astronomy/telescopes/int/}}) mounted on the Isaac Newton Telescope (INT). We reduce the SNIFS spectra using the procedure detailed in \cite{Tucker_2022}, and the IDS, spectra via the standard reduction pipeline \citep{muller_bravo_2023}.

\subsubsection{Near Infrared} \label{sec:GB_NIR_redu}
NIR spectra covering $\sim0.7-2.5$~\micron\ were obtained by the Hawai\okina i Infrared Supernova Study \citep{Medler25_HISS} using SpeX \citep{Rayner_2003} on the 3.0-m NASA Infrared Telescope Facility (IRTF) and the Near-Infrared Echellette Spectrometer \citep[NIRES;][]{Wilson_2004} on the $10$-m Keck-II telescope. 

The IRTF spectra were taken in ShortXD mode, with a slit size of $0.8\arcsec \times 15\arcsec$ using ABBA dithering. We reduce these observations with the \texttt{Spextool} software package \citep{Cushing_2004} applying telluric corrections that use an AV0 star that was at a similar airmass to \ixf. More details on the reduction procedure can be found in \citet{Medler25_HISS, Hoogendam25_epr} and \citet{Hoogendam25_pxl}. 

The NIRES spectra were also taken using ABBA dither patterns with 300s individual exposure times. We reduce these Keck-II/NIRES observations in a similar manner to the IRTF/SpeX observations, using the Keck-II/NIRES version of the \texttt{Spextool} package \citep[for more information, see][]{Medler25_HISS}.

An additional spectrum was obtained with GTC using the Espectrógrafo Multiobjeto Infra-Rojo spectroscope \citep[EMIR;][]{Grazon_2022}. We reduce the EMIR observations with our own pipeline based on PyEmir \citep{Pascual_2010, Cardiel_2019, Pascual_2019}.

\section{Spectral Energy distribution} \label{sec:SED}

Figure~\ref{fig:SED_plots} presents the $0.32 -  30.0$~\micron\ spectral energy distribution (SED) of \ixf, spanning from $+252.67$ to $+719.96$~d. The SED's remain relatively flat between $0.32$ and $1.2$~\micron\ and are dominated by forbidden line transitions, which evolve in both strength and velocity as different regions of the ejecta become exposed. 

During the first two epochs, the SEDs show strong emission from the first overtone and fundamental ro-vibrational bands of CO. These emission features weaken substantially by $+600.21$~d indicating a decrease in the temperature of the CO region to below $1000$~K \citep{Liljegren_2020}. Future papers in this series will discuss the modeling of the CO features and derived CO properties in \ixf.

Across the IR regime above $1.5$~\micron, the flux continuum rises to a peak at $\sim\!10.0$~\micron\ in the $+252.67$~d spectrum indicating the presence of warm dust. As \ixf\ evolves the IR continuum flux declines in strength alongside the optical, suggesting that the dust mass does not significantly increase during these epochs. By $+600.21$~d, the $10.0$~\micron\ peak dominates the MIR spectrum over any emission feature, while a secondary peak at $18.0$~\micron\ emerges by $+719.96$~d. These two MIR features suggest a significant contribution to the flux from Si-rich dust at these later epochs \citep{Chiar_2006, VanBreemen_2011, Shahbandeh_2023}, see section \ref{sec:dust} for further discussion.

\section{Identifications of Emission lines} \label{sec:Line_spec}

The identification of spectral features in the late-time spectra of \ixf\ provides critical insight into the structure and composition of the ejecta. The dominant features in the $+252.67$ optical and ground-based NIR SED's of \ixf\ include several strong H emission lines, primarily from low-level transitions, as well as prominent O, Mg, and Ca features, as shown in the top panel of Fig.~\ref{fig:JWST_IDs}. Iron-group elements such as Fe, Ni, and Co, along with other $\alpha$ elements, dominate the emission features in the \jwst\ NIR and MIR regions. As \ixf\ evolves, the strength of all emission lines gradually declines as the SN component of the SED fades. The MIR emission lines become increasingly suppressed due to the growing contribution from the IR excess which dominates the SED's, particularly at $t \geq +600$~d. 

Below, we highlight the strongest lines of key elemental species identified in the nebular-phase spectra from both ground-based and \jwst\ observations. To support the analysis, we compile spectral lines from previous studies of late-time SNe~II spanning the optical, NIR, and MIR regimes \citep[e.g.,][]{Jerkstrand_2012, Davis_2019, Shahbandeh_2022}. The identified spectral lines appear in Tables~\ref{tab:GB_line_ids} and \ref{tab:JWST_line_ids}.

\subsection{Hydrogen} \label{sec:H_lines}
Throughout the evolution of \ixf, we identify strong hydrogen emission lines from multiple transition series across the optical to MIR spectrum, with several lines displaying strong blending. The observed H lines include: Balmer series ($n = 2$): H$\alpha~0.6563$~\micron, H$\beta~0.4861$~\micron, H$\gamma~0.4340$~\micron, and H$\delta~0.4102$~\micron; Paschen series ($n = 3$): Pa$\alpha~1.875$~\micron, Pa$\beta~1.282$~\micron, Pa$\gamma~1.094$~\micron, and Pa$\delta~1.005$~\micron; Brackett series ($n = 4$): Br$\alpha~4.051$~\micron, Br$\beta~2.626$~\micron, Br$\gamma~2.166$~\micron, Br$\delta~1.944$~\micron, and Br$\varepsilon~1.817$~\micron; Pfund series ($n = 5$): Pf$\alpha~7.460$~\micron, Pf$\beta~4.654$~\micron, Pf$\gamma~3.741$~\micron, and Pf$\delta~3.297$~\micron; Humphreys series ($n = 6$): Hu$\alpha~12.37$~\micron\ and Hu$\beta~7.503$~\micron. We also identify potential emission lines from higher-order transitions, such as H$_{n = 9 \rightarrow 7}~11.309$~\micron\ and H$_{n = 11 \rightarrow 8}~12.372$~\micron. 

The emission peaks from the higher-order Balmer series are absent between $+252.67 - +373.52$~d as these lines lie within absorption features dominated by heavy elements, particularly Fe, but emerge between $+373.52$~d and $+600.21$~d. The $\alpha$ through $\varepsilon$ transitions of the Paschen, Brackett, Pfund, and Humphreys series are detected in the $+252.67$ and $+373.53$~d spectra. However, at later epochs ($+600.21$ and $+719.96$~d), the higher-order transitions, especially the $\delta$ and $\varepsilon$ lines, become too weak to detect.

The hydrogen lines evolve from the standard P-Cygni profile observed at earlier epochs \citepalias{DerKacy_2025} and exhibit three distinct components centered around the rest-frame emission wavelength. The first is a strongly blueshifted peak; the second is a weak, narrow peak located slightly redward of the rest wavelength; and the third is a broad red shoulder. We discuss the origin of these features in detail in Section~\ref{sec:Line_vels}.

\subsection{Helium} \label{sec:He_lines}
In the optical, He lines are difficult to identify due to significant line blending and contamination from other emission features, particularly those from H, Fe, and Na. However, we tentatively identify several weak features commonly associated with He, including \HeI~$0.5876$~\micron, \HeI~$0.6678$~\micron, and \HeI~$0.7065$~\micron. In the NIR we identify the strong \HeI~$1.083$~\micron and \HeI~$2.058$~\micron.

\subsection{Carbon} \label{sec:C_lines}
We identify multiple carbon features, all from neutral C, at \CI~$0.94$~\micron, $1.176$~\micron, $1.181$~\micron, and ~$1.454$~\micron. While the \CI~$0.94$~\micron\ line blends with nearby O and Fe lines, the other NIR lines are either isolated or blended only with adjacent C lines. Spectral models of other late-time SNe~II \citep{Jerkstrand_2012} also identify these features which, together with the prominent CO emission observed in the spectra, supporting the identification of C in \ixf.

\subsection{Oxygen} \label{sec:O_lines}
Throughout the optical and NIR spectra, we identify several emission features associated with oxygen. The strongest O lines in the spectra of \ixf\ are the forbidden optical doublet [\OI]~$0.6300$ and $0.6363$~\micron. In addition to this doublet, we detect several other forbidden and permitted O lines, including [\OI]~$0.5577$~\micron, the \OI~$0.777$~\micron\ triplet, [\OI]~$0.9266$~\micron, and \OI~$1.1290$~\micron. The strength of all O features declines over time, with the NIR lines becoming undetectable between the $+373.52$ and $+600.21$~d spectra. We discuss the evolution of the [\OI]~$0.6300$ and $0.6363$~\micron\ feature in detail in Section~\ref{sec:Line_vels}.

\subsection{Neon} \label{sec:Ne_lines}
A prominent Ne line, associated with the [\NeII]~$12.813$~\micron\ transition, appears in the nebular phase MIRI-LRS spectra of \ixf. No other \NeII\ lines or other excitation states of Ne are identified in the spectra of \ixf. This line has also been detected in several MIR observations of other SNe~II \citep{Kotak_2006, Meikle_2007, Meikle_2011}. 

\subsection{Argon} \label{sec:Ar_lines}
We identify a strong emission feature associated with the [\ArII]~$6.985$~\micron\ line. Other lines of Ar, such as [\ArIII]~$8.991$~\micron, are not detected in the spectrum, which may indicate that a significant fraction of both noble gases, including Ar and Ne, remain in a neutral state. A similar phenomenon may have also been present in SN~2005af \citep{Kotak_2006}.

\subsection{Calcium} \label{sec:Ca_lines}
The spectra of \ixf\ show multiple Ca lines from both forbidden and permitted transitions. These include the permitted \CaII\ NIR triplet at $0.8498$, $0.8542$, and $0.8662$~\micron, the \CaII\ H\&K lines at $0.3934$ and $0.3969$~\micron, and the forbidden [\CaII]~$0.7292$ and $0.7324$~\micron\ lines. As \ixf\ evolves, the permitted Ca lines decrease in strength, while the forbidden [\CaII] lines grow more prominent relative to the continuum, consistent with a declining ejecta density.

\subsection{Iron} \label{sec:Fe_lines}
Strong Fe lines appear throughout the late-time optical-MIR spectra of \ixf. A blend of several \FeII\ lines are located between $\sim 0.32 -  0.5$~\micron, including \FeII~$0.4352$~\micron, \FeII~$0.4924$~\micron, \FeII~$0.5018$~\micron, and \FeII~$0.5169$~\micron. Additional forbidden and allowed Fe lines are scattered throughout the optical and IR regimes, including [\FeII]~$0.7155$~\micron, [\FeII]~$1.279$~\micron, [\FeII]~$1.321$~\micron, \FeII~$1.644$~\micron, [\FeII]~$1.809$~\micron, \FeII~$1.939$~\micron, \FeII~$3.108$~\micron, \FeII~$3.137$~\micron, and [\FeII]~$4.075$~\micron. We also tentatively identify several weak emission features associated with neutral Fe throughout the spectra of \ixf, including \FeI~$0.5270$~\micron, \FeI~$0.5328$~\micron, \FeI~$0.5587$~\micron, \FeI~$0.7187$~\micron, and \FeI~$1.980$~\micron.

\subsection{Cobalt} \label{sec:Co_lines}
We identify several forbidden Co lines in the spectra of \ixf, including [\CoII]~$1.019$~\micron, [\CoII]~$3.286$~\micron, [\CoII]~$10.521$~\micron, and [\CoI]~$12.255$~\micron.
We also observe weak features associated with the second ionization state of Co, such as [\CoIII]~$11.886$~\micron\ and [\CoIII]~$13.813$~\micron. However, these features are either very weak relative to the continuum or located near the edge of the instrument's spectral range where throughput decreases, obfuscating their exact strength and evolution.

\subsection{Nickel} \label{sec:Ni_lines}
In addition to the forbidden Co lines, we identify several forbidden lines from neutral and first ionization state Ni. These include: \Nineb~$3.120$~\micron, \Nineb~$11.308$~\micron, [\NiII]~$1.939$~\micron, [\NiII]~$1.958$~\micron, [\NiII]~$6.636$~\micron, and [\NiII]~$10.682$~\micron.

\subsection{Other Species} \label{sec:Other_lines}
We also detect several emission features associated with other elements  including: Na (\NaI~$0.5896$~\micron, \NaI~$2.206$~\micron), Mg (\MgI~$0.3832$~\micron, \MgI~$0.457$~\micron, \MgI~$0.518$~\micron, \MgI~$0.517$~\micron, \MgI~$1.488$~\micron, and \MgI~$1.502$~\micron), Si (\SiI~$1.203$~\micron, \SiI~$1.033$~\micron, \SiI~$1.066$~\micron, [\SiI]~$1.606$~\micron, and [\SiI]~$1.645$~\micron), S (\SI~$3.107$~\micron), Ba (\BaII~$0.6142$~\micron)  and Sr (\SrII~$1.004$~\micron, \SrII~$1.033$~\micron, and \SrII~$1.092$~\micron). These lines blend with other emission features, forming shoulders or broad peaks centered between the blended lines.

\begin{deluxetable}{ccccc}
    \caption{Identified nebular phase spectral lines present in the optical and ground-based NIR spectra}
    \label{tab:GB_line_ids}
    \tablehead{\colhead{Ion} & \colhead{Wavelength} & \colhead{Ion} & \colhead{Wavelength} \\
         & \colhead{[\mic]} & & \colhead{[\mic]} }
    \startdata
    \multicolumn{4}{c}{Optical spectrum, 0.35 - 1.0 \micron} \\
    \hline
    \MgI	&	0.38	&	\FeI	&	0.5587	\\
    \CaII	&	0.3934	&	\NaI	&	0.5896	\\
    \CaII	&	0.3969	&	\BaII	&	0.6142	\\
    H$\delta$	&	0.410.2	&	$[$\OI$]$	&	0.6300	\\
    H$\gamma$	&	0.4340	&	$[$\OI$]$	&	0.6363	\\
    \FeII	&	0.4352	&	$\alpha$	&	0.6563	\\
    \MgI]	&	0.457	&	[\FeII]	&	0.7155	\\
    H$\beta$	&	0.4861	&	\FeI	&	0.7187	\\
    \FeII	&	0.4924	&	[\CaII]	&	0.7292	\\
    \FeII	&	0.5018	&	[\CaII]	&	0.7324	\\
    \FeII	&	0.5169	&	\OI	&	0.777	\\
    \MgI	&	0.517	&	\CaII	&	0.8498	\\
    \MgI	&	0.518	&	\CaII	&	0.8542	\\
    \MgI	&	0.518	&	\CaII	&	0.8662	\\
    \FeI	&	0.5270	&	[\OI]	&	0.9266	\\
    \FeI	&	0.5328	&	\CI	&	0.94	\\
    $[$\OI$]$	&	0.5577					\\
    \hline
    \multicolumn{4}{c}{GB-NIR spectrum, 1.0 - 1.6 \micron} \\
    \hline
    \SrII	&	1.004	&	\SiI	&	1.203	\\
    Pa$\delta$	&	1.005	&	[\FeII]	&	1.279	\\
    $[$\CoII$]$	&	1.019	&	Pa$\beta$	&	1.282	\\
    \SiI	&	1.033	&	[\FeII]	&	1.321	\\
    \SrII	&	1.033	&	\CI	&	1.454	\\
    \SiI	&	1.066	&	\MgI	&	1.488	\\
    \HeI	&	1.0830	&	\MgI	&	1.502	\\
    \SrII	&	1.092	&	[\SiI]	&	1.606	\\
    Pa$\gamma$	&	1.094	&	\FeII	&	1.644	\\
    \OI	&	1.1290	&	[\SiI]	&	1.645	\\
    \CI	&	1.181	&				\\
    \enddata
\end{deluxetable}

\begin{deluxetable}{ccccc}
    \caption{Identified nebular phase spectral lines present in the \jwst\ NIR and MIR spectra}
    \label{tab:JWST_line_ids}
    \tablehead{\colhead{Ion} & \colhead{Wavelength} & \colhead{Ion} & \colhead{Wavelength} \\
         & \colhead{[\mic]} & & \colhead{[\mic]} }
    \startdata    
    \multicolumn{4}{c}{NIRSpec spectrum, 1.6 - 5.2 \micron} \\
    \hline
    \HeI	&	1.7002	&	Br$\beta$	&	2.626	\\
    \CI	&	1.176	&	\SI	&	3.107	\\
    $[$\FeII$]$	&	1.809	&	\FeII	&	3.108	\\
    Br$\varepsilon$	&	1.817	&	\Nineb	&	3.120	\\
    Pa$\alpha$	&	1.875	&	\ArII	&	3.137	\\
    \FeII	&	1.939	&	\FeII	&	3.137	\\
    $[$\NiII$]$	&	1.939	&	Pf$\delta$	&	3.297	\\
    Br$\delta$	&	1.944	&	$[$\CoII$]$	&	3.286	\\
    $[$\NiII$]$	&	1.958	&	Pf$\gamma$	&	3.741	\\
    \FeI	&	1.980	&	Br$\alpha$	&	4.051	\\
    \HeI	&	2.0581	&	$[$\FeII$]$	&	4.075	\\
    Br$\gamma$	&	2.166	&	Pf$\beta$	&	4.654	\\
    \NaI	&	2.206	&				\\
    \hline
    \multicolumn{4}{c}{MIRI spectrum, 5.2 - 14 \micron} \\
    \hline
    $[$\NiII$]$	&	6.636	&	H$_{n=9\rightarrow7}$	&	11.309	\\
    $[$\ArII$]$	&	6.985	&	$[$\CoIII$]$	&	11.886	\\
    Pf$\alpha$	&	7.460	&	$[$\CoI$]$	&	12.255	\\
    Hu$\beta$	&	7.503	&	Hu$\alpha$	&	12.37	\\
    $[$\CoII$]$	&	10.521	&	H$_{n=11\rightarrow8}$	&	12.372	\\
    $[$\NiII$]$	&	10.682	&	$[$\NeII$]$	&	12.813	\\
    \Nineb	&	11.308	&	$[$\CoIII$]$	&	13.813	
    \enddata
\end{deluxetable}

\subsection{Molecules} \label{sec:mols}
Several molecular emission features are observed in the NIR/MIR spectra of \ixf\ which were not present in the initial \jwst\ data \citepalias{DerKacy_2025}. Here, we examine the evolution of molecules in \ixf\ as it transitions from the SN-dominated phase to the dust-dominated phase, while deferring the spectroscopic modeling of \ixf's molecular features to a future paper in the series.

Two strong features appear in the spectra of \ixf\ corresponding to the first overtone and fundamental ro-vibrational bands of CO. Ground-based NIR observations show the emergence of the CO first overtone between $+81$~d \citep{Li_2025} and $+200$~d \citep{park_2025}. In the \jwst\ observations, these features dominate the NIR spectrum of \ixf\ from $+252.67$ to $+373.13$~d, before fading over the following $\sim 230$~d, see the CO regions in Fig.~\ref{fig:SED_plots}. By $+600.21$~d, the CO first overtone completely fades below the continuum level. However, the CO fundamental band remains visible even at $+717.13$~d, where the broad $\nu = 1 \rightarrow 0$ band head is still detectable above the continuum flux.

Throughout the evolution of \ixf, both the CO first overtone and fundamental possess distinct broad R- and P-branch emission features, in contrast to the spiky nature of the CO features commonly observed in young stellar objects \citep[e.g.][]{Ilee_2018, Fedriani_2020}. These broad shapes arise from the high ejecta velocity blending the higher order CO transitions into the $\mathrm{\pi_0}$ and $\mathrm{\pi_1}$ transitions.   

Emission features in the region between $\sim 7.5 - 9.5$~$\mu$m are typically associated with the SiO fundamental band \citep{Liu_1994, Kotak_2006}. However, this region in \ixf\ does not show the strong emission features expected from SiO vibrational bands at any epoch, instead possessing a rising IR continuum, see Fig.~\ref{fig:SED_plots}. 

\section{Spectral Comparison} \label{sec:Spec_comp_sec}

To evaluate the uniqueness of \ixf's infrared spectral features relative to other SNe~II, we compare the NIR and MIR spectroscopic observations of \ixf\ and several other objects. This comparison reveals both shared and unique features, constraining the structural diversity of these events. As \ixf\ evolves, the IR continuum grows stronger and the SN flux fades drowning out the features from atomic lines. To accurately contrast \ixf\ with the other SNe~II, we remove the IR continuum by fitting a blackbody at each epoch. 
A single blackbody, while not physically representative of the underlying physics, works well to capture the shape of the continuum at earlier times, while at later times the dust becomes optically thin preventing a single blackbody from capturing the full IR continuum shape.

\subsection{NIR Comparison} \label{NIR_comp}

\begin{figure*}
    \centering
    \includegraphics[width=\linewidth]{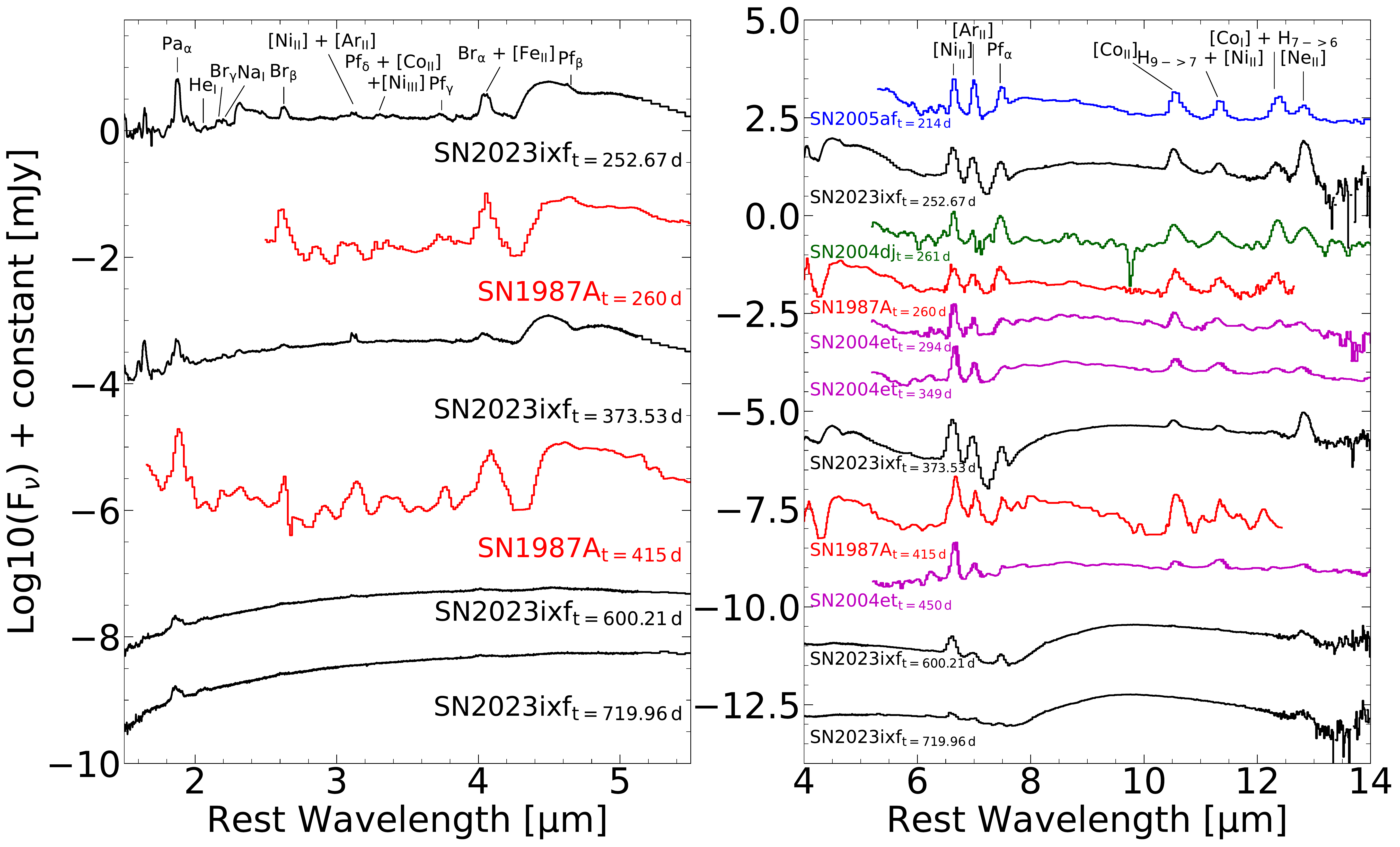}
    \caption{Spectral comparison of \ixf\ with several SNe~II observed at infrared wavelengths. All spectra have been corrected for redshift and extinction. To aid in comparison, the spectra of \ixf\ are continuum subtracted with blackbody functions of 850~K, 550~K, 360~K, and 300~K from the $+252.67$, $+373.52$, $+600.21$, and $+719.96$~d data, respectively.
    Left: Comparison of the NIR spectra, between $1.5 - 5.5$~\micron, of SN~1987A and \ixf. Right: Comparison of the \ixf\ MIRI spectra with SN~1987A \citep{Wooden_1993}, SN~2004dj \cite{Szalai_2011}, SN~2004et \citep{Kotak_2009}, and SN~2005af \citep{Kotak_2006}. Both SN~1987A and SN~2004et were digitized from \citet{Wooden_1993} and \citet{Kotak_2009}, respectively.}
    \label{fig:JWST_comps}
\end{figure*}

Historically, observations of SNe~II in the NIR at late times ($t > 200$~d) have been sparse because of their low luminosities and the lack of telescopes capable of observing these wavelengths. Until recently (\citealt{Shahbandeh_2024}, \citetalias{DerKacy_2025}, \citealt{Baron_2025}, \citealt{Mera_2025}), SN~1987A was the only SN~II with NIR spectroscopic data between $2.5 - 5$~\micron, obtained from the Kuiper Airborne Observatory \citep{Wooden_1993}. This limited sample prevents robust conclusions regarding the uniqueness of the NIR emission lines and molecular features in \ixf, especially considering its dusty progenitor, which differs drastically from the blue supergiant progenitor of SN~1987A \citep{Chevalier_1987, Woosley_1988, Arnett_1989, Podsiadlowski_1992}. The comparison of the late-time NIR spectra of SN~1987A and \ixf\ is shown in the left-hand panel of Fig.~\ref{fig:JWST_comps}.

Even more than a year after explosion, both SN~1987A and \ixf\ exhibit several strong H lines throughout the NIR region. The strongest of these lines are the Pa$\alpha$, Br$\alpha$, and Br$\beta$ transitions. Additional weaker spectral features associated with Br$\gamma$ and Pf$\beta$ also appear in both datasets, although they are less prominent in \ixf\ due to the strong contribution from the warm dust. A feature located at $\sim3.2$~\micron, identified as the \Nineb~$3.120$~\micron\ line in the SN~1987A spectrum \citep{Wooden_1993}, is present in the $+415$~d spectrum of SN~1987A and in the $+253$ to $+600$~d spectra of \ixf. While it appears as a single, blended emission feature with a blue shoulder in SN~1987A, the same feature in \ixf\ is resolved into a double-peaked structure. However, the difference in shape between the two SNe is likely due to the higher resolution of the \jwst\ spectra. 

In contrast to \ixf, SN~1987A does not exhibit a rising continuum; instead displaying a relatively flat flux across the NIR region. At $> 1$~year post-explosion, after removing the infrared excess in \ixf, both SNe display similar CO fundamental features at $\sim 5$~$\mu$m and a weak first overtone feature.
 Interestingly, the ratio of the CO fundamental peak to the Br$\alpha$ line is much higher in \ixf\ than in SN~1987A and a more prominent Pf$\beta$ is also observed in SN~1987A, see Fig.~\ref{fig:JWST_comps}. This difference is particularly pronounced at later times, where the CO fundamental in \ixf\ is dominant over the Br$\alpha$ emission, while in SN~1987A, both features are of similar intensity. The disparity in CO and Br$\alpha$ intensities may suggest that \ixf\ produced significantly more CO than SN~1987A at these epochs. If this is the case, and if we assume that the majority of C and O is bound in CO, then \ixf\ would have a larger C/O core than SN~1987A and thus a larger potential initial progenitor mass. However, given the unique status of SN~1987A's progenitor \citep{Woosley_1988, Arnett_1989}, a large C/O core in \ixf\ is only tentatively suggested.

\subsection{MIR Comparison} \label{MIR_comp}

In contrast to the NIR, several SNe~II beyond just SN~1987A have been observed at $\lambda > 5$~\micron\ thanks to Spitzer-era programs, including SN~2004dj \citep{Szalai_2011}, SN~2004et \citep{Kotak_2009}, and SN~2005af \citep{Kotak_2006}. The comparison with \ixf\ is shown in the right hand panel of Fig.~\ref{fig:JWST_comps}. 
The emission lines of \ixf\ appear weaker than those in other SNe~II, due to the strong IR excess that enhances the continuum, reducing the relative line strength especially at longer wavelengths. Once the continuum is subtracted the line strengths of \ixf\ become comparable to the other SNe.

Strong emission features are consistently found across all SNe. These features are associated with Co and Ni lines as well as the $\alpha$-elements Ne and Ar. In addition, \ixf\ lacks a strong emission of SiO between $7 - 9.5$~\micron unlike other SNe, especially SN~1987A. The line strength ratios of the [\NiII]~$6.636$~\micron, [\ArII]~$6.985$~\micron, and Pf$\alpha ~7.460$~\micron\ features of \ixf\ are consistent with the other SNe with the exception of SN~2004dj, which possesses a weaker [\ArII]~$6.985$~\micron\ feature.

The [\NeII]~$12.813$~\micron\ feature is significantly stronger in \ixf\ at all phases relative to the other SNe. While it has been suggested that this line can be powered by interaction \citep{Tinyanont_2025}, we do not see signatures of interactions such as narrow lines or flat-tops in this line suggesting the [\NeII]~$12.813$~\micron\ originates from within the ejecta. The presence of a strong [\NeII]~$12.813$~\micron\ feature indicates that the progenitor of \ixf\ possessed a relatively large pre-explosion Ne mass \citep{Jerkstrand_2012}, which has been associated with a higher progenitor mass \citep{Woosley_1995, Woosley_2002}. Nebular phase modeling of SN~2004et showed that the [\NiII]~$12.729$~\micron\ line becomes significantly stronger as the progenitor mass increases \citep{Jerkstrand_2012}. Thus, the strong [\NeII]~$12.813$~\micron\ line observed in \ixf\ suggests a relatively high mass progenitor, potentially larger than the $15$~\Msol\ progenitor of SN~2004et \citep{Jerkstrand_2012}.

The most notable difference between \ixf\ and other SNe in the MIR is its exceptionally strong infrared continuum, with NEOWISE $3.4$ and $4.6$~\micron\ photometry identifying \ixf\ as one of the brightest SN~II observed in the MIR \citep{Vandyk_2024}.
This strong infrared continuum dominates the spectra of \ixf\ over any other emission feature and is significantly stronger than in any other SN~II. \ixf's infrared excess emerges prior to $+252.67$~d, several hundred days earlier than the other SNe shown in Fig.~\ref{fig:JWST_comps}, e.g. SN~2004et developed a strong IR excess after $+400$~d \citep{Kotak_2006, Kotak_2009}. The strong IR excess is solely observed in the later epochs, $t \geq +250$~d \citepalias{DerKacy_2025}, suggesting that a large amount of dust, either pre-existing or freshly formed, is present. While comprehensive modeling of the dust component is left for future papers in this series, we discuss the origins of the infrared excess in Section~\ref{sec:dust}.

 \begin{figure*}
     \centering
     \includegraphics[width=\linewidth]{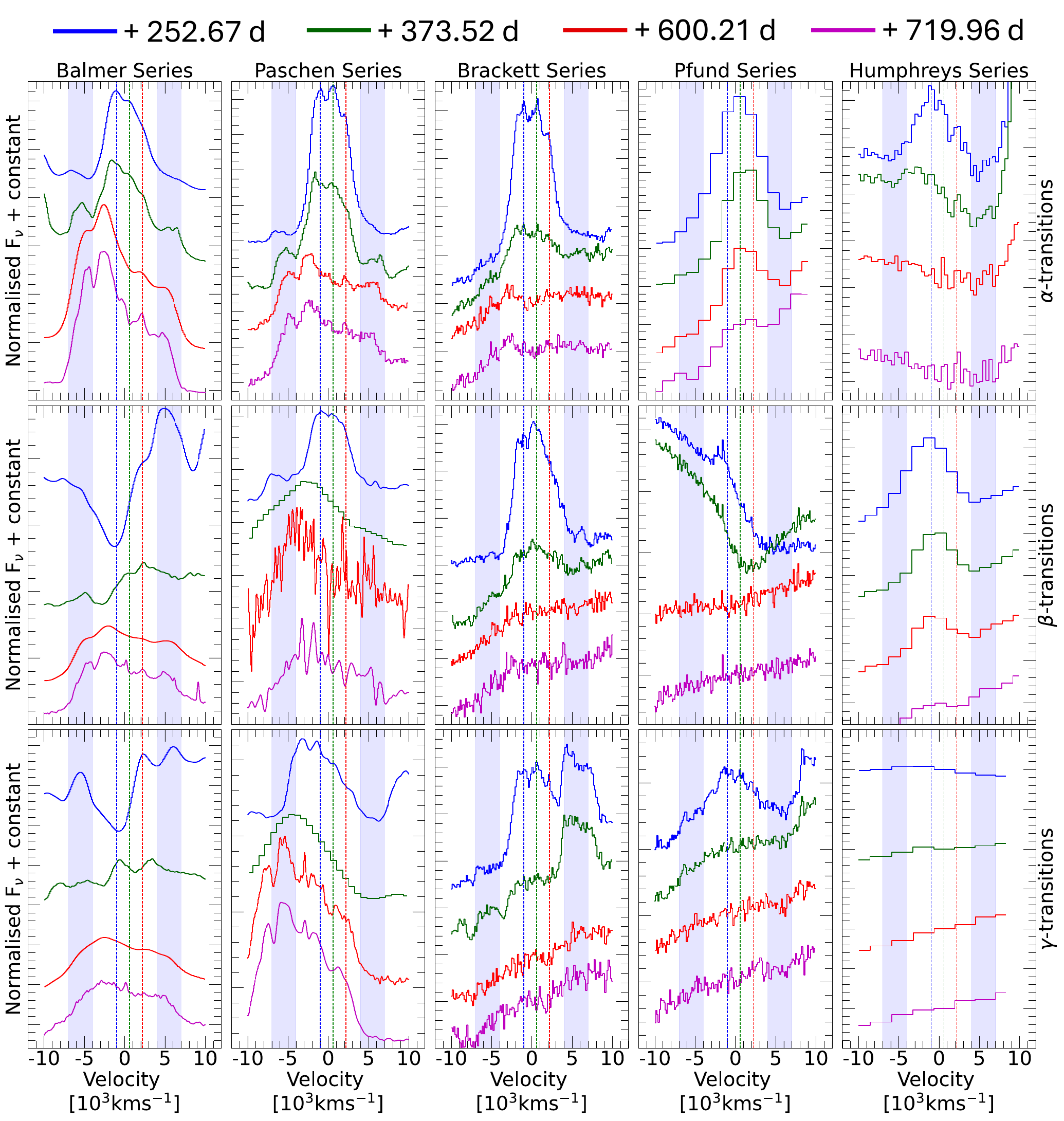}
     \caption{Evolution of the $\alpha$ (top), $\beta$ (middle), and $\gamma$ (bottom) transitions of the different hydrogen series located between $0.32 -  14.0$~\micron\ at $+252.67$ (blue), $+373.52$ (green), $+600.21$ (red), and $719.96$~d (purple). Several of the isolated hydrogen emission lines, especially the $\alpha$ transitions, show a triple-peaked profile consisting of a blueshifted peak ($-1,100$~\kms), a slightly redshifted peak ($+500$~\kms), and a strongly redshifted shoulder ($+2,200$~\kms), shown by the vertical dashed blue, green and red lines, respectively. Additional high velocity components, the shaded blue regions, emerge between $+252.67$ and $+372.53$~d located between $\pm 4,000 - 8,000$~\kms.}
     \label{fig:H_ABC_Evo}
 \end{figure*}

\section{Line Velocities} \label{sec:Line_vels}

The shapes of nebular emission lines are known to encode important information about the physics and evolution of the explosion. Additionally, the shape of spectral line profiles offer a powerful diagnostic for identifying signatures of newly formed dust.

\subsection{Hydrogen Line Diversity}
\label{sec:H_vels}

The H profiles in late time spectra of SNe~II have often been used as a key identifier of an asymmetric explosion \citep{Leonard_2002, Utrobin_2009} or the formation of dust within the ejecta \citep{Sahu_2006,Kotak_2009,Maguire_2010}. Newly formed dust attenuates the flux from the receding side of the SN, resulting in a suppression of the redshifted flux and a slightly blueshifted asymmetrical emission peak. This was first identified in SN~1987A \citep{Lucy_1989} but has since been observed in multiple SNe \citep{Lucy_1989,Kotak_2009,Maguire_2010,Stritzinger_2012,Inserra_2011,Smith_2012, Gall_2014}.  Many H lines are clearly present throughout the evolution of \ixf and the evolution of the $\alpha$, $\beta$, and $\gamma$ H transitions from the Balmer to Humphreys series is shown in Fig.~\ref{fig:H_ABC_Evo}.

Several of the H features seen in the  $+252.67$~d spectrum display a split three-peaked central profile, e.g. the \Ha, Pa$\alpha$, Pa~$\beta$, Br$\alpha$, Br$\beta$, and Br$\gamma$, see Fig.~\ref{fig:H_ABC_Evo}. The three peaks correspond to peak shifts of $-1,100$~\kms, $600$~\kms, and $2,200$~\kms, identified as the blue, green and red dashed vertical lines in Fig.~\ref{fig:H_ABC_Evo}, respectively. These multiple peaks are not always observed in the higher order transitions, either due to weak line strengths, e.g. Pa~$\gamma$ or blending with emission features from other atomic species or molecules, e.g. \Hb\ and Pf$\beta$.

The multi-peaked emission features are also visible at $+600.21$ and $+719.96$~d although the blueshifted peak has shifted more blue-ward to $\sim -2,500$~\kms, as seen in the \Ha\ and Pa$\alpha$ profiles in the top row of Fig.~\ref{fig:H_ABC_Evo}. A similar time-evolving shift is not seen in the two redward peaks. 
At these later epochs, line blending from other emission features is not significant, hence these peaks are visible in the higher order transitions, see the bottom two spectra of the \Hb\ and \Hc\ plots of  Fig.~\ref{fig:H_ABC_Evo}. The origin of the three shifted \Ha\ peaks in \ixf\ has been associated with strong clumping within ejecta and the surrounding swept-up CSM \citep[e.g.,][]{Singh_2024, Ferrari_2024, Kumar_2025}.

In addition to the evolving  multi-peak central emission feature, many of the H profiles exhibit two high-velocity features in the wings of the spectra. These features are boxy in shape and located at between $\pm 4,000 - 8,000$~\kms, and are shown by the highlighted  blue regions in Fig.~\ref{fig:H_ABC_Evo}. These structures are most clearly observed in the \Ha\ and Pa$\alpha$ transitions. 

The distinct high-velocity hydrogen features could be produced by the interaction of a fast-moving, asymmetric shock-front with a low density CSM. While the central triple-peaked features can be explained by a slower moving shock-front interacting with dense equatorial CSM \citep{Singh_2024, Kumar_2025}. These complex structure can significantly influence both the molecule and dust formation rates.

\subsection{Hydrogen \texorpdfstring{$\alpha$}--series Evolution} \label{sec:alpha_evos}

The study of the $\alpha$-series transitions, as the strongest transitions in each H-series, allows us to understand properties of the bulk of the H-envelope.

\subsubsection{\texorpdfstring{\Ha}\ Evolution} \label{sec:Hsplit}

\begin{figure}
    \centering
    \includegraphics[width=\columnwidth]{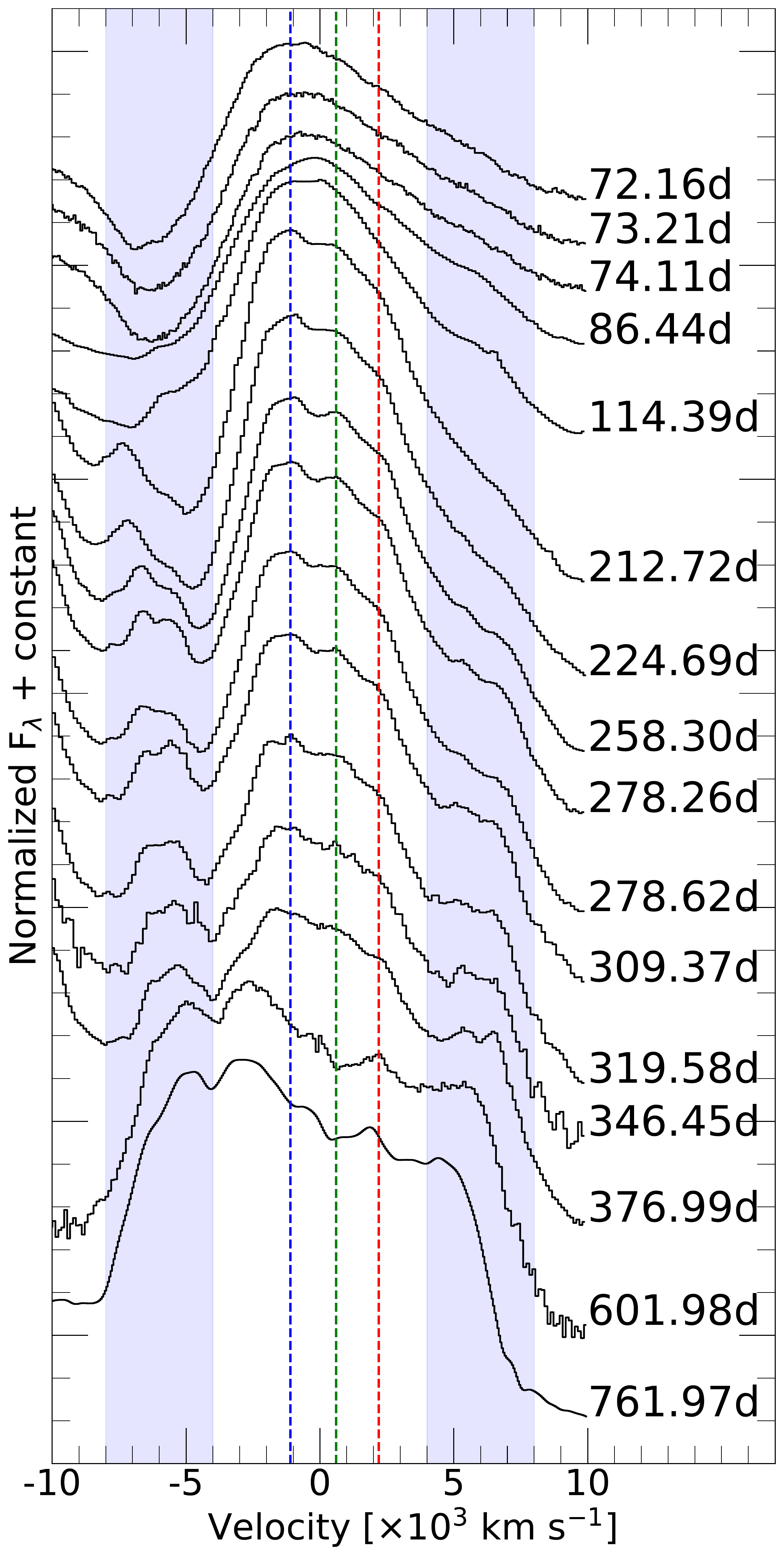}
    \caption{Evolution of the \Ha\ profile from a blueshifted P-Cygni like feature at $+ 72$~d to an asymmetric multi-peak feature at $+ 761.97$~d, with high velocity components that emerge around $+258.30$~d. Earlier spectra are presented in  \citepalias{DerKacy_2025}. All spectra have been corrected for redshift, extinction, and have been continuum subtracted.
    The blue, green and red vertical lines denote the three shifted peaks located at $\sim -1,100$~\kms, $\sim 600$~\kms, and $\sim 2,200$~\kms respectively. The blue shaded regions highlight the emerging high-velocity hydrogen features.
    }
    \label{fig:70+_H_vel_evo}
\end{figure}

Examining individual features with high cadence data may be able to  provide additional insight into the physics of the explosion which may be missed with sparsely sampled observations. The evolution of the \Ha\ feature between $70$ and $720$~d is shown in Fig.~\ref{fig:70+_H_vel_evo}, with the full optical spectra shown in Appendix~\ref{app:A}. 

Prior to $+72.16$~d the \Ha\ emission feature evolved from a sharp narrow flash ionization emission line \citep{Zimmerman_2023, Jacobson-Gal_2023} to a mostly smooth broad P-Cygni like emission profile \citetext{\citealp{Singh_2024, Zheng_2025}; \citetalias{DerKacy_2025}}.
The temporal evolution of \Ha\ in Fig.~\ref{fig:70+_H_vel_evo}, reveals a smooth broad emission peak with an initial blueshift of $\sim -1000$~\kms\ at $+72.16$~d that declines to rest velocity by $+86.44$~d as the ejecta becomes optically thin \citep{Anderson14}, which is accompanied by a decline in FWHM from $8,000 \pm 300$ to $5,900 \pm 300$~\kms.

Over the $\sim 140$ days between $+86.44$ and $+212.72$~d, the three peaks discussed in Section~\ref{sec:H_vels} emerge within the \Ha\ profile. These three peaks have velocities of $\sim -1,100$~\kms, $\sim 600$~\kms, and $\sim 2,200$~\kms, and are denoted by the blue, green and red dashed lines in Fig.~\ref{fig:70+_H_vel_evo}. They show no noticeable velocity evolution between their emergence and the \jwst\ epoch at $+252.67$~d. 

 A similar \Ha\ evolution has been observed in the nebular phase spectra several other SNe~II \citep[e.g.][]{Bevan_2016, Niculescu_2022}, including SNe~1987A \citep{Catchpole_1987, Hanuschik_1988, Lucy_1989}, 1999em \citep{Leonard_2002}, and 2017gmr \citep{Andrews_2019}. As well as the mechanism discussed in Section \ref{sec:H_vels}, multiple other origins have been proposed to explain multi-peaked \Ha\ features. 

Blanketing from an \FeII\ line has been suggested to cause the asymmetric shape of the \Ha\ \citep{Hoeflich_1988}, with addition line blanketing from the Br$\delta$ and Br$\varepsilon$ lines being invoked to explain the shape of the Pa$\alpha$ line profile. However, this line blanketing can not be invoked to explain the shape of all the H profiles in \ixf\ as several emission lines are isolated from any lines strong enough to cause blanketing, e.g. Br$\beta$ and Br$\gamma$.

Ejecta asymmetry and the mixing out of \Nifs\ has also been invoked to explain sloped \Ha\ emission features \citep{Elmhamdi_2003}. However, significant amounts of \Nifs\ mixed within the ejecta would result in ionized CO$^{+}$ which appears as an emission peak at $2.26$~\micron\ \citep{Spyromilio_1988,1990Ap&SS.171..213S,Rho_2021}, which is not observed in \ixf.   

Between $+114.39$ and $+212.72$~d, the \Ha\ in \ixf\ develops clear high-velocity blueshifted component, with a weaker high-velocity redshifted component emerging by $+258.30$~d. These features, which are the same as those mentioned in Section \ref{sec:H_vels}, are highlighted by the shaded blue regions in Fig.~\ref{fig:70+_H_vel_evo}. They are roughly symmetrical around the rest wavelength of \Ha, with initial velocities of $\sim 8,000 \pm 500$~\kms\ when they first emerge. There after, their velocity declines to $\sim 5,000 \pm 500$~\kms by $+761.97$~d. This is similar to the late time high-velocity \Ha\ profiles seen in SN~1993J \citep{Matheson_2000}, and as discussed in the previous section they may be produced by interaction between a fast moving shock front and low density CSM. 

Finally, starting at $+212.72$~d, as \Ha\ evolves, the red side of the profile becomes attenuated, creating an asymmetrical, sloped emission peak (see Fig.~\ref{fig:70+_H_vel_evo}). The suppression of the redshifted flux could result from blocking by emission from the receding material. This attenuation may be due to dust formation within the ejecta, as previously suggested for \ixf\ \citep{Singh_2024}.

\begin{figure*}
    \centering
    \includegraphics[width=\linewidth]{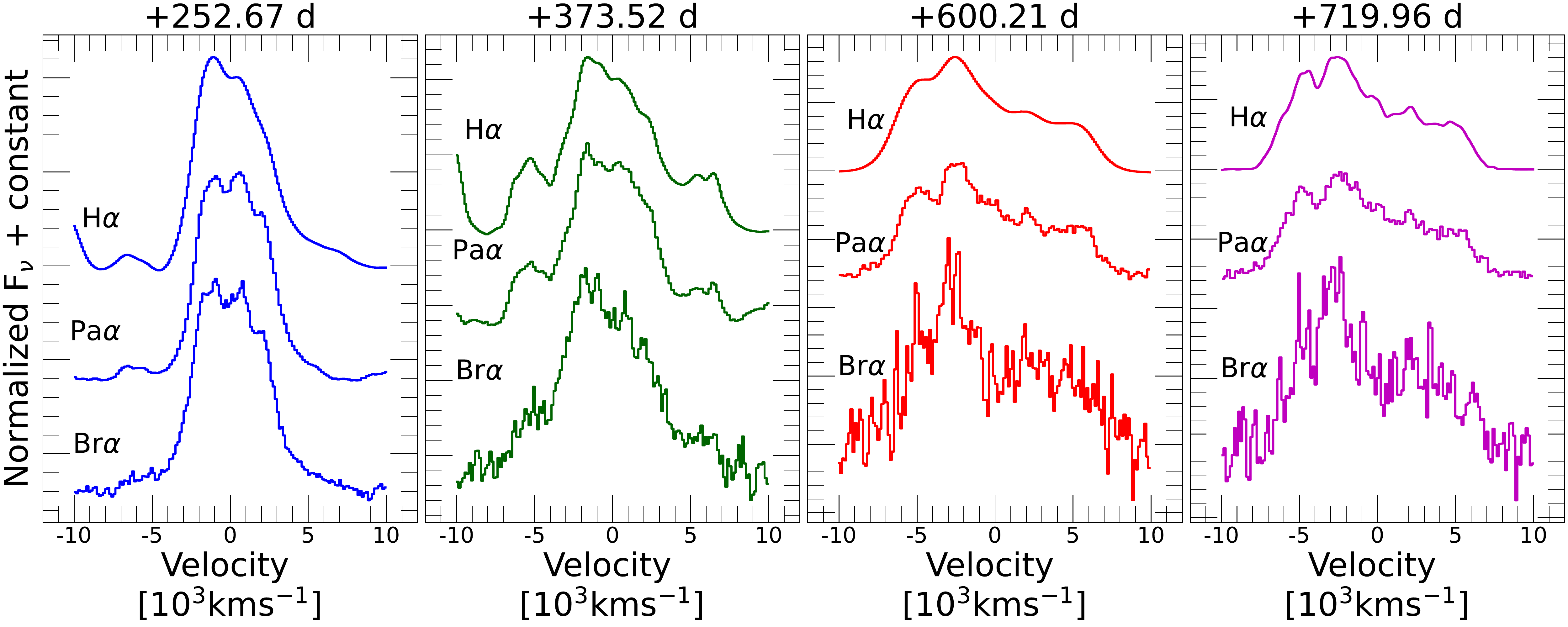}
    \caption{The evolution of the \Ha, Pa$\alpha$, and Br$\alpha$ emission features. All features have been continuum-subtracted and normalized to the flux at $0$~\kms. No wavelength-dependent attenuation of the redshifted flux is observed at a given epoch, although a  temporal-dependent attenuation is visible.
}
    \label{fig:A_evos}
\end{figure*}

\subsubsection{Panchromatic H \texorpdfstring{$\alpha$}--series} \label{sec:Pan_H}

While the study of \Ha\ is useful for understanding the evolution of the H envelope, panchromatic coverage of the H $\alpha$--series transitions enables a joint time- and wavelength-dependent analysis of the splitting and attenuation across multiple features.

The attenuation of the \Ha\ feature shown in Fig.~\ref{fig:70+_H_vel_evo} is also observed in other H emission features, see Fig.~\ref{fig:H_ABC_Evo}. The majority of these features exhibit strong attenuation of the redside flux relative to the blueside, most prominently in the $\alpha$-transitions. The normalized, continuum-subtracted emission features of the \Ha, Pa$\alpha$, and Br$\alpha$ transitions for each \jwst\ epoch are shown in Fig.~\ref{fig:A_evos}. For each emission feature, a flat continuum was fit across the bottom to remove the contribution from the IR excess, which significantly affects the shape of the redder emission lines.

All of the features exhibit similar structures, consisting of multiple peaks as highlighted in the previous sections. Additionally, the shapes of the profiles at each epoch are remarkably consistent, with attenuation of the red side of the features that increases over time. This wavelength-dependent analysis allows us to test for dust attenuation, which would one might think would typically affect the bluer emission features over the redder ones. In \ixf, we observe the same level of attenuation at a given epoch for all the features shown in Fig.~\ref{fig:A_evos}, regardless of wavelength. The absence of wavelength-dependent attenuation suggests that the shape of these features is not solely due to dust attenuation. However, as we discuss further in Section~\ref{sec:ejecta_dust}, this is in fact is heavily influenced by grain size and dust composition.

\subsection{Oxygen and Calcium Velocity Evolution} \label{sec:O+Ca}

\begin{figure}
    \centering
    \includegraphics[width=\linewidth]{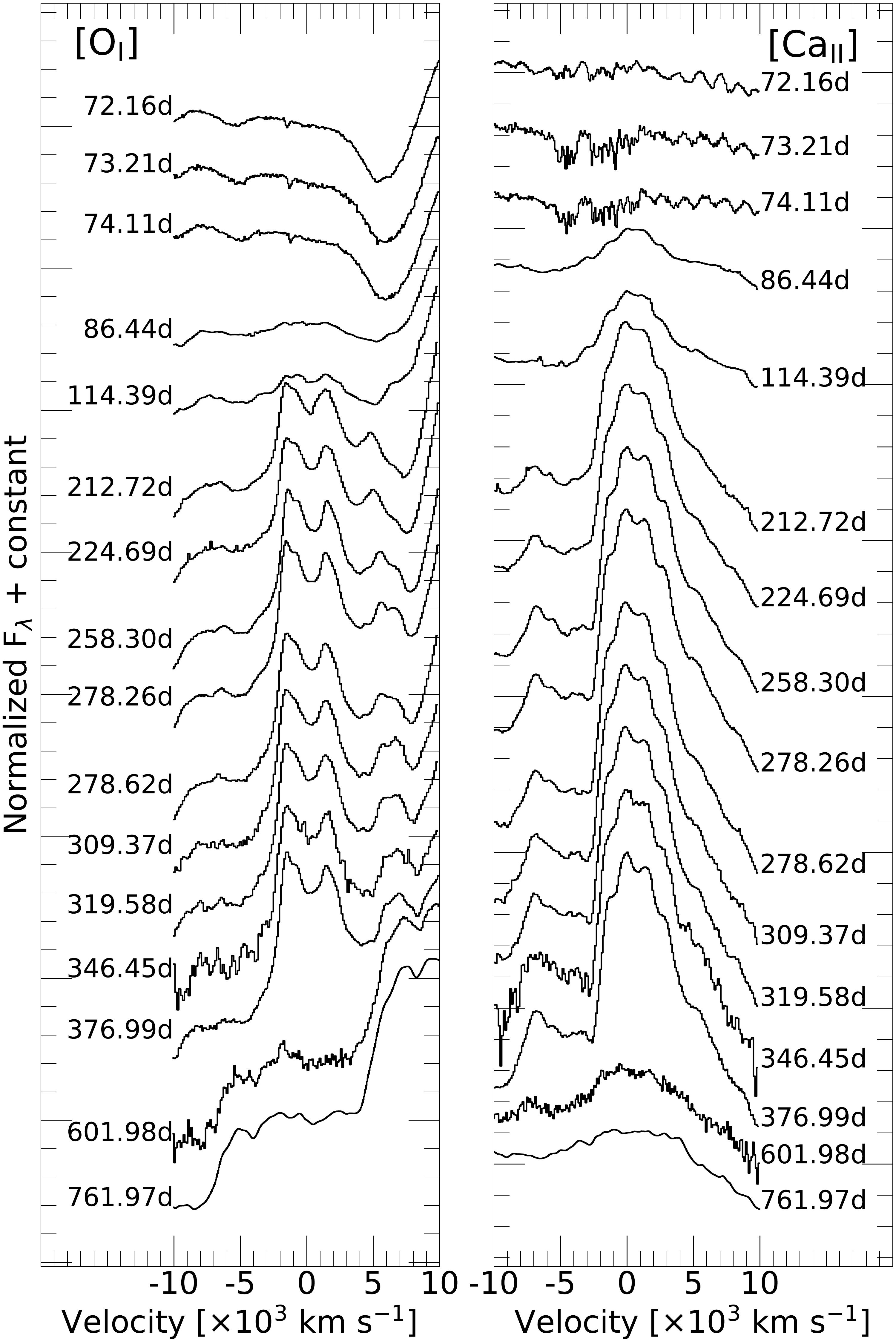}
    \caption{Spectral evolution of the [\OI]~$0.6300,~0.6363$~\micron\ doublet and [\CaII]~$0.7292~0.7324$~\micron. The features are plotted in velocity space relative to the rest-frame wavelengths of the [\OI]~$0.6300$~\micron\ and [\CaII]~$0.7292$~\micron\ lines, respectively. The [\OI]~$0.6300,~0.6363$~\micron\ features show a significant blueshift of $\sim 1,300 \pm 500$~\kms, similar to the blue peak observed in the split H emission peak, while the [\CaII]~$0.7292~0.7324$~\micron\ remains centered at the rest wavelength. The difference in the location of the peaks suggests that the O and Ca are located in drastically different regions of the SN ejecta.}
    \label{fig:O+Co_evo}
\end{figure}

The optical lines of O show a systematic blueshift of $\sim -1,300 \pm 500 $~\kms\ from the point they first appear in the spectra of \ixf\ (see Fig.~\ref{fig:O+Co_evo}). 
The blueshift in the optical O profiles is consistent with the shift observed in the H lines. However, the NIR \OI~$1.1290$~\micron\ feature does not show a similar blueshift, which may indicate that these lines form in different regions of the ejecta.

The optical [\CaII]~$0.7292,~0.7324$~\micron\ doublet and the \CaII\ NIR triplet exhibit no significant shifts from their rest wavelengths at any epoch shown in Fig.~\ref{fig:O+Co_evo}, or in other published observations \citep{Singh_2024, Li_2025}. However, the Ca features also develops a red side attenuation similar to the hydrogen emission features. This was also observed by \citep{Singh_2024}.

\subsection{Neon and Argon Velocity Evolution} \label{sec:Ar+Ne}
\begin{figure*}
    \centering
    \includegraphics[width=\linewidth]{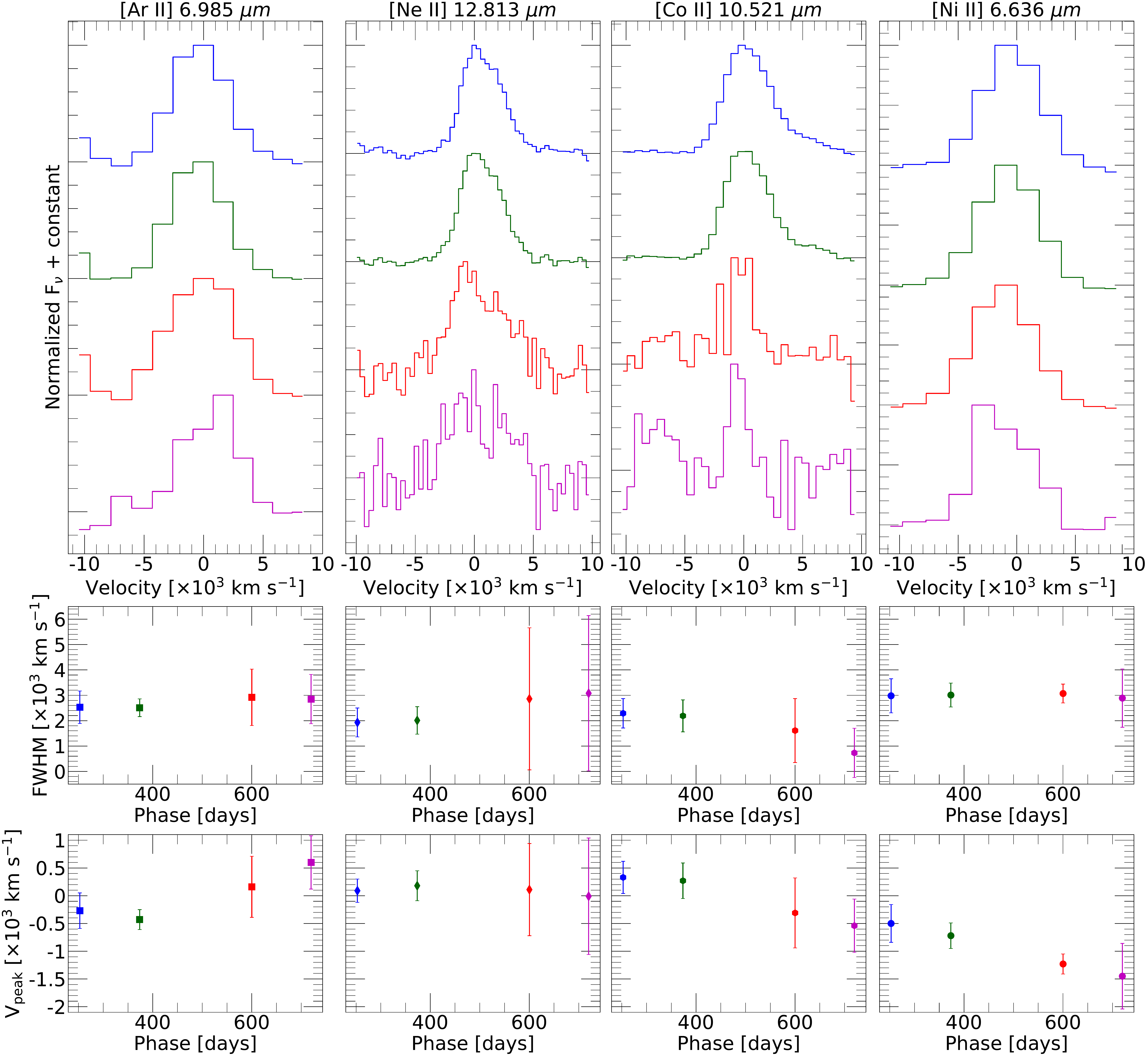}
    \caption{Velocity evolution of the [\NiII]~$6.636$~\micron, [\ArII]~$6.985$~\micron, [\CoII]~$10.521$~\micron, and [\NeII]~$12.813$~\micron. While the neon, argon, and cobalt lines are centered around $0$~\kms, the nickel emission feature is blueshifted by $\sim 1,000 \pm 500$~\kms at all epochs. All emission features have been corrected by removing a linear continuum across the bottom of the features to remove the influence of the IR excess, especially at late times when the lines are weaker. The FWHM and peak velocity determined by Gaussian fits are shown in the middle and bottom row, respectively}
    \label{fig:Ni_Co_Ar_Ne_vels}
\end{figure*}

The MIR contains lines from elements that lack emission features at shorter wavelength regions (e.g., Ar and Ne). The \jwst\ spectra of \ixf\ show two strong lines associated with Ar and Ne, located at $6.985$~\micron\ and $12.813$~\micron, respectively. The velocity evolution of these lines are presented in Fig.~\ref{fig:Ni_Co_Ar_Ne_vels}. 

The [\ArII]~$6.985$~\micron\ features maintain roughly the same shape and are centered around $0.0 \pm 0.5 \times 10^3$~\kms\ in all epochs except at $+713.73$~d, where the [\ArII]~$6.985$~\micron\ line shows a slight redshift of $\sim 2,000$~\kms\ in its central peak, see in the bottom panel left panel of Fig.~\ref{fig:Ni_Co_Ar_Ne_vels}.

In the first two epochs, the [\NeII]~$12.813$~\micron\ displays a slightly asymmetric emission feature with an enhanced flux shoulder at $\sim 1,900 \pm 500$~\kms. This could can be explained by the blending of the [\NeII] line with a weaker emission feature, which is supported by the shape of the [\NeII]~$12.813$~\micron\ at $+600.21$ and $+713.13$~d, where a second peak emerges coincident with the location of the red shoulder, see Fig.~\ref{fig:Ni_Co_Ar_Ne_vels}.

\subsection{Nickel and Cobalt Velocity Evolution} \label{sec:Ni+Co}

Deeper within the ejecta, Ni and Co form during the explosive nuclear burning of core-collapse. The [\CoII]~$10.521$~\micron\ line, shown in Fig.~\ref{fig:Ni_Co_Ar_Ne_vels}, exhibits nearly perfect symmetry around $0$~\kms\ throughout \ixf's evolution, suggesting that cobalt is centrally located within the ejecta with minimal mixing into the outer regions. 

In contrast, the velocity evolution of the [\NiII]~$6.636$~\micron\ line, also shown in Fig.~\ref{fig:Ni_Co_Ar_Ne_vels}, reveals a tentative blueshift in its emission peak of $\sim1,000 \pm 500$~\kms, evolving to $\sim 2,000 -3,000$~\kms at $+719.96$~d. The initial blueshift of the [\NiII]~$6.636$~\micron\ line mirrors peak velocity seen in the blue peak of the H lines in Fig.~\ref{fig:H_ABC_Evo} and the [\OI]~$0.6300, 0.6363$~\micron\ lines.

\subsection{Nature of the Asymmetrical Ejecta} \label{sec:Asym}

Several emission lines in \ixf, including those from H, O, and Ni, exhibit blue-shifted peaks. While some of this shift could be due to dust obscuring emission from the receding side of the ejecta (see Section~\ref{sec:dust}), the observed asymmetries also point to an intrinsically aspherical explosion geometry \citep{Singh_2024}. Aspherical ejecta are common for SNe~II and can naturally arise from jet-driven explosion mechanisms \citep[e.g.,][]{Khokhlov01, Scheck_2006, Modjaz_2008, Taubenberger_2009, Milisavljevic_2010, Andrews_2011a, Janka16, Tucker_2024_SN2023ufx}. In such scenarios, a significant fraction of the ejecta moves toward the observer, resulting in a net blue shift. The observed flux asymmetry around the rest wavelengths of the lines provides a direct probe of the degree of asymmetry in the explosion.

In \ixf\ the shifts in emission peaks we observe could also result from an asymmetric distribution of \Nifs\ within the ejecta \citep{Gerardy_2000, Utrobin_2017, Bose_2019}. In this scenario, uneven heating of the surrounding O-rich material by \Nifs\ would lead to asymmetric oxygen emission features \citep{Gerardy_2000, Maeda_2008}. However, we disfavor this interpretation due to the symmetric profile of the [\CoII]~$10.521$\,\micron\ line and the absence of CO$^{+}$, which forms through mixing between the CO and \Nifs-rich regions.

The discrepancy between the blueshifted O lines and the symmetrical line profiles observed in other species, such as Ca, and Co, could also be explained as a combination of a high-density clump, or torus, of O-rich material interacting with the equatorial dense CSM moving along our line of sight and a bulk ejecta distributed around $0$~\kms \citep{Fang_2024, Ferrari_2024, Kumar_2025}. This ejecta structure was first observed in SN~1987A \citep{Li_1992} and has since been identified in several other SNe \citep{Milisavljevic_2010, Kuncarayakti_2020, Fang_2022}.

Although an aspherical explosion geometry explains the systematic blueshifts observed in the O and Ni emission profiles, it does not account for the evolving morphology of the H-series lines, which have a time dependent attenuation of the red-side flux and multiple peaks. This requires interaction with a clumpy CSM and possibly dust formation in the ejecta.

\section{Origin of the Dust Emission} \label{sec:dust}

While it is clear that dust plays a significant role in the evolution of \ixf, determining the origin of this dust is not straightforward. As noted in Section~\ref{sec:SED}, the $+252.67$~d spectrum of \ixf\ shows a strong, featureless IR excess. By $+373.52$~d, a broad continuum feature develops around $10.0$~\micron, which remains present in subsequent epochs. The IR continuum of \ixf\ does not change again significantly until $+719.96$~d, when a second peak emerges at $\sim$18.0~\micron. These peaks, at $\sim$$10.0$~\micron and $\sim$18.0~\micron, are indicative of silicate-rich dust \citep{Shahbandeh_2023}. 

The evolving SED of \ixf\ can be explained by several possible sources of dust \citep[see Fig.~5 in][]{Sarangi_2018}, including:  1) pre-existing dust in the CSM located beyond the explosion's evaporation radius \citep{Bode_1980, Dwek_1983, Andrews_2011b};  
2) freshly formed dust in the cool dense shell (CDS) between the forward and reverse shocks \citep{Pozzo_2004, Smith_2008, Gall_2014}; and  
3) newly condensed dust within the SN ejecta itself \citep{Kotak_2005, Andrews_2011a, Szalai_2013}. Although determining the precise dust properties of \ixf\ requires detailed modeling, which we reserved for a future paper in this series, we discuss the plausibility of each dust source below.

\subsection{Pre-existing CSM Dust} \label{sec:pre_exist}
The first option is that the IR excess originates from pre-existing dust located outside the evaporation radius of the explosion. This dust is heated by the shock breakout emission and, as the dust evolves, transitions from optically thick to optically thin causing the shape of the IR emission to change \citep{Sarangi_2018, Shahbandeh_2023}. This dust would reside in the CSM of the progenitor, or the interaction region between CSM and the ISM \citep{Dragulin_2016}. 

At $+252.67$~d, the infared continuum is well approximated by a black body of a temperature of $688$~K, which is well below the evaporation temperature for both carbonates and silicates, hence it is produced from warm dust.
Assuming an optically thick and spherical dust distribution, a dust temperature of $688$~K would correspond to a radius of $\mathrm{r_{dust}} = 2.2 \times 10^{16}$~cm. A similar value, $\mathrm{r_{dust}} \sim 1.6 \times 10^{16}$~cm, was suggested from a $700$~K blackbody fit to the $3.4$~\micron\ NEOWISE photometry at $+212$~d \citep{Vandyk_2024}. 
For SNe II, the typical photospheric radius is $1 - 10 \times 10^{14}$~cm \citep[e.g.][]{Hoeflich_1988, Dessart_2005}, with photospheric temperatures of $5,000 - 10,000$~K. New dust forms once the ejecta drops below the evaporation temperature, and would reside close to the photosphere, assuming $T_{dust} \propto \sqrt{R_{ph}}$, while heated dust should be located further out, $\approx 100 ~R_{ph}$.

Moreover, the CSM distance derived from the X-ray emission, $< 7 \times 10^{15}$~cm is likely produced in the low density region ($100 - 10,000$ times the ISM density) and by the interaction between the reverse shock and the wind-like CSM \citep{Margutti_2012, Dragulin_2016, Grefenstette_2023, Panjkov_2024, Nayana_2025}.
High-velocity ejecta from the SNe ($\gtrsim 30,000$~\kms) would traverse the CSM with minimal deceleration, reaching radii of $\sim2 \times 10^{16}$~cm within a few months. This rapid interaction would be expected to cause dust evaporation, producing an infrared component near the dust sublimation temperature ($T \gtrsim 2000$~K), in contrast to the warm dust ($688$~K) temperatures that are actually observed.
Given this, the ejecta is unlikely to have reached the dust region, even by $+250$~d, placing the dust at distances similar to the estimate from the infrared continuum. 
In reality, the CSM is likely not a spherical shell and is likely clumpy, meaning that $2.2 \times 10^{16}$~cm should only be taken as a strict lower limit. An asymmetric clumpy CSM is also supported by the X-ray and radio observations, which require significant clumping to find agreement in the derived X-ray and radio CSM density profiles \citep{Panjkov_2024, Nayana_2025}. 
Future, detailed analysis is required to determine the exact location and nature of the dust. 

\subsection{Cold Dense Shell Dust} \label{sec:CDS_Dust}
The second mechanism involves the formation of dust within the CDS produced in the post-shocked CSM. As the forward shock moves through the surrounding CSM and the reverse shock traverses into the inner ejecta, material is built up, shocked and heated. This postshock gas then rapidly cools, through a combination of free-free and molecular/atomic line cooling, forming a CDS sufficiently cool and dense enough to facilitate the formation of dust \citep{Sarangi_2018}. This dust is heated by ionizing photons from the forward shock, with dust above $2000$~K rapidly subliming after formation.
It may be assumed that, given the temperature dependence, the CDS is unlikely to be the cause of the initial IR excess observed at $t < 250$~d as the heating from the forward shock would inhibit dust formation \citep{Sarangi_2018}.
At later times, as the shock luminosity declines, the gas temperature in the CDS becomes conducive to dust formation. 
However, for CDS dust to account for the observed attenuated line profiles (if they are not produced from a geometry effect) it would need to be mixed into the ejecta.

\subsection{Ejecta Dust} \label{sec:ejecta_dust}
The third location for dust formation is within the expanding SN ejecta interior to the reverse shock. Dust can be formed within the ejecta once the temperature drops below the condensation temperature. This is facilitated by emission from molecular ro-vibrational bands, features that are observed in the spectra as early as $+252.67$~d. While ejecta dust could explain the late-time IR excess, the ejecta temperature would be too high at earlier phases to allow for dust formation \citep{Singh_2024}. For example, in some instances, in situ dust formation is not expected to occur until at least  450~d after the explosion \citep{Sarangi22}.

One hallmark of dust formation in the ejecta is the attenuation of the red side of optical emission features \citep{Gall_2014}. This is clearly observed in the \Ha\-series and [\CaII]~$0.7292~0.7324$~\micron, line profiles. For \ixf, we also observe time-dependent attenuated profiles in the Pashen-$\alpha$ and Brackett-$\alpha$ features, suggesting that this is not a wavelength-dependent effect (at least between 0.5–4~$\micron$). While one might expect dust blocking to show a strong wavelength dependence, given the blue Rayleigh scattering, previous studies have examined this issue. \citet{Lucy89} considered the effects of scattering via dust albedo and concluded that, for SN~1987A, scattering was negligible. They also inferred a grain size of $a \le 0.05$~\micron. In the DAMOCLES code \citep{Bevan16,Bevan18}, scattering is modeled using the Henyey-Greenstein phase function, and no significant variations are observed across the Balmer lines \citep[see their Figure~9]{Bevan16}. While we acknowledge the possibility of dust producing a wavelength-dependent blocking effect, the literature does not find a strong wavelength dependence for the lines we observe in \ixf, and we defer further conclusions to future work in the series.

\subsection{The Full Picture} \label{sec:Full_Dust}
In reality, a combination of dust formation processes may be required to explain the observed properties of \ixf. The strong IR excess at $+252.67$ suggests the presence of dust either in the CSM or CDS. 
Given the dusty nature of the progenitor, significant heating of pre-existing dust in the CSM might be expected.

The evolution of the line profiles may also indicate that some dust may be present within the ejecta. This dust may form as the ejecta cools below the condensation temperature or originate in the CDS and subsequently mix into the ejecta \citep{Bevan_2019, Singh_2024}. The latter scenario implies a clumpy dust structure.  However,  these profiles could also be produced by geometry effects and detailed  modeling is needed to distinguish between these possibilities. 

Overall, we tentatively suggest that the IR excess in \ixf\ results from a combination of pre-existing dust in the CDS or pre-existing in the CSM, which dominates at early times, and newly formed dust, which becomes more prominent at later epochs. This newly formed dust likely originates from either within the ejecta, in a clumpy CDS that mixes into the ejecta, or a combination of both. We will examine this is more detail in the next papers of the series.

\section{Conclusions} \label{sec:Conc_sec}

Here we present the panchromatic dataset of \ixf\, spanning from $0.32 -  30.0$~\micron\ between $+252.67 - +719.96$~d past explosion, with optical data spanning from $+72.16$ to $+224.69$~d. This is the most complete \jwst\ time series of a SN~II to date. In this second paper of our series, we concentrate on presenting this exquisite dataset, as well as identifying the predominant spectral lines and molecular bands that form in the ejecta. The data presented here were obtained from the NIRSpec, and MIRI-LRS, and MIRI-imager instruments, and supplemented with data from a variety of ground based telescopes, allowing us to construct $0.32 -  30$~\micron\ panchromatic SED's. The main results from our work are as follows:

\begin{itemize}

\item The \jwst\ spectra of \ixf\ reveal a rich array of atomic species, including rare members of the hydrogen Brackett, Pfund, and Humphreys series, as well as forbidden lines from Co, Ni, Fe, Ne, and Ar.  C, O, Mg, Na, and Ca lines are also present in \ixf\ at shorter wavelengths (see \S~\ref{sec:Line_spec}).

\item \ixf\ shows significant features of CO formation, visible by the first overtone and fundamental ro-vibrational bands, which were not present in the earlier \jwst\ spectrum \citepalias{DerKacy_2025}. While the first overtone fades by $+600.21$~d, the fundamental band remains detectable at all epochs, indicating the CO region is dropping in temperature below 1000~K (see \S~\ref{sec:mols}).

\item The splitting of late-time H lines from the Balmer, Pashen, and Brackett series into multiple features suggests that \ixf\ was likely a highly asymmetric explosion with a disk-like CSM. Additional high-velocity features, linked to an aspherical shock front, are also observed (see \S~\ref{sec:alpha_evos}).

\item Several emission lines in \ixf\ (e.g., from H, O, and Ni) exhibit blueshifted peaks, consistent with an aspherical, clumpy explosion geometry. A high density clumpy or toroidal distribution of O-rich material expanding along the line of sight, combined with Ca- and \Nifs-rich regions aligned perpendicular to the observer, explains the observed blueshifts and the lack of velocity shifts in Ca and Co lines (see \S~\ref{sec:Asym}).

\item The SED of \ixf\ shows a time evolving IR dust component which emerges by $+252.67$~d, with distinct features indicative of Si-rich dust appearing at $10~\micron$ by $+373.52$~d and $18~\micron$ by $+719.96$~d (see \S~\ref{sec:SED}).

\item The initial ($+252.67$~d) featureless IR continuum may be explained by the presence of pre-existing dust located outside the evaporation radius of the \ixf, which is consistent with the dusty nature of the RSG progenitor star (see \S~\ref{sec:dust}).

\item The time-evolving obscuration of the red side of \Ha, Pashen-$\alpha$, Brackett-$\alpha$, and [\CaII]~$0.7292~0.7324$~\micron\ may be explained by the start of dust formation, or as a result of an aspherical explosion geometry. If the attenuation is caused by dust, it may have formed within the ejecta or in a clumpy, cold, dense shell that mixes into the ejecta (see \S~\ref{sec:dust}). 

\item We tentatively suggest the IR excess observed in \ixf\ arises from a combination of  pre-existing, likely Si-rich, dust in the CSM or CDS, dominant at early times, and newly formed either dust within the ejecta or in a clumpy CDS that mixes into the ejecta (see \S~\ref{sec:dust}). Although, detailed models are required to explore this further. 

\end{itemize}

While it is beyond the scope of this paper to determine the exact grain size, mass, and location of the dust, such early emission of dust suggests that some amount of pre-existing dust ejected by the RSG progenitor can survive within the CSM and survive the forward shock of the explosion. However, disentangling the exact location of the dust (CSM vs CDS vs ejecta), is left for a more detailed analysis, which will be presented in future papers of this series.

Overall, \jwst\ has allowed the MIR region of the electromagnetic spectrum to be explored in unprecedented detail. 
In this paper, the second in the series, we have presented nebular phase panchromatic data of \ixf\ and demonstrated how this data can be used to explore the line formation processes in the various wavelength regions in the ejecta. In future papers, we will focus on both molecule and dust modeling of the respective features, determining their properties, mass, and location, and examine how this links to the progenitor system. \ixf\ is becoming one of the best observed SNe~II to date, and future scheduled \jwst\ and ground based observations will ensure this truly amazing dataset continues to improve.

\begin{acknowledgments}
K.M., E.B., C.A., J.D., M.S.,  P.H., and A.V.F. acknowledge support from NASA grants JWST-GO-02114,
JWST-GO-02122, JWST-GO-04522, JWST-GO-04217, JWST-GO-04436,
JWST-GO-03726, JWST-GO-05057, JWST-GO-05290, JWST-GO-06023,
JWST-GO-06677, JWST-GO-06213, JWST-GO-06583. Support for
programs \#2114, \#2122, \#3726, \#4217, \#4436, \#4522,  \#5057,
\#6023, \#6213, \#6583, and \#6677
were provided by NASA through a grant from the Space Telescope Science
Institute, which is operated by the Association of Universities for Research in
Astronomy, Inc., under NASA contract NAS 5-03127.

Some of this material is based upon work supported by the National Science Foundation Graduate Research Fellowship Program under Grant Nos. 1842402 and 2236415, and NSF award AST-230639 to P.H.. Any opinions, findings, conclusions, or recommendations expressed in this material are those of the author(s) and do not necessarily reflect the views of the National Science Foundation. 

D.O.J. acknowledges support from NSF grants AST-2407632 and AST-2429450, NASA grant 80NSSC24M0023, and HST/JWST grants HST-GO-17128.028, HST-GO-16269.012, and JWST-GO-05324.031, awarded by the Space Telescope Science Institute (STScI), which is operated by the Association of Universities for Research in Astronomy, Inc., for NASA, under contract NAS5-26555.

M.D. Stritzinger is funded by the Independent Research Fund Denmark (IRFD, grant number  10.46540/2032-00022B) and by an Aarhus University Research Foundation Nova project (AUFF-E-2023-9-28).

A.V.F. is grateful to the Christopher R. Redlich Fund for additional support.

L.G. acknowledges financial support from AGAUR, CSIC, MCIN and AEI 10.13039/501100011033 under projects PID2023-151307NB-I00, PIE 20215AT016, CEX2020-001058-M, ILINK23001, COOPB2304, and 2021-SGR-01270.

Based on observations made with the Gran Telescopio Canarias (GTC), installed at the Spanish Observatorio del Roque de los Muchachos of the Instituto de Astrofísica de Canarias, on the island of La Palma.

The Shappee group at the University of Hawai‘i is supported with funds from NSF (grants AST-2407205) and NASA (grants HST-GO-17087, 80NSSC24K0521, 80NSSC24K0490, 80NSSC23K1431).

JTH was supported by NASA grant 80NSSC23K1431.

Parts of this research were supported by the Australian Research Council Centre of Excellence for Gravitational Wave Discovery (OzGrav), through project number CE230100016.

SZ received support from the NKFIH OTKA K142534 grant.

Parts of this research were supported by the Australian Research Council Centre of Excellence for Gravitational Wave Discovery (OzGrav), through project number CE230100016.

D.M.\ acknowledges support from the National Science Foundation (NSF) through grants PHY-2209451 and AST-2206532.

\end{acknowledgments}

\vspace{5mm}
\facilities{JWST(NIRSpec and MIRI), INT(IDS), IRFT(Spex), GTC(OSIRIS+ and EMIR), Keck:I(LRIS), Keck:II(NIRES), UH:2.2m(SNIFS)}

\software{\texttt{Astropy} \citep{Astropy_2013}, \texttt{Dust-extinction} \citep{Extinction_2023a, Extinction_2023b}, \jwst\ reduction notebooks \citep{law_2025}, \jwst\ Science Calibration Pipeline (ver. 1.18.0; \citep{Bushouse_2022_JWST_reduc}), \texttt{Matplotlib} \citep{matplotlib_2007}, \texttt{Numpy} \citep{numpy_2020}, \texttt{SciPy} \citep{SciPy_2020}, \texttt{Sextractor} \citep{sextractor_2020}}

\renewcommand{\thetable}{A\arabic{table}}
\setcounter{table}{0}
\renewcommand{\thefigure}{A\arabic{figure}}
\renewcommand{\theHfigure}{A\arabic{figure}}
\setcounter{figure}{0}
\appendix

\section{Observation Log and GB-Spectra} \label{app:A}
Here we provide detailed logs of all of the observations, analysis, and results used in this work. The JWST, ground-base optical and NIR spectral observations are presented in Tables \ref{tab:JWST_spec_info}, and \ref{tab:GB_spec_info} respectively. 

\begin{figure*}
    \centering
    \includegraphics[width=1.6\columnwidth]{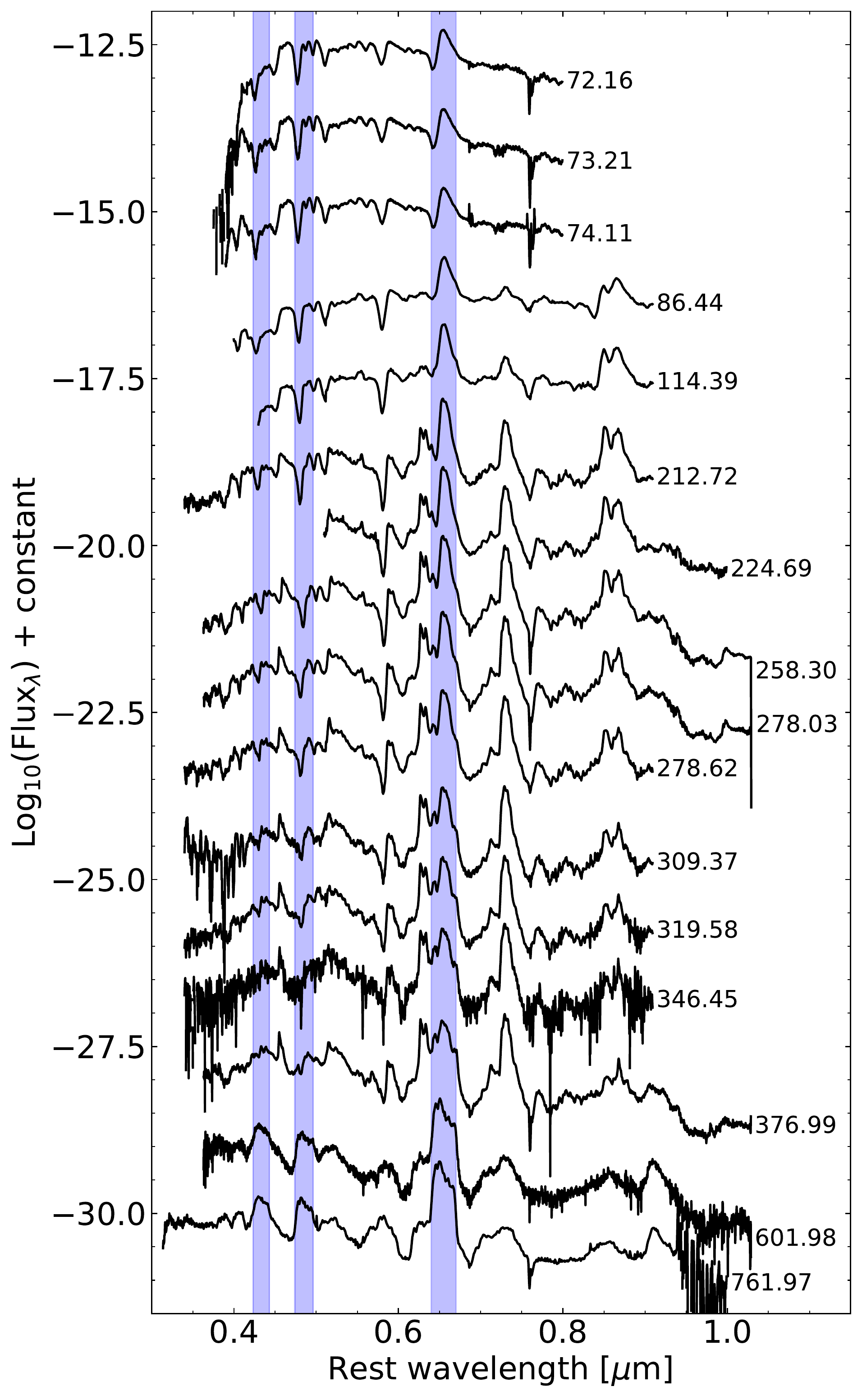}
    \caption{The optical spectral time series of \ixf\ used in this work. All spectra have been corrected for redshift and extinction. The phase relative to the explosion date is given for each spectrum on the right. The emission regions of the \Ha, \Hb, and \Hc\ lines are highlighted by the blue regions. }
    \label{app:All_opt_spec}
\end{figure*}

\begin{figure*}
    \centering
    \includegraphics[width=1.6\columnwidth]{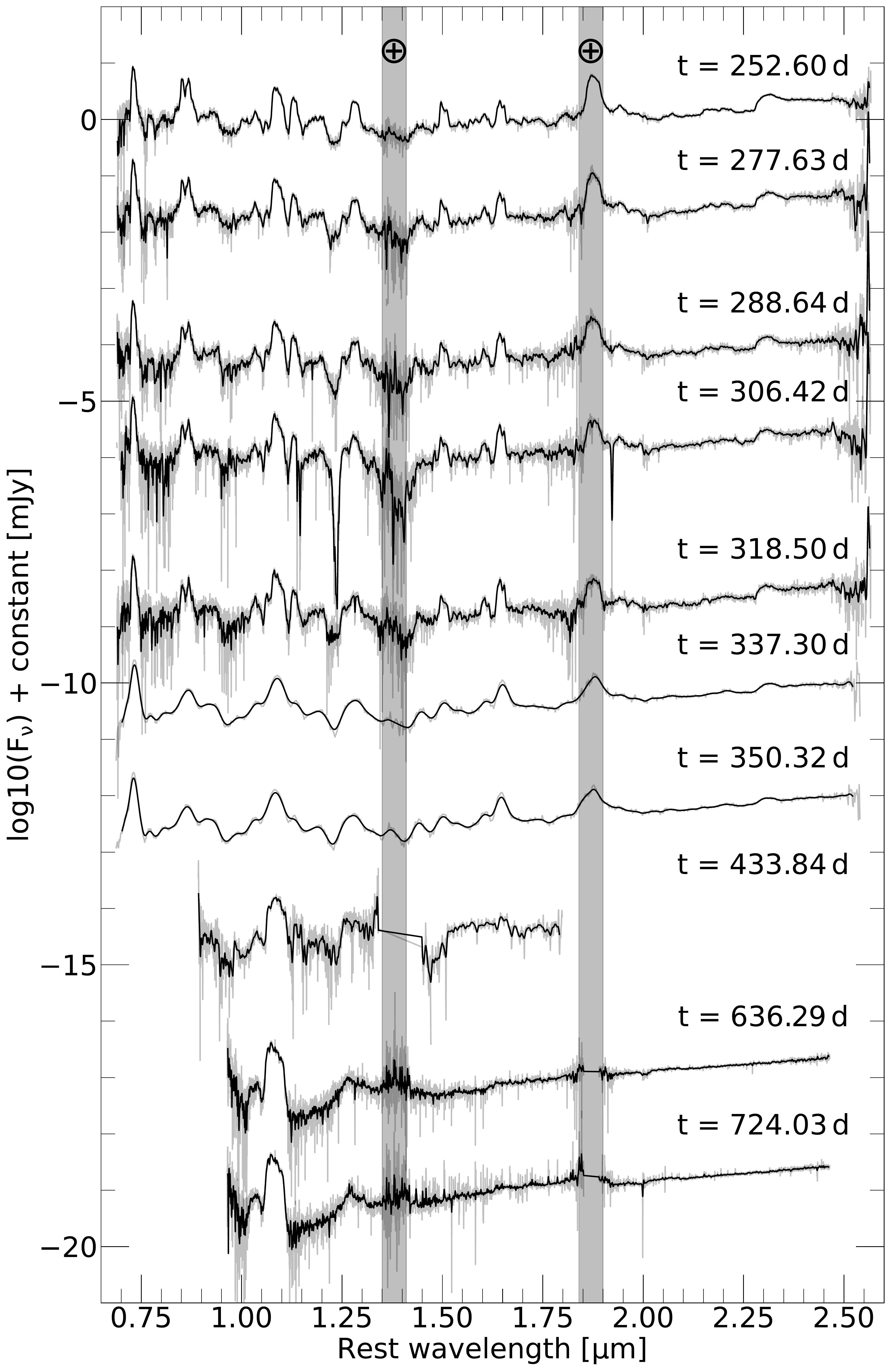}
    \caption{IRTF/SpeX and Keck/NIRES spectra of \ixf\ between $250$ and $725$~d. The telluric regions are highlighted by the gray regions. As \ixf\ evolved, the CO first overtone feature, $2.2 - 2.5$~\micron, declined in strength, and a strong IR excess developed in the NIR continuum above $\sim 1.2$~\micron, corresponding to hot dust.}
    \label{app:All_NIR_spec}
\end{figure*}

In addition, we show the full wavelength coverage of the ground based optical and NIR spectra of \ixf\ obtained between $+72.16 - 761.97$~d. These spectra are shown in Figs.~\ref{app:All_opt_spec} and \ref{app:All_NIR_spec}, respectively.

\begin{deluxetable}{ccccc}
    \tablecaption{Log of \jwst\ spectroscopic observations}
    \label{tab:JWST_spec_info}
    \tablehead{\colhead{Parameter} & \colhead{Visit 2} & \colhead{Visit 3} & \colhead{Visit 4} & \colhead{Visit 5} }
    \startdata
    \multicolumn{5}{c}{NIRSpec Spectral Observations} \\
    \hline
    Mode & Fixed Slit & Fixed Slit & Fixed Slit & Fixed Slit \\
    Slit & S400A1 & S400A1 & S400A1 & S400A1 \\
    \multirow{2}{*}{Grating / Filter} & G235M-F170LP/ & G235M-F170LP/ & G235M-F170LP/ & G235M-F170LP/ \\
     & G395M-F290LP & G395M-F290LP & G395M-F290LP & G395M-F290LP \\
    Exp Time / dither (s) & 63.90/63.90 & 157.38/157.38 & 163.61/126.22 & 163.61/126.22 \\
    Phase (days) & 252.66/252.65 & 373.52/373.51 & 600.21/600.20 & 722.90/722.98 \\
    $T_{\rm obs}$ (MJD) & 60335.61/60335.60 & 60456.57/60456.56 & 60683.44/60683.42 & 60806.23/60806.22 \\
    \hline
    \multicolumn{5}{c}{MIRI Spectral Observations} \\
    \hline
    Mode & LRS & LRS & LRS & LRS \\
    Exp Time / Dither (s) & 255.30 & 505.06 & 277.50 & 277.50 \\
    Phase (days) & 252.69 & 373.54 & 600.23 & 714.08 \\
    $T_{\rm obs}$ (MJD) & 60335.64 & 60456.58 & 60683.46 & 714.08 \\
    \enddata
\end{deluxetable}

\begin{deluxetable}{cccc}
    \tablecaption{Log of GB-optical and NIR spectra of \ixf obtained between $+250 - +765$~d. All phases are given in rest frame}
    \label{tab:GB_spec_info}
    \tablehead{\colhead{Obs. Date} & \colhead{Phase} & \colhead{Exp. Time} & \colhead{Telescope/Instrument} \\
    \colhead{[$MJD$]} & \colhead{[$days$]} & \colhead{[$s$]} & }
    \startdata
    \\
    \multicolumn{4}{c}{\Large{Optical-spectra}} \\
    \\
    \hline
    60154.97 & 72.16 & 600 & INT/IDS \\
    60155.02 & 73.21 & 600 & INT/IDS \\
    60156.92 & 74.11 & 2 $\times$ 600 & INT/IDS \\
    60169.26 & 86.44 & 2 $\times$ 1800 & UH88/SNIFS \\
    60197.23 & 114.39 & 1800 & UH88/SNIFS \\
    60295.64 & 212.72 & 1800 & UH88/SNIFS \\
    60307.62 & 224.69 & 1800 & UH88/SNIFS \\ 
    60340 & 258.39 & 2 $\times$ 900 & GTC/OSIRIS+ \\
    60361 & 278.03 & 4 $\times$ 450 & GTC/OSIRIS+ \\
    60361.59 & 278.62 & 2$\times$ 1800 & UH88/SNIFS \\
    60392.37 & 309.37 & 1800/1200 & UH88/SNIFS \\
    60402.59 & 319.58 & 1800 & UH88/SNIFS \\
    60429.48 & 346.45 & 1800 & UH88/SNIFS \\
    60459 & 375.95 & 2 $\times$ 600 & GTC/OSIRIS+ \\
    60685 & 601.77 & 4 $\times$ 500 & GTC/OSIRIS+ \\
    60845.33 & 761.97 & 6 $\times$ 1800 / 12 $\times$ 900 & Keck-I/LRIS \\
    \hline 
    \\
    \multicolumn{4}{c}{\Large{NIR-spectra}} \\
    \\
    \hline
    60335.55 & 252.60 & 44 $\times$ 120 & IRTF/SpeX \\
    60360.60 & 277.63 & 14 $\times$ 120 & IRTF/SpeX \\
    60371.62 & 288.64 & 8 $\times$ 120 & IRTF/SpeX \\
    60389.42 & 306.42 & 18 $\times$ 120 & IRTF/SpeX \\
    60401.50 & 318.50 & 30 $\times$ 120 & IRTF/SpeX \\
    60420.32 & 337.30 & 14 $\times$ 120$^{a}$ & IRTF/SpeX \\
    60433.35 & 350.32 & 10 $\times$ 120$^{a}$ & IRTF/SpeX \\
    60516.94  & 433.84 & 121 & GTC/EMIR \\
    60719.55 & 636.29 & 12 $\times$ 300.0 & Keck-II/NIRES \\
    60807.36 & 724.03 & 22 $\times$ 300.0 & Keck-II/NIRES 
    \enddata
    \tablecomments{$^{a}$ - Denotes observations taken in PRISM mode of IRTF/SpeX}
\end{deluxetable}

\bibliography{Refs}{}

\begin{thebibliography}{}
\expandafter\ifx\csname natexlab\endcsname\relax\def\natexlab#1{#1}\fi
\providecommand{\url}[1]{\href{#1}{#1}}
\providecommand{\dodoi}[1]{doi:~\href{http://doi.org/#1}{\nolinkurl{#1}}}
\providecommand{\doeprint}[1]{\href{http://ascl.net/#1}{\nolinkurl{http://ascl.net/#1}}}
\providecommand{\doarXiv}[1]{\href{https://arxiv.org/abs/#1}{\nolinkurl{https://arxiv.org/abs/#1}}}

\bibitem[{{Abac} {et~al.}(2025){Abac}, {Abbott}, {Abouelfettouh}, {Acernese}, {Ackley}, {Adhicary}, {Adhikari}, {Adhikari}, {Adkins}, {Agarwal}, {Agathos}, {Aghaei Abchouyeh}, {Aguiar}, {Aguilar}, {Aiello}, {Ain}, {Akutsu}, {Albanesi}, {Alfaidi}, {Al-Jodah}, {All{\'e}n{\'e}}, {Allocca}, {Al-Shammari}, {Altin}, {Alvarez-Lopez}, {Amato}, {Amez-Droz}, {Amorosi}, {Amra}, {Ananyeva}, {Anderson}, {Anderson}, {Andia}, {Ando}, {Andrade}, {Andres}, {Andr{\'e}s-Carcasona}, {Andri{\'c}}, {Anglin}, {Ansoldi}, {Antelis}, {Antier}, {Aoumi}, {Appavuravther}, {Appert}, {Apple}, {Arai}, {Araya}, {Araya}, {Areeda}, {Argianas}, {Aritomi}, {Armato}, {Arnaud}, {Arogeti}, {Aronson}, {Ashton}, {Aso}, {Assiduo}, {Assis de Souza Melo}, {Aston}, {Astone}, {Attadio}, {Aubin}, {Aultoneal}, {Avallone}, {Babak}, {Badaracco}, {Badger}, {Bae}, {Bagnasco}, {Bagui}, {Baier}, {Baiotti}, {Bajpai}, {Baka}, {Ball}, {Ballardin}, {Ballmer}, {Banagiri}, {Banerjee}, {Bankar}, {Baral}, {Barayoga}, {Barish}, {Barker}, {Barneo}, {Barone}, {Barr},
  {Barsotti}, {Barsuglia}, {Barta}, {Bartoletti}, {Barton}, {Bartos}, {Basak}, {Basalaev}, {Bassiri}, {Basti}, {Bates}, {Bawaj}, {Baxi}, {Bayley}, {Baylor}, {Baynard}, {Bazzan}, {Bedakihale}, {Beirnaert}, {Bejger}, {Belardinelli}, {Bell}, {Benedetto}, {Benoit}, {Bentley}, {Ben Yaala}, {Bera}, {Berbel}, {Bergamin}, {Berger}, {Bernuzzi}, {Beroiz}, {Bersanetti}, {Bertolini}, {Betzwieser}, {Beveridge}, {Bevins}, {Bhandare}, {Bhardwaj}, {Bhatt}, {Bhattacharjee}, {Bhaumik}, {Bhowmick}, {Bianchi}, {Bilenko}, {Billingsley}, {Binetti}, {Bini}, {Birnholtz}, {Biscoveanu}, {Bisht}, {Bitossi}, {Bizouard}, {Blackburn}, {Blagg}, {Blair}, {Blair}, {Bobba}, {Bode}, {Boileau}, {Boldrini}, {Bolingbroke}, {Bolliand}, {Bonavena}, {Bondarescu}, {Bondu}, {Bonilla}, {Bonilla}, {Bonino}, {Bonnand}, {Booker}, {Borchers}, {Boschi}, {Bose}, {Bossilkov}, {Boudart}, {Boudon}, {Bozzi}, {Bradaschia}, {Brady}, {Braglia}, {Branch}, {Branchesi}, {Brandt}, {Braun}, {Breschi}, {Briant}, {Brillet}, {Brinkmann}, {Brockill}, {Brockmueller},
  {Brooks}, {Brown}, {Brown}, {Brozzetti}, {Brunett}, {Bruno}, {Bruntz}, {Bryant}, {Bucci}, {Buchanan}, {Bulashenko}, {Bulik}, {Bulten}, {Buonanno}, {Burtnyk}, {Buscicchio}, {Buskulic}, {Buy}, {Byer}, \& {Cabourn Davies}}]{Abac_2025}
{Abac}, A.~G., {Abbott}, R., {Abouelfettouh}, I., {et~al.} 2025, \apj, 985, 183, \dodoi{10.3847/1538-4357/adc681}

\bibitem[{{Aldering} {et~al.}(2002){Aldering}, {Adam}, {Antilogus}, {Astier}, {Bacon}, {Bongard}, {Bonnaud}, {Copin}, {Hardin}, {Henault}, {Howell}, {Lemonnier}, {Levy}, {Loken}, {Nugent}, {Pain}, {Pecontal}, {Pecontal}, {Perlmutter}, {Quimby}, {Schahmaneche}, {Smadja}, \& {Wood-Vasey}}]{Aldering_2002}
{Aldering}, G., {Adam}, G., {Antilogus}, P., {et~al.} 2002, in Society of Photo-Optical Instrumentation Engineers (SPIE) Conference Series, Vol. 4836, Survey and Other Telescope Technologies and Discoveries, ed. J.~A. {Tyson} \& S.~{Wolff}, 61--72, \dodoi{10.1117/12.458107}

\bibitem[{{Alvarez} {et~al.}(1998){Alvarez}, {Rodr{\'\i}guez Espinosa}, \& {S{\'a}nchez}}]{Alverez_1998}
{Alvarez}, P., {Rodr{\'\i}guez Espinosa}, J.~M., \& {S{\'a}nchez}, F. 1998, \nar, 42, 553, \dodoi{10.1016/S1387-6473(98)00071-2}

\bibitem[{{Anderson} {et~al.}(2014){Anderson}, {Dessart}, {Gutierrez}, {Hamuy}, {Morrell}, {Phillips}, {Folatelli}, {Stritzinger}, {Freedman}, {Gonz{\'a}lez-Gait{\'a}n}, {McCarthy}, {Suntzeff}, \& {Thomas-Osip}}]{Anderson14}
{Anderson}, J.~P., {Dessart}, L., {Gutierrez}, C.~P., {et~al.} 2014, \mnras, 441, 671, \dodoi{10.1093/mnras/stu610}

\bibitem[{{Andrews} {et~al.}(2011{\natexlab{a}}){Andrews}, {Sugerman}, {Clayton}, {Gallagher}, {Barlow}, {Clem}, {Ercolano}, {Fabbri}, {Meixner}, {Otsuka}, {Welch}, \& {Wesson}}]{Andrews_2011a}
{Andrews}, J.~E., {Sugerman}, B.~E.~K., {Clayton}, G.~C., {et~al.} 2011{\natexlab{a}}, \apj, 731, 47, \dodoi{10.1088/0004-637X/731/1/47}

\bibitem[{{Andrews} {et~al.}(2011{\natexlab{b}}){Andrews}, {Clayton}, {Wesson}, {Sugerman}, {Barlow}, {Clem}, {Ercolano}, {Fabbri}, {Gallagher}, {Landolt}, {Meixner}, {Otsuka}, {Riebel}, \& {Welch}}]{Andrews_2011b}
{Andrews}, J.~E., {Clayton}, G.~C., {Wesson}, R., {et~al.} 2011{\natexlab{b}}, \aj, 142, 45, \dodoi{10.1088/0004-6256/142/2/45}

\bibitem[{{Andrews} {et~al.}(2019){Andrews}, {Sand}, {Valenti}, {Smith}, {Dastidar}, {Sahu}, {Misra}, {Singh}, {Hiramatsu}, {Brown}, {Hosseinzadeh}, {Wyatt}, {Vinko}, {Anupama}, {Arcavi}, {Ashall}, {Benetti}, {Berton}, {Bostroem}, {Bulla}, {Burke}, {Chen}, {Chomiuk}, {Cikota}, {Congiu}, {Cseh}, {Davis}, {Elias-Rosa}, {Faran}, {Fraser}, {Galbany}, {Gall}, {Gal-Yam}, {Gangopadhyay}, {Gromadzki}, {Haislip}, {Howell}, {Hsiao}, {Inserra}, {Kankare}, {Kuncarayakti}, {Kouprianov}, {Kumar}, {Li}, {Lin}, {Maguire}, {Mazzali}, {McCully}, {Milne}, {Mo}, {Morrell}, {Nicholl}, {Ochner}, {Olivares}, {Pastorello}, {Patat}, {Phillips}, {Pignata}, {Prentice}, {Reguitti}, {Reichart}, {Rodr{\'\i}guez}, {Rui}, {Sanwal}, {S{\'a}rneczky}, {Shahbandeh}, {Singh}, {Smartt}, {Strader}, {Stritzinger}, {Szak{\'a}ts}, {Tartaglia}, {Wang}, {Wang}, {Wang}, {Wheeler}, {Xiang}, {Yaron}, {Young}, \& {Zhang}}]{Andrews_2019}
{Andrews}, J.~E., {Sand}, D.~J., {Valenti}, S., {et~al.} 2019, \apj, 885, 43, \dodoi{10.3847/1538-4357/ab43e3}

\bibitem[{{Arnett} {et~al.}(1989){Arnett}, {Bahcall}, {Kirshner}, \& {Woosley}}]{Arnett_1989}
{Arnett}, W.~D., {Bahcall}, J.~N., {Kirshner}, R.~P., \& {Woosley}, S.~E. 1989, \araa, 27, 629, \dodoi{10.1146/annurev.aa.27.090189.003213}

\bibitem[{{Ashall} {et~al.}(2023{\natexlab{a}}){Ashall}, {Baron}, {DerKacy}, {Hoeflich}, {Shahbandeh}, {Baade}, {Brown}, {Burns}, {Engesser}, {Fox}, {Galbany}, {Guolo}, {Hsiao}, {Kumar}, {Lu}, {Mazzali}, {Medler}, {Mera Evans}, {Morrell}, {Phillips}, {Rest}, {Stritzinger}, {Strolger}, {Suntzeff}, {Temim}, {Tinyanont}, {Tucker}, {Wang}, \& {de Jaeger}}]{Ashall2023_cycle2_23ixf}
{Ashall}, C., {Baron}, E., {DerKacy}, J.~M., {et~al.} 2023{\natexlab{a}}, {Dust Our Luck? Measuring Molecule and Dust Formation in M101's Hydrogen-rich SN 2023ixf}, JWST Proposal. Cycle 2, ID. \#4575

\bibitem[{{Ashall} {et~al.}(2023{\natexlab{b}}){Ashall}, {Baron}, {DerKacy}, {Hoeflich}, {Shahbandeh}, {Baade}, {Brown}, {Burns}, {Engesser}, {Fox}, {Galbany}, {Guolo}, {Hsiao}, {Kumar}, {Lu}, {Mazzali}, {Medler}, {Mera Evans}, {Morrell}, {Phillips}, {Rest}, {Stritzinger}, {Strolger}, {Suntzeff}, {Temim}, {Tinyanont}, {Tucker}, {Wang}, \& {de Jaeger}}]{Ashall2023_cycle1_23ixf}
---. 2023{\natexlab{b}}, {Dust Our Luck? Measuring Molecule and Dust Formation in M101's Hydrogen-rich SN 2023ixf}, JWST Proposal. Cycle 1, ID. \#4522

\bibitem[{{Ashall} {et~al.}(2024{\natexlab{a}}){Ashall}, {Hoeflich}, {Baron}, {Shahbandeh}, {DerKacy}, {Medler}, {Shappee}, {Tucker}, {Fereidouni}, {Mera}, {Andrews}, {Baade}, {Bostroem}, {Brown}, {Burns}, {Burrow}, {Cikota}, {de Jaeger}, {Do}, {Dong}, {Dominguez}, {Fox}, {Galbany}, {Hsiao}, {Krisciunas}, {Khaghani}, {Kumar}, {Lu}, {Maund}, {Mazzali}, {Morrell}, {Patat}, {Pfeffer}, {Phillips}, {Schmidt}, {Stangl}, {Stevens}, {Stritzinger}, {Suntzeff}, {Telesco}, {Wang}, \& {Yang}}]{Ashall_2024}
{Ashall}, C., {Hoeflich}, P., {Baron}, E., {et~al.} 2024{\natexlab{a}}, \apj, 975, 203, \dodoi{10.3847/1538-4357/ad6608}

\bibitem[{{Ashall} {et~al.}(2024{\natexlab{b}}){Ashall}, {Hoeflich}, {Shahbandeh}, {Baade}, {Baron}, {Brown}, {Burns}, {DerKacy}, {Engesser}, {Fox}, {Galbany}, {Hsiao}, {Johansson}, {Krisciunas}, {Kumar}, {Lu}, {Matsuura}, {Mazzali}, {Medler}, {Mera Evans}, {Phillips}, {Rest}, {Sarangi}, {Stritzinger}, {Strolger}, {Suntzeff}, {Szalai}, {Temim}, {Tinyanont}, {Tucker}, {Van Dyk}, {Wang}, {Wesson}, {Zsiros}, \& {de Jaeger}}]{Ashall2024_cycle3_23ixf}
{Ashall}, C., {Hoeflich}, P.~A., {Shahbandeh}, M., {et~al.} 2024{\natexlab{b}}, {Building the Legacy of Supernova 2023ixf: How Does Molecule Formation Lead to Dust?}, JWST Proposal. Cycle 3, ID. \#5290

\bibitem[{{Astropy Collaboration} {et~al.}(2013){Astropy Collaboration}, {Robitaille}, {Tollerud}, {Greenfield}, {Droettboom}, {Bray}, {Aldcroft}, {Davis}, {Ginsburg}, {Price-Whelan}, {Kerzendorf}, {Conley}, {Crighton}, {Barbary}, {Muna}, {Ferguson}, {Grollier}, {Parikh}, {Nair}, {Unther}, {Deil}, {Woillez}, {Conseil}, {Kramer}, {Turner}, {Singer}, {Fox}, {Weaver}, {Zabalza}, {Edwards}, {Azalee Bostroem}, {Burke}, {Casey}, {Crawford}, {Dencheva}, {Ely}, {Jenness}, {Labrie}, {Lim}, {Pierfederici}, {Pontzen}, {Ptak}, {Refsdal}, {Servillat}, \& {Streicher}}]{Astropy_2013}
{Astropy Collaboration}, {Robitaille}, T.~P., {Tollerud}, E.~J., {et~al.} 2013, \aap, 558, A33, \dodoi{10.1051/0004-6361/201322068}

\bibitem[{{Baron} {et~al.}(2025)}]{Baron_2025}
{Baron}, E., {et~al.} 2025, \apj, submitted

\bibitem[{{Berger} {et~al.}(2023){Berger}, {Keating}, {Margutti}, {Maeda}, {Alexander}, {Cendes}, {Eftekhari}, {Gurwell}, {Hiramatsu}, {Ho}, {Laskar}, {Rao}, \& {Williams}}]{Berger_2023}
{Berger}, E., {Keating}, G.~K., {Margutti}, R., {et~al.} 2023, \apjl, 951, L31, \dodoi{10.3847/2041-8213/ace0c4}

\bibitem[{{Bersten} {et~al.}(2024){Bersten}, {Orellana}, {Folatelli}, {Martinez}, {Piccirilli}, {Regna}, {Rom{\'a}n Aguilar}, \& {Ertini}}]{Bersten_2024}
{Bersten}, M.~C., {Orellana}, M., {Folatelli}, G., {et~al.} 2024, \aap, 681, L18, \dodoi{10.1051/0004-6361/202348183}

\bibitem[{{Bevan}(2018)}]{Bevan18}
{Bevan}, A. 2018, \mnras, 480, 4659, \dodoi{10.1093/mnras/sty2094}

\bibitem[{{Bevan} \& {Barlow}(2016{\natexlab{a}})}]{Bevan_2016}
{Bevan}, A., \& {Barlow}, M.~J. 2016{\natexlab{a}}, \mnras, 456, 1269, \dodoi{10.1093/mnras/stv2651}

\bibitem[{{Bevan} \& {Barlow}(2016{\natexlab{b}})}]{Bevan16}
---. 2016{\natexlab{b}}, \mnras, 456, 1269, \dodoi{10.1093/mnras/stv2651}

\bibitem[{{Bevan} {et~al.}(2019){Bevan}, {Wesson}, {Barlow}, {De Looze}, {Andrews}, {Clayton}, {Krafton}, {Matsuura}, \& {Milisavljevic}}]{Bevan_2019}
{Bevan}, A., {Wesson}, R., {Barlow}, M.~J., {et~al.} 2019, \mnras, 485, 5192, \dodoi{10.1093/mnras/stz679}

\bibitem[{{Biscaro} \& {Cherchneff}(2014)}]{Biscaro_2014}
{Biscaro}, C., \& {Cherchneff}, I. 2014, \aap, 564, A25, \dodoi{10.1051/0004-6361/201322932}

\bibitem[{{Bode} \& {Evans}(1980)}]{Bode_1980}
{Bode}, M.~F., \& {Evans}, A. 1980, \mnras, 193, 21P, \dodoi{10.1093/mnras/193.1.21P}

\bibitem[{{B{\"o}ker} {et~al.}(2023){B{\"o}ker}, {Beck}, {Birkmann}, {Giardino}, {Keyes}, {Kumari}, {Muzerolle}, {Rawle}, {Zeidler}, {Abul-Huda}, {Alves de Oliveira}, {Arribas}, {Bechtold}, {Bhatawdekar}, {Bonaventura}, {Bunker}, {Cameron}, {Carniani}, {Charlot}, {Curti}, {Espinoza}, {Ferruit}, {Franx}, {Jakobsen}, {Karakla}, {L{\'o}pez-Caniego}, {L{\"u}tzgendorf}, {Maiolino}, {Manjavacas}, {Marston}, {Moseley}, {Ogle}, {Perna}, {Pe{\~n}a-Guerrero}, {Pirzkal}, {Plesha}, {Proffitt}, {Rauscher}, {Rix}, {Rodr{\'\i}guez del Pino}, {Rustamkulov}, {Sabbi}, {Sing}, {Sirianni}, {te Plate}, {{\'U}beda}, {Wahlgren}, {Wislowski}, {Wu}, \& {Willott}}]{Boker_2023}
{B{\"o}ker}, T., {Beck}, T.~L., {Birkmann}, S.~M., {et~al.} 2023, \pasp, 135, 038001, \dodoi{10.1088/1538-3873/acb846}

\bibitem[{{Bose} {et~al.}(2019){Bose}, {Dong}, {Elias-Rosa}, {Shappee}, {Bersier}, {Benetti}, {Stritzinger}, {Grupe}, {Kochanek}, {Prieto}, {Chen}, {Kuncarayakti}, {Mattila}, {Morales-Garoffolo}, {Morrell}, {Onori}, {Reynolds}, {Siviero}, {Somero}, {Stanek}, {Terreran}, {Thompson}, {Tomasella}, {Ashall}, {Gall}, {Gromadzki}, \& {Holoien}}]{Bose_2019}
{Bose}, S., {Dong}, S., {Elias-Rosa}, N., {et~al.} 2019, \apjl, 873, L3, \dodoi{10.3847/2041-8213/ab0558}

\bibitem[{{Bostroem} {et~al.}(2023){Bostroem}, {Pearson}, {Shrestha}, {Sand}, {Valenti}, {Jha}, {Andrews}, {Smith}, {Terreran}, {Green}, {Dong}, {Lundquist}, {Haislip}, {Hoang}, {Hosseinzadeh}, {Janzen}, {Jencson}, {Kouprianov}, {Paraskeva}, {Meza Retamal}, {Reichart}, {Arcavi}, {Bonanos}, {Coughlin}, {Dobson}, {Farah}, {Galbany}, {Guti{\'e}rrez}, {Hawley}, {Hebb}, {Hiramatsu}, {Howell}, {Iijima}, {Ilyin}, {Jhass}, {McCully}, {Moran}, {Morris}, {Mura}, {M{\"u}ller-Bravo}, {Munday}, {Newsome}, {Pabst}, {Ochner}, {Gonzalez}, {Pastorello}, {Pellegrino}, {Piscarreta}, {Ravi}, {Reguitti}, {Salo}, {Vink{\'o}}, {de Vos}, {Wheeler}, {Williams}, \& {Wyatt}}]{Bostroem_2023}
{Bostroem}, K.~A., {Pearson}, J., {Shrestha}, M., {et~al.} 2023, \apjl, 956, L5, \dodoi{10.3847/2041-8213/acf9a4}

\bibitem[{{Burrow} {et~al.}(2020){Burrow}, {Baron}, {Ashall}, {Burns}, {Morrell}, {Stritzinger}, {Brown}, {Folatelli}, {Freedman}, {Galbany}, {Hoeflich}, {Hsiao}, {Krisciunas}, {Phillips}, {Piro}, {Suntzeff}, \& {Uddin}}]{sextractor_2020}
{Burrow}, A., {Baron}, E., {Ashall}, C., {et~al.} 2020, \apj, 901, 154, \dodoi{10.3847/1538-4357/abafa2}

\bibitem[{{Bushouse} {et~al.}(2022){Bushouse}, {Eisenhamer}, {Dencheva}, {Davies}, {Greenfield}, {Morrison}, {Hodge}, {Simon}, {Grumm}, {Droettboom}, {Slavich}, {Sosey}, {Pauly}, {Miller}, {Jedrzejewski}, {Hack}, {Davis}, {Crawford}, {Law}, {Gordon}, {Regan}, {Cara}, {MacDonald}, {Bradley}, {Shanahan}, \& {Jamieson}}]{Bushouse_2022_JWST_reduc}
{Bushouse}, H., {Eisenhamer}, J., {Dencheva}, N., {et~al.} 2022, {JWST Calibration Pipeline}, Tech. rep., \dodoi{10.5281/zenodo.7038885}

\bibitem[{{Cardiel} {et~al.}(2019){Cardiel}, {Pascual}, {Gallego}, {Cabello}, {Garz{\'o}n}, {Balcells}, {Castro-Rodr{\'\i}guez}, {Dom{\'\i}nguez-Palmero}, {Hammersley}, {Laporte}, {Patrick}, {Pell{\'o}}, {Prieto}, \& {Streblyanska}}]{Cardiel_2019}
{Cardiel}, N., {Pascual}, S., {Gallego}, J., {et~al.} 2019, in Astronomical Society of the Pacific Conference Series, Vol. 523, Astronomical Data Analysis Software and Systems XXVII, ed. P.~J. {Teuben}, M.~W. {Pound}, B.~A. {Thomas}, \& E.~M. {Warner}, 317

\bibitem[{{Catchpole} {et~al.}(1987){Catchpole}, {Glass}, {Karovska}, {Nisenson}, {Papaliolios}, {Standley}, {Flower}, {Seargent}, \& {McNaught}}]{Catchpole_1987}
{Catchpole}, R., {Glass}, I., {Karovska}, M., {et~al.} 1987, \iaucirc, 4457, 1

\bibitem[{{Chandra} {et~al.}(2024){Chandra}, {Chevalier}, {Maeda}, {Ray}, \& {Nayana}}]{Chandra_2023}
{Chandra}, P., {Chevalier}, R.~A., {Maeda}, K., {Ray}, A.~K., \& {Nayana}, A.~J. 2024, \apjl, 963, L4, \dodoi{10.3847/2041-8213/ad275d}

\bibitem[{{Chevalier} \& {Fransson}(1987)}]{Chevalier_1987}
{Chevalier}, R.~A., \& {Fransson}, C. 1987, \nat, 328, 44, \dodoi{10.1038/328044a0}

\bibitem[{{Chiar} \& {Tielens}(2006)}]{Chiar_2006}
{Chiar}, J.~E., \& {Tielens}, A.~G.~G.~M. 2006, \apj, 637, 774, \dodoi{10.1086/498406}

\bibitem[{{Clayton} {et~al.}(2001){Clayton}, {Deneault}, \& {Meyer}}]{Clayton_2001}
{Clayton}, D.~D., {Deneault}, E. A.~N., \& {Meyer}, B.~S. 2001, \apj, 562, 480, \dodoi{10.1086/323467}

\bibitem[{{Clayton} {et~al.}(2025){Clayton}, {Wesson}, {Fox}, {Shahbandeh}, {Filippenko}, {Nickson}, {Engesser}, {Van Dyk}, {Zheng}, {Brink}, {Yang}, {Temim}, {Smith}, {Andrews}, {Ashall}, {De Looze}, {Derkacy}, {Dessart}, {Dulude}, {Dwek}, {Foley}, {Gezari}, {Gomez}, {Gonzaga}, {Indukuri}, {Jencson}, {Johansson}, {Kasliwal}, {Lane}, {Lau}, {Law}, {Marston}, {Milisavljevic}, {O'Steen}, {Pierel}, {Rest}, {Sarangi}, {Siebert}, {Skrutskie}, {Strolger}, {Szalai}, {Tinyanont}, {Wang}, {Williams}, {Xiao}, \& {Zsiros}}]{Clayton_2025}
{Clayton}, G.~C., {Wesson}, R., {Fox}, O.~D., {et~al.} 2025, arXiv e-prints, arXiv:2505.01574, \dodoi{10.48550/arXiv.2505.01574}

\bibitem[{{Cushing} {et~al.}(2004){Cushing}, {Vacca}, \& {Rayner}}]{Cushing_2004}
{Cushing}, M.~C., {Vacca}, W.~D., \& {Rayner}, J.~T. 2004, \pasp, 116, 362, \dodoi{10.1086/382907}

\bibitem[{{Davis} {et~al.}(2019){Davis}, {Hsiao}, {Ashall}, {Hoeflich}, {Phillips}, {Marion}, {Kirshner}, {Morrell}, {Sand}, {Burns}, {Contreras}, {Stritzinger}, {Anderson}, {Baron}, {Diamond}, {Guti{\'e}rrez}, {Hamuy}, {Holmbo}, {Kasliwal}, {Krisciunas}, {Kumar}, {Lu}, {Pessi}, {Piro}, {Prieto}, {Shahbandeh}, \& {Suntzeff}}]{Davis_2019}
{Davis}, S., {Hsiao}, E.~Y., {Ashall}, C., {et~al.} 2019, \apj, 887, 4, \dodoi{10.3847/1538-4357/ab4c40}

\bibitem[{{DerKacy} {et~al.}(2025)}]{DerKacy_2025}
{DerKacy}, J., {et~al.} 2025, \apj, submitted

\bibitem[{{Dessart} \& {Hillier}(2005)}]{Dessart_2005}
{Dessart}, L., \& {Hillier}, D.~J. 2005, \aap, 437, 667, \dodoi{10.1051/0004-6361:20042525}

\bibitem[{{Dragulin} \& {Hoeflich}(2016)}]{Dragulin_2016}
{Dragulin}, P., \& {Hoeflich}, P. 2016, \apj, 818, 26, \dodoi{10.3847/0004-637X/818/1/26}

\bibitem[{{Dwek}(1983)}]{Dwek_1983}
{Dwek}, E. 1983, \apj, 274, 175, \dodoi{10.1086/161435}

\bibitem[{Dwek(2006)}]{Dwek_2006}
Dwek, E. 2006, Science, 313, 178, \dodoi{10.1126/science.1130423}

\bibitem[{{Dwek} {et~al.}(2007){Dwek}, {Galliano}, \& {Jones}}]{Dwek_2007}
{Dwek}, E., {Galliano}, F., \& {Jones}, A.~P. 2007, \apj, 662, 927, \dodoi{10.1086/518430}

\bibitem[{{Dwek} \& {Scalo}(1980)}]{Dwek_1980}
{Dwek}, E., \& {Scalo}, J.~M. 1980, \apj, 239, 193, \dodoi{10.1086/158100}

\bibitem[{{Dyer} {et~al.}(2024){Dyer}, {Ackley}, {Jim{\'e}nez-Ibarra}, {Lyman}, {Ulaczyk}, {Steeghs}, {Galloway}, {Dhillon}, {O'Brien}, {Ramsay}, {Noysena}, {Kotak}, {Breton}, {Nuttall}, {Pall{\'e}}, {Pollacco}, {Killestein}, {Kumar}, {O'Neill}, {Kelsey}, {Godson}, \& {Jarvis}}]{Dyer24_GOTO}
{Dyer}, M.~J., {Ackley}, K., {Jim{\'e}nez-Ibarra}, F., {et~al.} 2024, in Society of Photo-Optical Instrumentation Engineers (SPIE) Conference Series, Vol. 13094, Ground-based and Airborne Telescopes X, ed. H.~K. {Marshall}, J.~{Spyromilio}, \& T.~{Usuda}, 130941X, \dodoi{10.1117/12.3018305}

\bibitem[{{Elmhamdi} {et~al.}(2003){Elmhamdi}, {Danziger}, {Chugai}, {Pastorello}, {Turatto}, {Cappellaro}, {Altavilla}, {Benetti}, {Patat}, \& {Salvo}}]{Elmhamdi_2003}
{Elmhamdi}, A., {Danziger}, I.~J., {Chugai}, N., {et~al.} 2003, \mnras, 338, 939, \dodoi{10.1046/j.1365-8711.2003.06150.x}

\bibitem[{{Ercolano} {et~al.}(2007){Ercolano}, {Barlow}, \& {Sugerman}}]{Ercolano_2007}
{Ercolano}, B., {Barlow}, M.~J., \& {Sugerman}, B.~E.~K. 2007, \mnras, 375, 753, \dodoi{10.1111/j.1365-2966.2006.11336.x}

\bibitem[{{Fan} {et~al.}(2003){Fan}, {Strauss}, {Schneider}, {Becker}, {White}, {Haiman}, {Gregg}, {Pentericci}, {Grebel}, {Narayanan}, {Loh}, {Richards}, {Gunn}, {Lupton}, {Knapp}, {Ivezi{\'c}}, {Brandt}, {Collinge}, {Hao}, {Harbeck}, {Prada}, {Schaye}, {Strateva}, {Zakamska}, {Anderson}, {Brinkmann}, {Bahcall}, {Lamb}, {Okamura}, {Szalay}, \& {York}}]{Fan_2003}
{Fan}, X., {Strauss}, M.~A., {Schneider}, D.~P., {et~al.} 2003, \aj, 125, 1649, \dodoi{10.1086/368246}

\bibitem[{{Fang} {et~al.}(2024){Fang}, {Maeda}, {Kuncarayakti}, \& {Nagao}}]{Fang_2024}
{Fang}, Q., {Maeda}, K., {Kuncarayakti}, H., \& {Nagao}, T. 2024, Nature Astronomy, 8, 111, \dodoi{10.1038/s41550-023-02120-8}

\bibitem[{{Fang} {et~al.}(2022){Fang}, {Maeda}, {Kuncarayakti}, {Tanaka}, {Kawabata}, {Hattori}, {Aoki}, {Moriya}, \& {Yamanaka}}]{Fang_2022}
{Fang}, Q., {Maeda}, K., {Kuncarayakti}, H., {et~al.} 2022, \apj, 928, 151, \dodoi{10.3847/1538-4357/ac4f60}

\bibitem[{{Fedriani} {et~al.}(2020){Fedriani}, {Caratti o Garatti}, {Koutoulaki}, {Garcia-Lopez}, {Natta}, {Cesaroni}, {Oudmaijer}, {Coffey}, {Ray}, \& {Stecklum}}]{Fedriani_2020}
{Fedriani}, R., {Caratti o Garatti}, A., {Koutoulaki}, M., {et~al.} 2020, \aap, 633, A128, \dodoi{10.1051/0004-6361/201936748}

\bibitem[{{Ferrari} {et~al.}(2024){Ferrari}, {Folatelli}, {Ertini}, {Kuncarayakti}, \& {Andrews}}]{Ferrari_2024}
{Ferrari}, L., {Folatelli}, G., {Ertini}, K., {Kuncarayakti}, H., \& {Andrews}, J.~E. 2024, \aap, 687, L20, \dodoi{10.1051/0004-6361/202450440}

\bibitem[{{Ferrarotti} \& {Gail}(2006)}]{Ferrarotti_2006}
{Ferrarotti}, A.~S., \& {Gail}, H.~P. 2006, \aap, 447, 553, \dodoi{10.1051/0004-6361:20041198}

\bibitem[{{Flinner} {et~al.}(2023){Flinner}, {Tucker}, {Beacom}, \& {Shappee}}]{Flinner_2023}
{Flinner}, N., {Tucker}, M.~A., {Beacom}, J.~F., \& {Shappee}, B.~J. 2023, Research Notes of the American Astronomical Society, 7, 174, \dodoi{10.3847/2515-5172/acefc4}

\bibitem[{{Folatelli} {et~al.}(2025){Folatelli}, {Ferrari}, {Ertini}, {Kuncarayakti}, \& {Maeda}}]{Folatelli_2025}
{Folatelli}, G., {Ferrari}, L., {Ertini}, K., {Kuncarayakti}, H., \& {Maeda}, K. 2025, \aap, 698, A213, \dodoi{10.1051/0004-6361/202554128}

\bibitem[{{Fox} {et~al.}(2024){Fox}, {Szalai}, {Van Dyk}, {Jencson}, {Shahbandeh}, {Filippenko}, {Gomez}, \& {Sarangi}}]{Fox_2024}
{Fox}, O., {Szalai}, T., {Van Dyk}, S.~D., {et~al.} 2024, {Building A Modern Sample of Dusty Supernovae with JWST}, JWST Proposal. Cycle 2, ID. \#3921

\bibitem[{{Gall} {et~al.}(2011){Gall}, {Hjorth}, \& {Andersen}}]{Gall_2011}
{Gall}, C., {Hjorth}, J., \& {Andersen}, A.~C. 2011, \aapr, 19, 43, \dodoi{10.1007/s00159-011-0043-7}

\bibitem[{{Gall} {et~al.}(2014){Gall}, {Hjorth}, {Watson}, {Dwek}, {Maund}, {Fox}, {Leloudas}, {Malesani}, \& {Day-Jones}}]{Gall_2014}
{Gall}, C., {Hjorth}, J., {Watson}, D., {et~al.} 2014, \nat, 511, 326, \dodoi{10.1038/nature13558}

\bibitem[{{Garz{\'o}n} {et~al.}(2022){Garz{\'o}n}, {Balcells}, {Gallego}, {Gry}, {Guzm{\'a}n}, {Hammersley}, {Herrero}, {Mu{\~n}oz-Tu{\~n}{\'o}n}, {Pell{\'o}}, {Prieto}, {Bourrec}, {Cabello}, {Cardiel}, {Gonz{\'a}lez-Fern{\'a}ndez}, {Laporte}, {Milliard}, {Pascual}, {Patrick}, {Patr{\'o}n}, {Ram{\'\i}rez-Alegr{\'\i}a}, \& {Streblyanska}}]{Grazon_2022}
{Garz{\'o}n}, F., {Balcells}, M., {Gallego}, J., {et~al.} 2022, \aap, 667, A107, \dodoi{10.1051/0004-6361/202244729}

\bibitem[{{Gerardy} {et~al.}(2000){Gerardy}, {Fesen}, {H{\"o}flich}, \& {Wheeler}}]{Gerardy_2000}
{Gerardy}, C.~L., {Fesen}, R.~A., {H{\"o}flich}, P., \& {Wheeler}, J.~C. 2000, \aj, 119, 2968, \dodoi{10.1086/301390}

\bibitem[{{Gordon}(2023)}]{Extinction_2023b}
{Gordon}, K. 2023, {karllark/dust\_extinction: OneRelationForAllWaves}, v1.2, Zenodo,  Zenodo, \dodoi{10.5281/zenodo.7799360}

\bibitem[{{Gordon} {et~al.}(2023){Gordon}, {Clayton}, {Decleir}, {Fitzpatrick}, {Massa}, {Misselt}, \& {Tollerud}}]{Extinction_2023a}
{Gordon}, K.~D., {Clayton}, G.~C., {Decleir}, M., {et~al.} 2023, \apj, 950, 86, \dodoi{10.3847/1538-4357/accb59}

\bibitem[{{Grefenstette} {et~al.}(2023){Grefenstette}, {Brightman}, {Earnshaw}, {Harrison}, \& {Margutti}}]{Grefenstette_2023}
{Grefenstette}, B.~W., {Brightman}, M., {Earnshaw}, H.~P., {Harrison}, F.~A., \& {Margutti}, R. 2023, \apjl, 952, L3, \dodoi{10.3847/2041-8213/acdf4e}

\bibitem[{{Groot} {et~al.}(2022){Groot}, {Bloemen}, {Vreeswijk}, {Jonker}, {Pieterse}, {Engels}, {Michiels}, {Bakker}, {Hahn}, {Raskin}, {Morren}, {Navarro}, {Elswijk}, {ter Horst}, {Schuil}, {Kragt}, {Lesman}, {de Haan}, {Bekema}, {de Haan}, {Klein-Wolt}, {Blagorodnova}, {Johnston}, \& {Le Poole}}]{Groot22_BlackGEM}
{Groot}, P.~J., {Bloemen}, S., {Vreeswijk}, P.~M., {et~al.} 2022, in Society of Photo-Optical Instrumentation Engineers (SPIE) Conference Series, Vol. 12182, Ground-based and Airborne Telescopes IX, ed. H.~K. {Marshall}, J.~{Spyromilio}, \& T.~{Usuda}, 121821V, \dodoi{10.1117/12.2630160}

\bibitem[{{Guetta} {et~al.}(2023){Guetta}, {Langella}, {Gagliardini}, \& {Della Valle}}]{Guetta_2023}
{Guetta}, D., {Langella}, A., {Gagliardini}, S., \& {Della Valle}, M. 2023, \apjl, 955, L9, \dodoi{10.3847/2041-8213/acf573}

\bibitem[{{Hanuschik}(1988)}]{Hanuschik_1988}
{Hanuschik}, R.~W. 1988, \pasa, 7, 446, \dodoi{10.1017/S1323358000022621}

\bibitem[{{Harris} {et~al.}(2020){Harris}, {Millman}, {van der Walt}, {Gommers}, {Virtanen}, {Cournapeau}, {Wieser}, {Taylor}, {Berg}, {Smith}, {Kern}, {Picus}, {Hoyer}, {van Kerkwijk}, {Brett}, {Haldane}, {del R{\'\i}o}, {Wiebe}, {Peterson}, {G{\'e}rard-Marchant}, {Sheppard}, {Reddy}, {Weckesser}, {Abbasi}, {Gohlke}, \& {Oliphant}}]{numpy_2020}
{Harris}, C.~R., {Millman}, K.~J., {van der Walt}, S.~J., {et~al.} 2020, \nat, 585, 357, \dodoi{10.1038/s41586-020-2649-2}

\bibitem[{{Hiramatsu} {et~al.}(2023){Hiramatsu}, {Tsuna}, {Berger}, {Itagaki}, {Goldberg}, {Gomez}, {Kishalay}, {Hosseinzadeh}, {Bostroem}, {Brown}, {Arcavi}, {Bieryla}, {Blanchard}, {Esquerdo}, {Farah}, {Howell}, {Matsumoto}, {McCully}, {Newsome}, {Gonzalez}, {Pellegrino}, {Rhee}, {Terreran}, {Vink{\'o}}, \& {Wheeler}}]{Hiramatsu_2023}
{Hiramatsu}, D., {Tsuna}, D., {Berger}, E., {et~al.} 2023, \apjl, 955, L8, \dodoi{10.3847/2041-8213/acf299}

\bibitem[{{Hoeflich}(1988)}]{Hoeflich_1988}
{Hoeflich}, P. 1988, \pasa, 7, 434, \dodoi{10.1017/S1323358000022608}

\bibitem[{{Hoogendam} {et~al.}(2025{\natexlab{a}}){Hoogendam}, {Jones}, {Ashall}, {Shappee}, {Foley}, {Tucker}, {Huber}, {Auchettl}, {Desai}, {Do}, {Hinkle}, {Romagnoli}, {Shi}, {Syncatto}, {Angus}, {Chambers}, {Coulter}, {Davis}, {de Boer}, {Gagliano}, {Kong}, {Lin}, {Lowe}, {Magnier}, {Minguez}, {Pan}, {Patra}, {Severson}, {Taggart}, {Wasserman}, \& {Yadavalli}}]{Hoogendam25_epr}
{Hoogendam}, W.~B., {Jones}, D.~O., {Ashall}, C., {et~al.} 2025{\natexlab{a}}, arXiv e-prints, arXiv:2502.17556, \dodoi{10.48550/arXiv.2502.17556}

\bibitem[{{Hoogendam} {et~al.}(2025{\natexlab{b}}){Hoogendam}, {Ashall}, {Jones}, {Shappee}, {Tucker}, {Huber}, {Auchettl}, {Desai}, {Hinkle}, {Kong}, {Romagnoli}, {Shi}, {Syncatto}, \& {Kilpatrick}}]{Hoogendam25_pxl}
{Hoogendam}, W.~B., {Ashall}, C., {Jones}, D.~O., {et~al.} 2025{\natexlab{b}}, arXiv e-prints, arXiv:2505.04610, \dodoi{10.48550/arXiv.2505.04610}

\bibitem[{{Hosseinzadeh} {et~al.}(2023){Hosseinzadeh}, {Farah}, {Shrestha}, {Sand}, {Dong}, {Brown}, {Bostroem}, {Valenti}, {Jha}, {Andrews}, {Arcavi}, {Haislip}, {Hiramatsu}, {Hoang}, {Howell}, {Janzen}, {Jencson}, {Kouprianov}, {Lundquist}, {McCully}, {Meza Retamal}, {Modjaz}, {Newsome}, {Padilla Gonzalez}, {Pearson}, {Pellegrino}, {Ravi}, {Reichart}, {Smith}, {Terreran}, \& {Vink{\'o}}}]{Hosseinzadeh_2023}
{Hosseinzadeh}, G., {Farah}, J., {Shrestha}, M., {et~al.} 2023, \apjl, 953, L16, \dodoi{10.3847/2041-8213/ace4c4}

\bibitem[{{Huber} {et~al.}(2015){Huber}, {Chambers}, {Flewelling}, {Willman}, {Primak}, {Schultz}, {Gibson}, {Magnier}, {Waters}, {Tonry}, {Wainscoat}, {Smith}, {Wright}, {Smartt}, {Foley}, {Jha}, {Rest}, \& {Scolnic}}]{Huber15}
{Huber}, M., {Chambers}, K.~C., {Flewelling}, H., {et~al.} 2015, The Astronomer's Telegram, 7153, 1

\bibitem[{{Hughes} {et~al.}(1997){Hughes}, {Dunlop}, \& {Rawlings}}]{Hughes_1997}
{Hughes}, D.~H., {Dunlop}, J.~S., \& {Rawlings}, S. 1997, \mnras, 289, 766, \dodoi{10.1093/mnras/289.4.766}

\bibitem[{{Hunter}(2007)}]{matplotlib_2007}
{Hunter}, J.~D. 2007, Computing in Science and Engineering, 9, 90, \dodoi{10.1109/MCSE.2007.55}

\bibitem[{{Ilee} {et~al.}(2018){Ilee}, {Oudmaijer}, {Wheelwright}, \& {Pomohaci}}]{Ilee_2018}
{Ilee}, J.~D., {Oudmaijer}, R.~D., {Wheelwright}, H.~E., \& {Pomohaci}, R. 2018, \mnras, 477, 3360, \dodoi{10.1093/mnras/sty863}

\bibitem[{{Inserra} {et~al.}(2011){Inserra}, {Turatto}, {Pastorello}, {Benetti}, {Cappellaro}, {Pumo}, {Zampieri}, {Agnoletto}, {Bufano}, {Botticella}, {Della Valle}, {Elias Rosa}, {Iijima}, {Spiro}, \& {Valenti}}]{Inserra_2011}
{Inserra}, C., {Turatto}, M., {Pastorello}, A., {et~al.} 2011, \mnras, 417, 261, \dodoi{10.1111/j.1365-2966.2011.19128.x}

\bibitem[{{Itagaki}(2023)}]{Itagaki_2023}
{Itagaki}, K. 2023, Transient Name Server Discovery Report, 2023-1158, 1

\bibitem[{{Iwata} {et~al.}(2025){Iwata}, {Akimoto}, {Matsuoka}, {Maeda}, {Yonekura}, {Tominaga}, {Moriya}, {Fujisawa}, {Niinuma}, {Yoon}, {Lee}, {Jung}, \& {Byun}}]{Iwata_2025}
{Iwata}, Y., {Akimoto}, M., {Matsuoka}, T., {et~al.} 2025, \apj, 978, 138, \dodoi{10.3847/1538-4357/ad9a62}

\bibitem[{{Jacobson-Gal{\'a}n} {et~al.}(2023){Jacobson-Gal{\'a}n}, {Dessart}, {Margutti}, {Chornock}, {Foley}, {Kilpatrick}, {Jones}, {Taggart}, {Angus}, {Bhattacharjee}, {Braff}, {Brethauer}, {Burgasser}, {Cao}, {Carlile}, {Chambers}, {Coulter}, {Dominguez-Ruiz}, {Dickinson}, {de Boer}, {Gagliano}, {Gall}, {Gao}, {Gates}, {Gomez}, {Guolo}, {Halford}, {Hjorth}, {Huber}, {Johnson}, {Karpoor}, {Laskar}, {LeBaron}, {Li}, {Lin}, {Loch}, {Lynam}, {Magnier}, {Maloney}, {Matthews}, {McDonald}, {Miao}, {Milisavljevic}, {Pan}, {Pradyumna}, {Ransome}, {Rees}, {Rest}, {Rojas-Bravo}, {Sandford}, {Ascencio}, {Sanjaripour}, {Savino}, {Sears}, {Sharei}, {Smartt}, {Softich}, {Theissen}, {Tinyanont}, {Tohfa}, {Villar}, {Wang}, {Wainscoat}, {Westerling}, {Wiston}, {Wozniak}, {Yadavalli}, \& {Zenati}}]{Jacobson-Gal_2023}
{Jacobson-Gal{\'a}n}, W.~V., {Dessart}, L., {Margutti}, R., {et~al.} 2023, \apjl, 954, L42, \dodoi{10.3847/2041-8213/acf2ec}

\bibitem[{{Jakobsen} {et~al.}(2022){Jakobsen}, {Ferruit}, {Alves de Oliveira}, {Arribas}, {Bagnasco}, {Barho}, {Beck}, {Birkmann}, {B{\"o}ker}, {Bunker}, {Charlot}, {de Jong}, {de Marchi}, {Ehrenwinkler}, {Falcolini}, {Fels}, {Franx}, {Franz}, {Funke}, {Giardino}, {Gnata}, {Holota}, {Honnen}, {Jensen}, {Jentsch}, {Johnson}, {Jollet}, {Karl}, {Kling}, {K{\"o}hler}, {Kolm}, {Kumari}, {Lander}, {Lemke}, {L{\'o}pez-Caniego}, {L{\"u}tzgendorf}, {Maiolino}, {Manjavacas}, {Marston}, {Maschmann}, {Maurer}, {Messerschmidt}, {Moseley}, {Mosner}, {Mott}, {Muzerolle}, {Pirzkal}, {Pittet}, {Plitzke}, {Posselt}, {Rapp}, {Rauscher}, {Rawle}, {Rix}, {R{\"o}del}, {Rumler}, {Sabbi}, {Salvignol}, {Schmid}, {Sirianni}, {Smith}, {Strada}, {te Plate}, {Valenti}, {Wettemann}, {Wiehe}, {Wiesmayer}, {Willott}, {Wright}, {Zeidler}, \& {Zincke}}]{Jakobsen_2022}
{Jakobsen}, P., {Ferruit}, P., {Alves de Oliveira}, C., {et~al.} 2022, \aap, 661, A80, \dodoi{10.1051/0004-6361/202142663}

\bibitem[{{Janka}(2012)}]{Janka_2012}
{Janka}, H.-T. 2012, Annual Review of Nuclear and Particle Science, 62, 407, \dodoi{10.1146/annurev-nucl-102711-094901}

\bibitem[{{Janka} {et~al.}(2016){Janka}, {Melson}, \& {Summa}}]{Janka16}
{Janka}, H.-T., {Melson}, T., \& {Summa}, A. 2016, Annual Review of Nuclear and Particle Science, 66, 341, \dodoi{10.1146/annurev-nucl-102115-044747}

\bibitem[{{Jencson} {et~al.}(2023){Jencson}, {Pearson}, {Beasor}, {Lau}, {Andrews}, {Bostroem}, {Dong}, {Engesser}, {Gomez}, {Guolo}, {Hoang}, {Hosseinzadeh}, {Jha}, {Karambelkar}, {Kasliwal}, {Lundquist}, {Meza Retamal}, {Rest}, {Sand}, {Shahbandeh}, {Shrestha}, {Smith}, {Strader}, {Valenti}, {Wang}, \& {Zenati}}]{Jencson_2023}
{Jencson}, J.~E., {Pearson}, J., {Beasor}, E.~R., {et~al.} 2023, \apjl, 952, L30, \dodoi{10.3847/2041-8213/ace618}

\bibitem[{{Jerkstrand} {et~al.}(2012){Jerkstrand}, {Fransson}, {Maguire}, {Smartt}, {Ergon}, \& {Spyromilio}}]{Jerkstrand_2012}
{Jerkstrand}, A., {Fransson}, C., {Maguire}, K., {et~al.} 2012, \aap, 546, A28, \dodoi{10.1051/0004-6361/201219528}

\bibitem[{{Kendrew} {et~al.}(2015){Kendrew}, {Scheithauer}, {Bouchet}, {Amiaux}, {Azzollini}, {Bouwman}, {Chen}, {Dubreuil}, {Fischer}, {Glasse}, {Greene}, {Lagage}, {Lahuis}, {Ronayette}, {Wright}, \& {Wright}}]{Kendrew_2015}
{Kendrew}, S., {Scheithauer}, S., {Bouchet}, P., {et~al.} 2015, \pasp, 127, 623, \dodoi{10.1086/682255}

\bibitem[{{Khokhlov} \& {H{\"o}flich}(2001)}]{Khokhlov01}
{Khokhlov}, A., \& {H{\"o}flich}, P. 2001, in American Institute of Physics Conference Series, Vol. 556, Explosive Phenomena in Astrophysical Compact Objects, ed. H.-Y. {Chang}, C.-H. {Lee}, M.~{Rho}, \& I.~{Yi} (AIP), 301--312, \dodoi{10.1063/1.1368287}

\bibitem[{{Kilpatrick} {et~al.}(2023){Kilpatrick}, {Foley}, {Jacobson-Gal{\'a}n}, {Piro}, {Smartt}, {Drout}, {Gagliano}, {Gall}, {Hjorth}, {Jones}, {Mandel}, {Margutti}, {Ramirez-Ruiz}, {Ransome}, {Villar}, {Coulter}, {Gao}, {Matthews}, {Taggart}, \& {Zenati}}]{Kilpatrick_2023}
{Kilpatrick}, C.~D., {Foley}, R.~J., {Jacobson-Gal{\'a}n}, W.~V., {et~al.} 2023, \apjl, 952, L23, \dodoi{10.3847/2041-8213/ace4ca}

\bibitem[{{Kotak} {et~al.}(2005){Kotak}, {Meikle}, {van Dyk}, {H{\"o}flich}, \& {Mattila}}]{Kotak_2005}
{Kotak}, R., {Meikle}, P., {van Dyk}, S.~D., {H{\"o}flich}, P.~A., \& {Mattila}, S. 2005, \apjl, 628, L123, \dodoi{10.1086/432719}

\bibitem[{{Kotak} {et~al.}(2006){Kotak}, {Meikle}, {Pozzo}, {van Dyk}, {Farrah}, {Fesen}, {Filippenko}, {Foley}, {Fransson}, {Gerardy}, {H{\"o}flich}, {Lundqvist}, {Mattila}, {Sollerman}, \& {Wheeler}}]{Kotak_2006}
{Kotak}, R., {Meikle}, P., {Pozzo}, M., {et~al.} 2006, \apjl, 651, L117, \dodoi{10.1086/509655}

\bibitem[{{Kotak} {et~al.}(2009){Kotak}, {Meikle}, {Farrah}, {Gerardy}, {Foley}, {Van Dyk}, {Fransson}, {Lundqvist}, {Sollerman}, {Fesen}, {Filippenko}, {Mattila}, {Silverman}, {Andersen}, {H{\"o}flich}, {Pozzo}, \& {Wheeler}}]{Kotak_2009}
{Kotak}, R., {Meikle}, W.~P.~S., {Farrah}, D., {et~al.} 2009, \apj, 704, 306, \dodoi{10.1088/0004-637X/704/1/306}

\bibitem[{{Kumar} {et~al.}(2025){Kumar}, {Dastidar}, {Maund}, {Singleton}, \& {Sun}}]{Kumar_2025}
{Kumar}, A., {Dastidar}, R., {Maund}, J.~R., {Singleton}, A.~J., \& {Sun}, N.-C. 2025, \mnras, 538, 659, \dodoi{10.1093/mnras/staf312}

\bibitem[{{Kuncarayakti} {et~al.}(2020){Kuncarayakti}, {Folatelli}, {Maeda}, {Dessart}, {Jerkstrand}, {Anderson}, {Aoki}, {Bersten}, {Ferrari}, {Galbany}, {Garc{\'\i}a}, {Guti{\'e}rrez}, {Hattori}, {Kawabata}, {Kravtsov}, {Lyman}, {Mattila}, {Olivares E.}, {S{\'a}nchez}, \& {Van Dyk}}]{Kuncarayakti_2020}
{Kuncarayakti}, H., {Folatelli}, G., {Maeda}, K., {et~al.} 2020, \apj, 902, 139, \dodoi{10.3847/1538-4357/abb4e7}

\bibitem[{{Lantz} {et~al.}(2004){Lantz}, {Aldering}, {Antilogus}, {Bonnaud}, {Capoani}, {Castera}, {Copin}, {Dubet}, {Gangler}, {Henault}, {Lemonnier}, {Pain}, {Pecontal}, {Pecontal}, \& {Smadja}}]{Lantz_2004}
{Lantz}, B., {Aldering}, G., {Antilogus}, P., {et~al.} 2004, in Society of Photo-Optical Instrumentation Engineers (SPIE) Conference Series, Vol. 5249, Optical Design and Engineering, ed. L.~{Mazuray}, P.~J. {Rogers}, \& R.~{Wartmann}, 146--155, \dodoi{10.1117/12.512493}

\bibitem[{Law {et~al.}(2025)Law, Diaz, Sosey, Clarke, Coe, Cooper, Cracraft, Glidic, Gough, Hayes, Henry, Hilbert, Karatay, LaMassa, Larson, Manjavacas, Muzerolle, Nickson, Nikolov, Sunnquist, Wong, \& Zeidler}]{law_2025}
Law, D., Diaz, R., Sosey, M., {et~al.} 2025, JWST Pipeline Notebooks, Tech. rep., \dodoi{10.5281/zenodo.15571377}

\bibitem[{{Leonard} {et~al.}(2002){Leonard}, {Filippenko}, {Gates}, {Li}, {Eastman}, {Barth}, {Bus}, {Chornock}, {Coil}, {Frink}, {Grady}, {Harris}, {Malkan}, {Matheson}, {Quirrenbach}, \& {Treffers}}]{Leonard_2002}
{Leonard}, D.~C., {Filippenko}, A.~V., {Gates}, E.~L., {et~al.} 2002, \pasp, 114, 35, \dodoi{10.1086/324785}

\bibitem[{{Li} {et~al.}(2025){Li}, {Wang}, {Yang}, {Pastorello}, {Reguitti}, {Valerin}, {Ochner}, {Cai}, {Iijima}, {Munari}, {Salmaso}, {Farina}, {Cazzola}, {Trabacchin}, {Fiscale}, {Ciroi}, {Mura}, {Siviero}, {Cabras}, {Pabst}, {Taubenberger}, {Vogl}, {Fiorin}, {Liu}, {Chen}, {Xiang}, {Mo}, {Li}, {Wang}, {Zhang}, {Zhai}, {Mirzaqulov}, {Ehgamberdiev}, {Filippenko}, {Yan}, {Hu}, {Ma}, {Xia}, {Gao}, \& {Li}}]{Li_2025}
{Li}, G., {Wang}, X., {Yang}, Y., {et~al.} 2025, arXiv e-prints, arXiv:2504.03856, \dodoi{10.48550/arXiv.2504.03856}

\bibitem[{{Li} \& {McCray}(1992)}]{Li_1992}
{Li}, H., \& {McCray}, R. 1992, \apj, 387, 309, \dodoi{10.1086/171082}

\bibitem[{{Liljegren} {et~al.}(2020){Liljegren}, {Jerkstrand}, \& {Grumer}}]{Liljegren_2020}
{Liljegren}, S., {Jerkstrand}, A., \& {Grumer}, J. 2020, \aap, 642, A135, \dodoi{10.1051/0004-6361/202038116}

\bibitem[{{Liu} {et~al.}(2023){Liu}, {Chen}, {Er}, {Zeimann}, {Vink{\'o}}, {Wheeler}, {Cooper}, {Davis}, {Farrow}, {Gebhardt}, {Guo}, {Hill}, {House}, {Kollatschny}, {Kong}, {Kumar}, {Liu}, {Tuttle}, {Endl}, {Duke}, {Cochran}, {Zhang}, \& {Liu}}]{Liu_2023}
{Liu}, C., {Chen}, X., {Er}, X., {et~al.} 2023, \apjl, 958, L37, \dodoi{10.3847/2041-8213/ad0da8}

\bibitem[{{Liu} \& {Dalgarno}(1994)}]{Liu_1994}
{Liu}, W., \& {Dalgarno}, A. 1994, \apj, 428, 769, \dodoi{10.1086/174285}

\bibitem[{{Liu} \& {Dalgarno}(1995)}]{Liu_1995}
---. 1995, \apj, 454, 472, \dodoi{10.1086/176498}

\bibitem[{{Liu} {et~al.}(1992){Liu}, {Dalgarno}, \& {Lepp}}]{Liu_1992}
{Liu}, W., {Dalgarno}, A., \& {Lepp}, S. 1992, \apj, 396, 679, \dodoi{10.1086/171749}

\bibitem[{{Lucy} {et~al.}(1989{\natexlab{a}}){Lucy}, {Danziger}, {Gouiffes}, \& {Bouchet}}]{Lucy_1989}
{Lucy}, L.~B., {Danziger}, I.~J., {Gouiffes}, C., \& {Bouchet}, P. 1989{\natexlab{a}}, in IAU Colloq. 120: Structure and Dynamics of the Interstellar Medium, ed. G.~{Tenorio-Tagle}, M.~{Moles}, \& J.~{Melnick}, Vol. 350, 164, \dodoi{10.1007/BFb0114861}

\bibitem[{{Lucy} {et~al.}(1989{\natexlab{b}}){Lucy}, {Danziger}, {Gouiffes}, \& {Bouchet}}]{Lucy89}
---. 1989{\natexlab{b}}, in IAU Colloq. 120: Structure and Dynamics of the Interstellar Medium, ed. G.~{Tenorio-Tagle}, M.~{Moles}, \& J.~{Melnick}, Vol. 350, 164, \dodoi{10.1007/BFb0114861}

\bibitem[{{Maeda} {et~al.}(2008){Maeda}, {Kawabata}, {Mazzali}, {Tanaka}, {Valenti}, {Nomoto}, {Hattori}, {Deng}, {Pian}, {Taubenberger}, {Iye}, {Matheson}, {Filippenko}, {Aoki}, {Kosugi}, {Ohyama}, {Sasaki}, \& {Takata}}]{Maeda_2008}
{Maeda}, K., {Kawabata}, K., {Mazzali}, P.~A., {et~al.} 2008, Science, 319, 1220, \dodoi{10.1126/science.1149437}

\bibitem[{{Maguire} {et~al.}(2010){Maguire}, {Di Carlo}, {Smartt}, {Pastorello}, {Tsvetkov}, {Benetti}, {Spiro}, {Arkharov}, {Beccari}, {Botticella}, {Cappellaro}, {Cristallo}, {Dolci}, {Elias-Rosa}, {Fiaschi}, {Gorshanov}, {Harutyunyan}, {Larionov}, {Navasardyan}, {Pietrinferni}, {Raimondo}, {di Rico}, {Valenti}, {Valentini}, \& {Zampieri}}]{Maguire_2010}
{Maguire}, K., {Di Carlo}, E., {Smartt}, S.~J., {et~al.} 2010, \mnras, 404, 981, \dodoi{10.1111/j.1365-2966.2010.16332.x}

\bibitem[{{Maiolino} {et~al.}(2004){Maiolino}, {Schneider}, {Oliva}, {Bianchi}, {Ferrara}, {Mannucci}, {Pedani}, \& {Roca Sogorb}}]{Maiolino_2004}
{Maiolino}, R., {Schneider}, R., {Oliva}, E., {et~al.} 2004, \nat, 431, 533, \dodoi{10.1038/nature02930}

\bibitem[{{Margutti} {et~al.}(2012){Margutti}, {Soderberg}, {Chomiuk}, {Chevalier}, {Hurley}, {Milisavljevic}, {Foley}, {Hughes}, {Slane}, {Fransson}, {Moe}, {Barthelmy}, {Boynton}, {Briggs}, {Connaughton}, {Costa}, {Cummings}, {Del Monte}, {Enos}, {Fellows}, {Feroci}, {Fukazawa}, {Gehrels}, {Goldsten}, {Golovin}, {Hanabata}, {Harshman}, {Krimm}, {Litvak}, {Makishima}, {Marisaldi}, {Mitrofanov}, {Murakami}, {Ohno}, {Palmer}, {Sanin}, {Starr}, {Svinkin}, {Takahashi}, {Tashiro}, {Terada}, \& {Yamaoka}}]{Margutti_2012}
{Margutti}, R., {Soderberg}, A.~M., {Chomiuk}, L., {et~al.} 2012, \apj, 751, 134, \dodoi{10.1088/0004-637X/751/2/134}

\bibitem[{{Mart{\'\i}-Devesa} {et~al.}(2024){Mart{\'\i}-Devesa}, {Cheung}, {Di Lalla}, {Renaud}, {Principe}, {Omodei}, \& {Acero}}]{Marti-Devesa_2024}
{Mart{\'\i}-Devesa}, G., {Cheung}, C.~C., {Di Lalla}, N., {et~al.} 2024, \aap, 686, A254, \dodoi{10.1051/0004-6361/202349061}

\bibitem[{Matheson {et~al.}(2000)Matheson, Filippenko, Barth, Ho, Leonard, Bershady, Davis, Finley, Fisher, Gonz{\'a}lez, {et~al.}}]{Matheson_2000}
Matheson, T., Filippenko, A.~V., Barth, A.~J., {et~al.} 2000, The Astronomical Journal, 120, 1487

\bibitem[{{Maund}(2017)}]{Maund_2017}
{Maund}, J.~R. 2017, \mnras, 469, 2202, \dodoi{10.1093/mnras/stx879}

\bibitem[{{McCarthy} {et~al.}(1998){McCarthy}, {Cohen}, {Butcher}, {Cromer}, {Croner}, {Douglas}, {Goeden}, {Grewal}, {Lu}, {Petrie}, {Weng}, {Weber}, {Koch}, \& {Rodgers}}]{McCarthy_1998}
{McCarthy}, J.~K., {Cohen}, J.~G., {Butcher}, B., {et~al.} 1998, in Society of Photo-Optical Instrumentation Engineers (SPIE) Conference Series, Vol. 3355, Optical Astronomical Instrumentation, ed. S.~{D'Odorico}, 81--92, \dodoi{10.1117/12.316831}

\bibitem[{{Medler} {et~al.}(2025){Medler}, {Ashall}, {Shahbandeh}, {DerKacy}, {Hoogendam}, {Jones}, {Shappee}, {Hinkle}, {Pfeffer}, {Baron}, {Hoeflich}, \& {Hsiao}}]{Medler25_HISS}
{Medler}, K., {Ashall}, C., {Shahbandeh}, M., {et~al.} 2025, arXiv e-prints, arXiv:2505.18507, \dodoi{10.48550/arXiv.2505.18507}

\bibitem[{{Meikle} {et~al.}(2007){Meikle}, {Mattila}, {Pastorello}, {Gerardy}, {Kotak}, {Sollerman}, {Van Dyk}, {Farrah}, {Filippenko}, {H{\"o}flich}, {Lundqvist}, {Pozzo}, \& {Wheeler}}]{Meikle_2007}
{Meikle}, W.~P.~S., {Mattila}, S., {Pastorello}, A., {et~al.} 2007, \apj, 665, 608, \dodoi{10.1086/519733}

\bibitem[{{Meikle} {et~al.}(2011){Meikle}, {Kotak}, {Farrah}, {Mattila}, {Van Dyk}, {Andersen}, {Fesen}, {Filippenko}, {Foley}, {Fransson}, {Gerardy}, {H{\"o}flich}, {Lundqvist}, {Pozzo}, {Sollerman}, \& {Wheeler}}]{Meikle_2011}
{Meikle}, W.~P.~S., {Kotak}, R., {Farrah}, D., {et~al.} 2011, \apj, 732, 109, \dodoi{10.1088/0004-637X/732/2/109}

\bibitem[{{Mera} {et~al.}(2025)}]{Mera_2025}
{Mera}, T., {et~al.} 2025, \apj, in prep.

\bibitem[{{Michel} {et~al.}(2025){Michel}, {Mazzali}, {Perley}, {Hinds}, \& {Wise}}]{Michel_2025}
{Michel}, P.~D., {Mazzali}, P.~A., {Perley}, D.~A., {Hinds}, K.~R., \& {Wise}, J.~L. 2025, \mnras, 539, 633, \dodoi{10.1093/mnras/staf443}

\bibitem[{{Milisavljevic} {et~al.}(2010){Milisavljevic}, {Fesen}, {Gerardy}, {Kirshner}, \& {Challis}}]{Milisavljevic_2010}
{Milisavljevic}, D., {Fesen}, R.~A., {Gerardy}, C.~L., {Kirshner}, R.~P., \& {Challis}, P. 2010, \apj, 709, 1343, \dodoi{10.1088/0004-637X/709/2/1343}

\bibitem[{{Modjaz} {et~al.}(2008){Modjaz}, {Kirshner}, {Blondin}, {Challis}, \& {Matheson}}]{Modjaz_2008}
{Modjaz}, M., {Kirshner}, R.~P., {Blondin}, S., {Challis}, P., \& {Matheson}, T. 2008, \apjl, 687, L9, \dodoi{10.1086/593135}

\bibitem[{{M{\"u}ller} {et~al.}(2016){M{\"u}ller}, {Heger}, {Liptai}, \& {Cameron}}]{Muller_2016}
{M{\"u}ller}, B., {Heger}, A., {Liptai}, D., \& {Cameron}, J.~B. 2016, \mnras, 460, 742, \dodoi{10.1093/mnras/stw1083}

\bibitem[{{M{\"u}ller Bravo}(2023)}]{muller_bravo_2023}
{M{\"u}ller Bravo}, T.~E. 2023, {temuller/idsred: First Release!}, v0.1.0,  Zenodo, \dodoi{10.5281/zenodo.7851772}

\bibitem[{{Nayana} {et~al.}(2025){Nayana}, {Margutti}, {Wiston}, {Chornock}, {Campana}, {Laskar}, {Murase}, {Krips}, {Migliori}, {Tsuna}, {Alexander}, {Chandra}, {Bietenholz}, {Berger}, {Chevalier}, {De Colle}, {Dessart}, {Diesing}, {Grefenstette}, {Jacobson-Gal{\'a}n}, {Maeda}, {Marcote}, {Matthews}, {Milisavljevic}, {Ray}, {Reguitti}, \& {Polzin}}]{Nayana_2025}
{Nayana}, A.~J., {Margutti}, R., {Wiston}, E., {et~al.} 2025, \apj, 985, 51, \dodoi{10.3847/1538-4357/adc2fb}

\bibitem[{{Neustadt} {et~al.}(2024){Neustadt}, {Kochanek}, \& {Smith}}]{Neustadt_2024}
{Neustadt}, J.~M.~M., {Kochanek}, C.~S., \& {Smith}, M.~R. 2024, \mnras, 527, 5366, \dodoi{10.1093/mnras/stad3073}

\bibitem[{{Niculescu-Duvaz} {et~al.}(2022){Niculescu-Duvaz}, {Barlow}, {Bevan}, {Wesson}, {Milisavljevic}, {De Looze}, {Clayton}, {Krafton}, {Matsuura}, \& {Brady}}]{Niculescu_2022}
{Niculescu-Duvaz}, M., {Barlow}, M.~J., {Bevan}, A., {et~al.} 2022, \mnras, 515, 4302, \dodoi{10.1093/mnras/stac1626}

\bibitem[{{Niu} {et~al.}(2023){Niu}, {Sun}, {Maund}, {Zhang}, {Zhao}, \& {Liu}}]{Niu_2023}
{Niu}, Z., {Sun}, N.-C., {Maund}, J.~R., {et~al.} 2023, \apjl, 955, L15, \dodoi{10.3847/2041-8213/acf4e3}

\bibitem[{{Oke} {et~al.}(1995){Oke}, {Cohen}, {Carr}, {Cromer}, {Dingizian}, {Harris}, {Labrecque}, {Lucinio}, {Schaal}, {Epps}, \& {Miller}}]{Oke_1995}
{Oke}, J.~B., {Cohen}, J.~G., {Carr}, M., {et~al.} 1995, \pasp, 107, 375, \dodoi{10.1086/133562}

\bibitem[{{Panjkov} {et~al.}(2024){Panjkov}, {Auchettl}, {Shappee}, {Do}, {Lopez}, \& {Beacom}}]{Panjkov_2024}
{Panjkov}, S., {Auchettl}, K., {Shappee}, B.~J., {et~al.} 2024, \pasa, 41, e059, \dodoi{10.1017/pasa.2024.66}

\bibitem[{{Park} {et~al.}(2025){Park}, {Rho}, {Yoon}, {Pearson}, {Shrestha}, {Tinyanont}, {Geballe}, {Foley}, {Ravi}, {Andrews}, {Sand}, {Bostroem}, {Ashall}, {Hoeflich}, {Valenti}, {Dong}, {Meza Retamal}, {Hoang}, {Mehta}, {Howell}, {Farah}, {Terreran}, {Padilla Gonzalez}, {Andrews}, {Newsome}, {Shahbandeh}, {Smith}, {Kang}, {Suntzeff}, {Baron}, {Medler}, {Mera Evans}, {DerKacy}, {Larison}, {Galbany}, \& {Jacobson-Galan}}]{park_2025}
{Park}, S.~H., {Rho}, J., {Yoon}, S.-C., {et~al.} 2025, arXiv e-prints, arXiv:2507.11877, \dodoi{10.48550/arXiv.2507.11877}

\bibitem[{{Pascual} {et~al.}(2019){Pascual}, {Cardiel}, {Garz{\'o}n}, {Castro-Rodr{\'\i}guez}, {Gonz{\'a}lez-Fern{\'a}ndez}, {Hammersley}, {Manjavacas}, \& {Miluzio}}]{Pascual_2019}
{Pascual}, S., {Cardiel}, N., {Garz{\'o}n}, F., {et~al.} 2019, in Astronomical Society of the Pacific Conference Series, Vol. 521, Astronomical Data Analysis Software and Systems XXVI, ed. M.~{Molinaro}, K.~{Shortridge}, \& F.~{Pasian}, 232

\bibitem[{{Pascual} {et~al.}(2010){Pascual}, {Gallego}, {Cardiel}, \& {Eliche-Moral}}]{Pascual_2010}
{Pascual}, S., {Gallego}, J., {Cardiel}, N., \& {Eliche-Moral}, M.~C. 2010, in Astronomical Society of the Pacific Conference Series, Vol. 434, Astronomical Data Analysis Software and Systems XIX, ed. Y.~{Mizumoto}, K.~I. {Morita}, \& M.~{Ohishi}, 353

\bibitem[{{Perley} \& {Gal-Yam}(2023)}]{Perley_2023}
{Perley}, D., \& {Gal-Yam}, A. 2023, Transient Name Server Classification Report, 2023-1164, 1

\bibitem[{{Perrin} {et~al.}(2014){Perrin}, {Sivaramakrishnan}, {Lajoie}, {Elliott}, {Pueyo}, {Ravindranath}, \& {Albert}}]{Perrin14}
{Perrin}, M.~D., {Sivaramakrishnan}, A., {Lajoie}, C.-P., {et~al.} 2014, in Society of Photo-Optical Instrumentation Engineers (SPIE) Conference Series, Vol. 9143, Space Telescopes and Instrumentation 2014: Optical, Infrared, and Millimeter Wave, ed. J.~M. {Oschmann}, Jr., M.~{Clampin}, G.~G. {Fazio}, \& H.~A. {MacEwen}, 91433X, \dodoi{10.1117/12.2056689}

\bibitem[{{Pledger} \& {Shara}(2023)}]{Pledger_2023}
{Pledger}, J.~L., \& {Shara}, M.~M. 2023, \apjl, 953, L14, \dodoi{10.3847/2041-8213/ace88b}

\bibitem[{{Podsiadlowski}(1992)}]{Podsiadlowski_1992}
{Podsiadlowski}, P. 1992, \pasp, 104, 717, \dodoi{10.1086/133043}

\bibitem[{{Pozzo} {et~al.}(2004){Pozzo}, {Meikle}, {Fassia}, {Geballe}, {Lundqvist}, {Chugai}, \& {Sollerman}}]{Pozzo_2004}
{Pozzo}, M., {Meikle}, W.~P.~S., {Fassia}, A., {et~al.} 2004, \mnras, 352, 457, \dodoi{10.1111/j.1365-2966.2004.07951.x}

\bibitem[{{Pozzo} {et~al.}(2007){Pozzo}, {Meikle}, {Rayner}, {Joseph}, {Filippenko}, {Foley}, {Li}, {Mattila}, \& {Sollerman}}]{Pozzo_2007}
{Pozzo}, M., {Meikle}, W.~P.~S., {Rayner}, J.~T., {et~al.} 2007, \mnras, 375, 416, \dodoi{10.1111/j.1365-2966.2006.11298.x}

\bibitem[{{Qin} {et~al.}(2024){Qin}, {Zhang}, {Bloom}, {Sollerman}, {Zimmerman}, {Irani}, {Schulze}, {Gal-Yam}, {Kasliwal}, {Coughlin}, {Perley}, {Fremling}, \& {Kulkarni}}]{Qin_2024}
{Qin}, Y.-J., {Zhang}, K., {Bloom}, J., {et~al.} 2024, \mnras, 534, 271, \dodoi{10.1093/mnras/stae2012}

\bibitem[{{Ransome} {et~al.}(2024){Ransome}, {Villar}, {Tartaglia}, {Gonzalez}, {Jacobson-Gal{\'a}n}, {Kilpatrick}, {Margutti}, {Foley}, {Grayling}, {Ni}, {Yarza}, {Ye}, {Auchettl}, {de Boer}, {Chambers}, {Coulter}, {Drout}, {Farias}, {Gall}, {Gao}, {Huber}, {Ibik}, {Jones}, {Khetan}, {Lin}, {Politsch}, {Raimundo}, {Rest}, {Wainscoat}, {Yadavalli}, \& {Zenati}}]{Ransome_2024}
{Ransome}, C.~L., {Villar}, V.~A., {Tartaglia}, A., {et~al.} 2024, \apj, 965, 93, \dodoi{10.3847/1538-4357/ad2df7}

\bibitem[{{Ravensburg} {et~al.}(2024){Ravensburg}, {Carenza}, {Eckner}, \& {Goobar}}]{Ravensburg_2024}
{Ravensburg}, E., {Carenza}, P., {Eckner}, C., \& {Goobar}, A. 2024, \prd, 109, 023018, \dodoi{10.1103/PhysRevD.109.023018}

\bibitem[{{Ravi} {et~al.}(2023){Ravi}, {Rho}, {Park}, {Park}, {Yoon}, {Geballe}, {Vink{\'o}}, {Tinyanont}, {Bostroem}, {Burke}, {Hiramatsu}, {Howell}, {McCully}, {Newsome}, {Padilla Gonzalez}, {Pellegrino}, {Cartier}, {Pritchard}, {Andersen}, {Blinnikov}, {Dong}, {Blanchard}, {Kilpatrick}, {Hoeflich}, {Valenti}, {Filippenko}, {Suntzeff}, {Seok}, {K{\"o}nyves-T{\'o}th}, {Foley}, {Siebert}, \& {Jones}}]{Ravi_2023}
{Ravi}, A.~P., {Rho}, J., {Park}, S., {et~al.} 2023, \apj, 950, 14, \dodoi{10.3847/1538-4357/accddc}

\bibitem[{{Rayner} {et~al.}(2003){Rayner}, {Toomey}, {Onaka}, {Denault}, {Stahlberger}, {Vacca}, {Cushing}, \& {Wang}}]{Rayner_2003}
{Rayner}, J.~T., {Toomey}, D.~W., {Onaka}, P.~M., {et~al.} 2003, \pasp, 115, 362, \dodoi{10.1086/367745}

\bibitem[{{Rho} {et~al.}(2021){Rho}, {Evans}, {Geballe}, {Banerjee}, {Hoeflich}, {Shahbandeh}, {Valenti}, {Yoon}, {Jin}, {Williamson}, {Modjaz}, {Hiramatsu}, {Howell}, {Pellegrino}, {Vink{\'o}}, {Cartier}, {Burke}, {McCully}, {An}, {Cha}, {Pritchard}, {Wang}, {Andrews}, {Galbany}, {Van Dyk}, {Graham}, {Blinnikov}, {Joshi}, {P{\'a}l}, {Kriskovics}, {Ordasi}, {Szakats}, {Vida}, {Chen}, {Li}, {Zhang}, \& {Yan}}]{Rho_2021}
{Rho}, J., {Evans}, A., {Geballe}, T.~R., {et~al.} 2021, \apj, 908, 232, \dodoi{10.3847/1538-4357/abd850}

\bibitem[{{Riess} {et~al.}(2022){Riess}, {Yuan}, {Macri}, {Scolnic}, {Brout}, {Casertano}, {Jones}, {Murakami}, {Anand}, {Breuval}, {Brink}, {Filippenko}, {Hoffmann}, {Jha}, {D'arcy Kenworthy}, {Mackenty}, {Stahl}, \& {Zheng}}]{Riess_2022}
{Riess}, A.~G., {Yuan}, W., {Macri}, L.~M., {et~al.} 2022, \apjl, 934, L7, \dodoi{10.3847/2041-8213/ac5c5b}

\bibitem[{{Roche} {et~al.}(1991){Roche}, {Aitken}, \& {Smith}}]{Roche_1991}
{Roche}, P.~F., {Aitken}, D.~K., \& {Smith}, C.~H. 1991, \mnras, 252, 39P, \dodoi{10.1093/mnras/252.1.39P}

\bibitem[{{Roche} {et~al.}(1989){Roche}, {Aitken}, {Smith}, \& {James}}]{Roche_1989}
{Roche}, P.~F., {Aitken}, D.~K., {Smith}, C.~H., \& {James}, S.~D. 1989, \nat, 337, 533, \dodoi{10.1038/337533a0}

\bibitem[{{Rockosi} {et~al.}(2010){Rockosi}, {Stover}, {Kibrick}, {Lockwood}, {Peck}, {Cowley}, {Bolte}, {Adkins}, {Alcott}, {Allen}, {Brown}, {Cabak}, {Deich}, {Hilyard}, {Kassis}, {Lanclos}, {Lewis}, {Pfister}, {Phillips}, {Robinson}, {Saylor}, {Thompson}, {Ward}, {Wei}, \& {Wright}}]{Rockosi_2010}
{Rockosi}, C., {Stover}, R., {Kibrick}, R., {et~al.} 2010, in Society of Photo-Optical Instrumentation Engineers (SPIE) Conference Series, Vol. 7735, Ground-based and Airborne Instrumentation for Astronomy III, ed. I.~S. {McLean}, S.~K. {Ramsay}, \& H.~{Takami}, 77350R, \dodoi{10.1117/12.856818}

\bibitem[{{Sahu} {et~al.}(2006){Sahu}, {Anupama}, {Srividya}, \& {Muneer}}]{Sahu_2006}
{Sahu}, D.~K., {Anupama}, G.~C., {Srividya}, S., \& {Muneer}, S. 2006, \mnras, 372, 1315, \dodoi{10.1111/j.1365-2966.2006.10937.x}

\bibitem[{{Sarangi}(2022)}]{Sarangi22}
{Sarangi}, A. 2022, \aap, 668, A57, \dodoi{10.1051/0004-6361/202244391}

\bibitem[{{Sarangi} \& {Cherchneff}(2015)}]{Sarangi_2015}
{Sarangi}, A., \& {Cherchneff}, I. 2015, \aap, 575, A95, \dodoi{10.1051/0004-6361/201424969}

\bibitem[{{Sarangi} {et~al.}(2018){Sarangi}, {Dwek}, \& {Arendt}}]{Sarangi_2018}
{Sarangi}, A., {Dwek}, E., \& {Arendt}, R.~G. 2018, \apj, 859, 66, \dodoi{10.3847/1538-4357/aabfc3}

\bibitem[{{Sarmah}(2024)}]{Sarmah_2023}
{Sarmah}, P. 2024, \jcap, 2024, 083, \dodoi{10.1088/1475-7516/2024/04/083}

\bibitem[{{Scheck} {et~al.}(2006){Scheck}, {Kifonidis}, {Janka}, \& {M{\"u}ller}}]{Scheck_2006}
{Scheck}, L., {Kifonidis}, K., {Janka}, H.~T., \& {M{\"u}ller}, E. 2006, \aap, 457, 963, \dodoi{10.1051/0004-6361:20064855}

\bibitem[{{Schlafly} \& {Finkbeiner}(2011)}]{Schlafly_2011}
{Schlafly}, E.~F., \& {Finkbeiner}, D.~P. 2011, \apj, 737, 103, \dodoi{10.1088/0004-637X/737/2/103}

\bibitem[{{Schneider} {et~al.}(2004){Schneider}, {Ferrara}, \& {Salvaterra}}]{Schneider_2004}
{Schneider}, R., {Ferrara}, A., \& {Salvaterra}, R. 2004, \mnras, 351, 1379, \dodoi{10.1111/j.1365-2966.2004.07876.x}

\bibitem[{{Shahbandeh} {et~al.}(2022){Shahbandeh}, {Hsiao}, {Ashall}, {Teffs}, {Hoeflich}, {Morrell}, {Phillips}, {Anderson}, {Baron}, {Burns}, {Contreras}, {Davis}, {Diamond}, {Folatelli}, {Galbany}, {Gall}, {Hachinger}, {Holmbo}, {Karamehmetoglu}, {Kasliwal}, {Kirshner}, {Krisciunas}, {Kumar}, {Lu}, {Marion}, {Mazzali}, {Piro}, {Sand}, {Stritzinger}, {Suntzeff}, {Taddia}, \& {Uddin}}]{Shahbandeh_2022}
{Shahbandeh}, M., {Hsiao}, E.~Y., {Ashall}, C., {et~al.} 2022, \apj, 925, 175, \dodoi{10.3847/1538-4357/ac4030}

\bibitem[{{Shahbandeh} {et~al.}(2023){Shahbandeh}, {Sarangi}, {Temim}, {Szalai}, {Fox}, {Tinyanont}, {Dwek}, {Dessart}, {Filippenko}, {Brink}, {Foley}, {Jencson}, {Pierel}, {Zs{\'\i}ros}, {Rest}, {Zheng}, {Andrews}, {Clayton}, {De}, {Engesser}, {Gezari}, {Gomez}, {Gonzaga}, {Johansson}, {Kasliwal}, {Lau}, {De Looze}, {Marston}, {Milisavljevic}, {O'Steen}, {Siebert}, {Skrutskie}, {Smith}, {Strolger}, {Van Dyk}, {Wang}, {Williams}, {Williams}, {Xiao}, \& {Yang}}]{Shahbandeh_2023}
{Shahbandeh}, M., {Sarangi}, A., {Temim}, T., {et~al.} 2023, \mnras, 523, 6048, \dodoi{10.1093/mnras/stad1681}

\bibitem[{{Shahbandeh} {et~al.}(2024){Shahbandeh}, {Ashall}, {Hoeflich}, {Baron}, {Fox}, {Mera}, {DerKacy}, {Stritzinger}, {Shappee}, {Law}, {Morrison}, {Pauly}, {Pierel}, {Medler}, {Andrews}, {Baade}, {Bostroem}, {Brown}, {Burns}, {Burrow}, {Cikota}, {Cross}, {Davis}, {de Jaeger}, {Do}, {Dong}, {Hsiao}, {Dominguez}, {Galbany}, {Janzen}, {Jencson}, {Hoang}, {Karamehmetoglu}, {Khaghani}, {Krisciunas}, {Kumar}, {Lu}, {Mazzali}, {Morrell}, {Patat}, {Pearson}, {Pfeffer}, {Wang}, {Yang}, {Cai}, {Camacho-Neves}, {Elias-Rosa}, {Lundquist}, {Maund}, {Phillips}, {Rest}, {Retamal}, {Stangl}, {Shrestha}, {Stevens}, {Suntzeff}, {Telesco}, {Tucker}, {Foley}, {Jha}, {Kwok}, {Larison}, {LeBaron}, {Moran}, {Rho}, {Salmaso}, {Schmidt}, \& {Tinyanont}}]{Shahbandeh_2024}
{Shahbandeh}, M., {Ashall}, C., {Hoeflich}, P., {et~al.} 2024, arXiv e-prints, arXiv:2401.14474, \dodoi{10.48550/arXiv.2401.14474}

\bibitem[{{Shahbandeh} {et~al.}(2025){Shahbandeh}, {Fox}, {Temim}, {Dwek}, {Sarangi}, {Smith}, {Dessart}, {Nickson}, {Engesser}, {Filippenko}, {Brink}, {Zheng}, {Szalai}, {Johansson}, {Rest}, {Van Dyk}, {Andrews}, {Ashall}, {Clayton}, {De Looze}, {DerKacy}, {Dulude}, {Foley}, {Gezari}, {Gomez}, {Gonzaga}, {Indukuri}, {Jencson}, {Kasliwal}, {Lane}, {Lau}, {Law}, {Marston}, {Milisavljevic}, {O'Steen}, {Pierel}, {Siebert}, {Skrutskie}, {Strolger}, {Tinyanont}, {Wang}, {Williams}, {Xiao}, {Yang}, \& {Zs{\'\i}ros}}]{Shahbandeh_2025}
{Shahbandeh}, M., {Fox}, O.~D., {Temim}, T., {et~al.} 2025, \apj, 985, 262, \dodoi{10.3847/1538-4357/adce77}

\bibitem[{{Shappee} {et~al.}(2014){Shappee}, {Prieto}, {Grupe}, {Kochanek}, {Stanek}, {De Rosa}, {Mathur}, {Zu}, {Peterson}, {Pogge}, {Komossa}, {Im}, {Jencson}, {Holoien}, {Basu}, {Beacom}, {Szczygie{\l}}, {Brimacombe}, {Adams}, {Campillay}, {Choi}, {Contreras}, {Dietrich}, {Dubberley}, {Elphick}, {Foale}, {Giustini}, {Gonzalez}, {Hawkins}, {Howell}, {Hsiao}, {Koss}, {Leighly}, {Morrell}, {Mudd}, {Mullins}, {Nugent}, {Parrent}, {Phillips}, {Pojmanski}, {Rosing}, {Ross}, {Sand}, {Terndrup}, {Valenti}, {Walker}, \& {Yoon}}]{Shappee14_ASASSN}
{Shappee}, B.~J., {Prieto}, J.~L., {Grupe}, D., {et~al.} 2014, \apj, 788, 48, \dodoi{10.1088/0004-637X/788/1/48}

\bibitem[{{Sharp} \& {Hoeflich}(1990)}]{1990Ap&SS.171..213S}
{Sharp}, C.~M., \& {Hoeflich}, P. 1990, \apss, 171, 213, \dodoi{10.1007/BF00646849}

\bibitem[{{Singh} {et~al.}(2024){Singh}, {Teja}, {Moriya}, {Maeda}, {Kawabata}, {Tanaka}, {Imazawa}, {Nakaoka}, {Gangopadhyay}, {Yamanaka}, {Swain}, {Sahu}, {Anupama}, {Kumar}, {Anche}, {Sano}, {Raj}, {Agnihotri}, {Bhalerao}, {Bisht}, {Bisht}, {Belwal}, {Chakrabarti}, {Fujii}, {Nagayama}, {Matsumoto}, {Hamada}, {Kawabata}, {Kumar}, {Kumar}, {Malkan}, {Smith}, {Sakagami}, {Taguchi}, {Tominaga}, \& {Watanabe}}]{Singh_2024}
{Singh}, A., {Teja}, R.~S., {Moriya}, T.~J., {et~al.} 2024, \apj, 975, 132, \dodoi{10.3847/1538-4357/ad7955}

\bibitem[{{Singh}(1975)}]{Singh_1975}
{Singh}, P.~D. 1975, \aap, 44, 411

\bibitem[{{Sluder} {et~al.}(2018){Sluder}, {Milosavljevi{\'c}}, \& {Montgomery}}]{Sluder_2018}
{Sluder}, A., {Milosavljevi{\'c}}, M., \& {Montgomery}, M.~H. 2018, \mnras, 480, 5580, \dodoi{10.1093/mnras/sty2060}

\bibitem[{{Smartt} {et~al.}(2009){Smartt}, {Eldridge}, {Crockett}, \& {Maund}}]{Smartt_2009}
{Smartt}, S.~J., {Eldridge}, J.~J., {Crockett}, R.~M., \& {Maund}, J.~R. 2009, \mnras, 395, 1409, \dodoi{10.1111/j.1365-2966.2009.14506.x}

\bibitem[{{Smith} {et~al.}(2008){Smith}, {Foley}, \& {Filippenko}}]{Smith_2008}
{Smith}, N., {Foley}, R.~J., \& {Filippenko}, A.~V. 2008, \apj, 680, 568, \dodoi{10.1086/587860}

\bibitem[{{Smith} {et~al.}(2023){Smith}, {Pearson}, {Sand}, {Ilyin}, {Bostroem}, {Hosseinzadeh}, \& {Shrestha}}]{Smith_2023}
{Smith}, N., {Pearson}, J., {Sand}, D.~J., {et~al.} 2023, \apj, 956, 46, \dodoi{10.3847/1538-4357/acf366}

\bibitem[{{Smith} {et~al.}(2012){Smith}, {Silverman}, {Filippenko}, {Cooper}, {Matheson}, {Bian}, {Weiner}, \& {Comerford}}]{Smith_2012}
{Smith}, N., {Silverman}, J.~M., {Filippenko}, A.~V., {et~al.} 2012, \aj, 143, 17, \dodoi{10.1088/0004-6256/143/1/17}

\bibitem[{{Snow} \& {Rideal}(1929)}]{Snow_1929}
{Snow}, C.~P., \& {Rideal}, E.~K. 1929, Proceedings of the Royal Society of London Series A, 125, 462, \dodoi{10.1098/rspa.1929.0179}

\bibitem[{{Soraisam} {et~al.}(2023){Soraisam}, {Szalai}, {Van Dyk}, {Andrews}, {Srinivasan}, {Chun}, {Matheson}, {Scicluna}, \& {Vasquez-Torres}}]{Soraisam_2023}
{Soraisam}, M.~D., {Szalai}, T., {Van Dyk}, S.~D., {et~al.} 2023, \apj, 957, 64, \dodoi{10.3847/1538-4357/acef22}

\bibitem[{{Spyromilio} \& {Leibundgut}(1996)}]{Spyromilio_1996}
{Spyromilio}, J., \& {Leibundgut}, B. 1996, \mnras, 283, L89, \dodoi{10.1093/mnras/283.3.L89}

\bibitem[{{Spyromilio} {et~al.}(1988){Spyromilio}, {Meikle}, {Learner}, \& {Allen}}]{Spyromilio_1988}
{Spyromilio}, J., {Meikle}, W.~P.~S., {Learner}, R.~C.~M., \& {Allen}, D.~A. 1988, \nat, 334, 327, \dodoi{10.1038/334327a0}

\bibitem[{{Stritzinger} {et~al.}(2012){Stritzinger}, {Taddia}, {Fransson}, {Fox}, {Morrell}, {Phillips}, {Sollerman}, {Anderson}, {Boldt}, {Brown}, {Campillay}, {Castellon}, {Contreras}, {Folatelli}, {Habergham}, {Hamuy}, {Hjorth}, {James}, {Krzeminski}, {Mattila}, {Persson}, \& {Roth}}]{Stritzinger_2012}
{Stritzinger}, M., {Taddia}, F., {Fransson}, C., {et~al.} 2012, \apj, 756, 173, \dodoi{10.1088/0004-637X/756/2/173}

\bibitem[{{Stritzinger} {et~al.}(2023){Stritzinger}, {Valerin}, {Elias-Rosa}, {Fraser}, {Galbany}, {Gutierrez}, {Kankare}, {Kotak}, {Moran}, {Lundqvist}, {Matilainen}, {Reguitti}, {Reynolds}, {Salmaso}, \& {Shappee}}]{Stritzinger_2023}
{Stritzinger}, M., {Valerin}, G., {Elias-Rosa}, N., {et~al.} 2023, Transient Name Server AstroNote, 145, 1

\bibitem[{{Szalai} \& {Van Dyk}(2023)}]{Szalai_2023}
{Szalai}, T., \& {Van Dyk}, S.~V. 2023, The Astronomer's Telegram, 16042, 1

\bibitem[{{Szalai} \& {Vink{\'o}}(2013)}]{Szalai_2013}
{Szalai}, T., \& {Vink{\'o}}, J. 2013, \aap, 549, A79, \dodoi{10.1051/0004-6361/201220015}

\bibitem[{{Szalai} {et~al.}(2011){Szalai}, {Vink{\'o}}, {Balog}, {G{\'a}sp{\'a}r}, {Block}, \& {Kiss}}]{Szalai_2011}
{Szalai}, T., {Vink{\'o}}, J., {Balog}, Z., {et~al.} 2011, \aap, 527, A61, \dodoi{10.1051/0004-6361/201015624}

\bibitem[{{Szalai} {et~al.}(2019){Szalai}, {Vink{\'o}}, {K{\"o}nyves-T{\'o}th}, {Nagy}, {Bostroem}, {S{\'a}rneczky}, {Brown}, {Pejcha}, {B{\'o}di}, {Cseh}, {Cs{\"o}rnyei}, {Dencs}, {Hanyecz}, {Ign{\'a}cz}, {Kalup}, {Kriskovics}, {Ordasi}, {P{\'a}l}, {Seli}, {S{\'o}dor}, {Szak{\'a}ts}, {Vida}, {Zsidi}, {Konkoly Team}, {Arcavi}, {Ashall}, {Burke}, {Galbany}, {Hiramatsu}, {Hosseinzadeh}, {Hsiao}, {Howell}, {McCully}, {Moran}, {Rho}, {Sand}, {Shahbandeh}, {Valenti}, {Wang}, {Wheeler}, \& {Supernova Project}}]{Szalai_2019}
{Szalai}, T., {Vink{\'o}}, J., {K{\"o}nyves-T{\'o}th}, R., {et~al.} 2019, \apj, 876, 19, \dodoi{10.3847/1538-4357/ab12d0}

\bibitem[{{Taubenberger} {et~al.}(2009){Taubenberger}, {Valenti}, {Benetti}, {Cappellaro}, {Della Valle}, {Elias-Rosa}, {Hachinger}, {Hillebrandt}, {Maeda}, {Mazzali}, {Pastorello}, {Patat}, {Sim}, \& {Turatto}}]{Taubenberger_2009}
{Taubenberger}, S., {Valenti}, S., {Benetti}, S., {et~al.} 2009, \mnras, 397, 677, \dodoi{10.1111/j.1365-2966.2009.15003.x}

\bibitem[{{Teja} {et~al.}(2023){Teja}, {Singh}, {Basu}, {Anupama}, {Sahu}, {Dutta}, {Swain}, {Nakaoka}, {Pathak}, {Bhalerao}, {Barway}, {Kumar}, {A.~J.}, {Imazawa}, {Kumar}, \& {Kawabata}}]{Teja_2023a}
{Teja}, R.~S., {Singh}, A., {Basu}, J., {et~al.} 2023, \apjl, 954, L12, \dodoi{10.3847/2041-8213/acef20}

\bibitem[{{Thwaites} {et~al.}(2023){Thwaites}, {Vandenbroucke}, {Santander}, \& {IceCube Collaboration}}]{Thwaites_2023}
{Thwaites}, J., {Vandenbroucke}, J., {Santander}, M., \& {IceCube Collaboration}. 2023, The Astronomer's Telegram, 16043, 1

\bibitem[{{Tinyanont} {et~al.}(2019){Tinyanont}, {Lau}, {Kasliwal}, {Maeda}, {Smith}, {Fox}, {Gehrz}, {De}, {Jencson}, {Bally}, \& {Masci}}]{Tinyanont_2019b}
{Tinyanont}, S., {Lau}, R.~M., {Kasliwal}, M.~M., {et~al.} 2019, \apj, 887, 75, \dodoi{10.3847/1538-4357/ab521b}

\bibitem[{{Tinyanont} {et~al.}(2025){Tinyanont}, {Fox}, {Shahbandeh}, {Temim}, {Williams}, {Wangnok}, {Rest}, {Lau}, {Maeda}, {Jencson}, {Auchettl}, {Filippenko}, {Larison}, {Ashall}, {Brink}, {Davis}, {Dessart}, {Foley}, {Galbany}, {Grayling}, {Johansson}, {Kasliwal}, {Lane}, {LeBaron}, {Milisavljevic}, {Rho}, {Sakon}, {Sarangi}, {Szalai}, {Taggart}, {Van Dyk}, {Wang}, {Yang}, {Zheng}, \& {Zs{\'\i}ros}}]{Tinyanont_2025}
{Tinyanont}, S., {Fox}, O.~D., {Shahbandeh}, M., {et~al.} 2025, \apj, 985, 198, \dodoi{10.3847/1538-4357/adccc0}

\bibitem[{{Tonry} {et~al.}(2018){Tonry}, {Denneau}, {Heinze}, {Stalder}, {Smith}, {Smartt}, {Stubbs}, {Weiland}, \& {Rest}}]{Tonry18}
{Tonry}, J.~L., {Denneau}, L., {Heinze}, A.~N., {et~al.} 2018, \pasp, 130, 064505, \dodoi{10.1088/1538-3873/aabadf}

\bibitem[{{Tucker} {et~al.}(2022){Tucker}, {Shappee}, {Huber}, {Payne}, {Do}, {Hinkle}, {de Jaeger}, {Ashall}, {Desai}, {Hoogendam}, {Aldering}, {Auchettl}, {Baranec}, {Bulger}, {Chambers}, {Chun}, {Hodapp}, {Lowe}, {McKay}, {Rampy}, {Rubin}, \& {Tonry}}]{Tucker_2022}
{Tucker}, M.~A., {Shappee}, B.~J., {Huber}, M.~E., {et~al.} 2022, \pasp, 134, 124502, \dodoi{10.1088/1538-3873/aca719}

\bibitem[{{Tucker} {et~al.}(2024){Tucker}, {Hinkle}, {Angus}, {Auchettl}, {Hoogendam}, {Shappee}, {Kochanek}, {Ashall}, {de Boer}, {Chambers}, {Desai}, {Do}, {Fulton}, {Gao}, {Herman}, {Huber}, {Lidman}, {Lin}, {Lowe}, {Magnier}, {Martin}, {M{\'\i}nguez}, {Nicholl}, {Pursiainen}, {Smartt}, {Smith}, {Srivastav}, {Tucker}, \& {Wainscoat}}]{Tucker_2024_SN2023ufx}
{Tucker}, M.~A., {Hinkle}, J., {Angus}, C.~R., {et~al.} 2024, \apj, 976, 178, \dodoi{10.3847/1538-4357/ad8448}

\bibitem[{{Utrobin} \& {Chugai}(2009)}]{Utrobin_2009}
{Utrobin}, V.~P., \& {Chugai}, N.~N. 2009, \aap, 506, 829, \dodoi{10.1051/0004-6361/200912273}

\bibitem[{{Utrobin} \& {Chugai}(2017)}]{Utrobin_2017}
---. 2017, \mnras, 472, 5004, \dodoi{10.1093/mnras/stx2415}

\bibitem[{{Valiante} {et~al.}(2009){Valiante}, {Schneider}, {Bianchi}, \& {Andersen}}]{Valiante_2009}
{Valiante}, R., {Schneider}, R., {Bianchi}, S., \& {Andersen}, A.~C. 2009, \mnras, 397, 1661, \dodoi{10.1111/j.1365-2966.2009.15076.x}

\bibitem[{{van Breemen} {et~al.}(2011){van Breemen}, {Min}, {Chiar}, {Waters}, {Kemper}, {Boogert}, {Cami}, {Decin}, {Knez}, {Sloan}, \& {Tielens}}]{VanBreemen_2011}
{van Breemen}, J.~M., {Min}, M., {Chiar}, J.~E., {et~al.} 2011, \aap, 526, A152, \dodoi{10.1051/0004-6361/200811142}

\bibitem[{{Van Dyk} {et~al.}(2024{\natexlab{a}}){Van Dyk}, {Szalai}, {Cutri}, {Kirkpatrick}, {Grillmair}, {Fajardo-Acosta}, {Masiero}, {Mainzer}, {Gelino}, {Vink{\'o}}, {Jo{\'o}}, {P{\'a}l}, {K{\"o}nyves-T{\'o}th}, {Kriskovics}, {Szak{\'a}ts}, {Vida}, {Zheng}, {Brink}, \& {Filippenko}}]{Vandyk_2024}
{Van Dyk}, S.~D., {Szalai}, T., {Cutri}, R.~M., {et~al.} 2024{\natexlab{a}}, \apj, 977, 98, \dodoi{10.3847/1538-4357/ad8cd8}

\bibitem[{{Van Dyk} {et~al.}(2024{\natexlab{b}}){Van Dyk}, {Srinivasan}, {Andrews}, {Soraisam}, {Szalai}, {Howell}, {Isaacson}, {Matheson}, {Petigura}, {Scicluna}, {Stephens}, {Van Zandt}, {Zheng}, {Chun}, \& {Fillippenko}}]{Vandyk_2024a}
{Van Dyk}, S.~D., {Srinivasan}, S., {Andrews}, J.~E., {et~al.} 2024{\natexlab{b}}, \apj, 968, 27, \dodoi{10.3847/1538-4357/ad414b}

\bibitem[{{Vasylyev} {et~al.}(2023){Vasylyev}, {Yang}, {Filippenko}, {Patra}, {Brink}, {Wang}, {Chornock}, {Margutti}, {Gates}, {Burgasser}, {Karpoor}, {LeBaron}, {Softich}, {Theissen}, {Wiston}, \& {Zheng}}]{Vasylyev_2023}
{Vasylyev}, S.~S., {Yang}, Y., {Filippenko}, A.~V., {et~al.} 2023, \apjl, 955, L37, \dodoi{10.3847/2041-8213/acf1a3}

\bibitem[{{Virtanen} {et~al.}(2020){Virtanen}, {Gommers}, {Oliphant}, {Haberland}, {Reddy}, {Cournapeau}, {Burovski}, {Peterson}, {Weckesser}, {Bright}, {van der Walt}, {Brett}, {Wilson}, {Millman}, {Mayorov}, {Nelson}, {Jones}, {Kern}, {Larson}, {Carey}, {Polat}, {Feng}, {Moore}, {VanderPlas}, {Laxalde}, {Perktold}, {Cimrman}, {Henriksen}, {Quintero}, {Harris}, {Archibald}, {Ribeiro}, {Pedregosa}, {van Mulbregt}, \& {SciPy 1. 0 Contributors}}]{SciPy_2020}
{Virtanen}, P., {Gommers}, R., {Oliphant}, T.~E., {et~al.} 2020, Nature Methods, 17, 261, \dodoi{10.1038/s41592-019-0686-2}

\bibitem[{{Wang} \& {Burrows}(2024)}]{Wang_2024}
{Wang}, T., \& {Burrows}, A. 2024, \apj, 962, 71, \dodoi{10.3847/1538-4357/ad12b8}

\bibitem[{{Watson} {et~al.}(2015){Watson}, {Christensen}, {Knudsen}, {Richard}, {Gallazzi}, \& {Micha{\l}owski}}]{Watson_2015}
{Watson}, D., {Christensen}, L., {Knudsen}, K.~K., {et~al.} 2015, \nat, 519, 327, \dodoi{10.1038/nature14164}

\bibitem[{{Wesson} {et~al.}(2015){Wesson}, {Barlow}, {Matsuura}, \& {Ercolano}}]{Wesson_2015}
{Wesson}, R., {Barlow}, M.~J., {Matsuura}, M., \& {Ercolano}, B. 2015, \mnras, 446, 2089, \dodoi{10.1093/mnras/stu2250}

\bibitem[{{Williams} {et~al.}(2014){Williams}, {Peterson}, {Murphy}, {Gilbert}, {Dalcanton}, {Dolphin}, \& {Jennings}}]{Williams_2014}
{Williams}, B.~F., {Peterson}, S., {Murphy}, J., {et~al.} 2014, \apj, 791, 105, \dodoi{10.1088/0004-637X/791/2/105}

\bibitem[{{Wilson} {et~al.}(2004){Wilson}, {Henderson}, {Herter}, {Matthews}, {Skrutskie}, {Adams}, {Moon}, {Smith}, {Gautier}, {Ressler}, {Soifer}, {Lin}, {Howard}, {LaMarr}, {Stolberg}, \& {Zink}}]{Wilson_2004}
{Wilson}, J.~C., {Henderson}, C.~P., {Herter}, T.~L., {et~al.} 2004, in Society of Photo-Optical Instrumentation Engineers (SPIE) Conference Series, Vol. 5492, Ground-based Instrumentation for Astronomy, ed. A.~F.~M. {Moorwood} \& M.~{Iye}, 1295--1305, \dodoi{10.1117/12.550925}

\bibitem[{{Wooden} {et~al.}(1993){Wooden}, {Rank}, {Bregman}, {Witteborn}, {Tielens}, {Cohen}, {Pinto}, \& {Axelrod}}]{Wooden_1993}
{Wooden}, D.~H., {Rank}, D.~M., {Bregman}, J.~D., {et~al.} 1993, \apjs, 88, 477, \dodoi{10.1086/191830}

\bibitem[{{Woosley} {et~al.}(2002){Woosley}, {Heger}, \& {Weaver}}]{Woosley_2002}
{Woosley}, S.~E., {Heger}, A., \& {Weaver}, T.~A. 2002, Reviews of Modern Physics, 74, 1015, \dodoi{10.1103/RevModPhys.74.1015}

\bibitem[{{Woosley} {et~al.}(1988){Woosley}, {Pinto}, \& {Ensman}}]{Woosley_1988}
{Woosley}, S.~E., {Pinto}, P.~A., \& {Ensman}, L. 1988, \apj, 324, 466, \dodoi{10.1086/165908}

\bibitem[{{Woosley} \& {Weaver}(1995)}]{Woosley_1995}
{Woosley}, S.~E., \& {Weaver}, T.~A. 1995, \apjs, 101, 181, \dodoi{10.1086/192237}

\bibitem[{{Xiang} {et~al.}(2024){Xiang}, {Mo}, {Wang}, {Wang}, {Zhang}, {Lin}, \& {Wang}}]{Xiang_2024}
{Xiang}, D., {Mo}, J., {Wang}, L., {et~al.} 2024, Science China Physics, Mechanics, and Astronomy, 67, 219514, \dodoi{10.1007/s11433-023-2267-0}

\bibitem[{{Yamanaka} {et~al.}(2023){Yamanaka}, {Fujii}, \& {Nagayama}}]{Yamanaka_2023}
{Yamanaka}, M., {Fujii}, M., \& {Nagayama}, T. 2023, \pasj, 75, L27, \dodoi{10.1093/pasj/psad051}

\bibitem[{{Zhang} {et~al.}(2023){Zhang}, {Lin}, {Wang}, {Zhao}, {Li}, {Liu}, {Yan}, {Xiang}, {Wang}, \& {Bai}}]{Zhang_2023}
{Zhang}, J., {Lin}, H., {Wang}, X., {et~al.} 2023, Science Bulletin, 68, 2548, \dodoi{10.1016/j.scib.2023.09.015}

\bibitem[{{Zheng} {et~al.}(2025){Zheng}, {Dessart}, {Filippenko}, {Yang}, {Brink}, {de Jaeger}, {Vasylyev}, {Van Dyk}, {Patra}, {Jacobson-Gal{\'a}n}, {Stewart}, {Alvarado}, {Arikatla}, {Beddow}, {Betz}, {Born}, {Bostow}, {Burgasser}, {Caceres}, {Carrasco}, {Chuang}, {DeGraw}, {Gates}, {Gendreau-Distler}, {Jacobus}, {Jennings}, {Karpoor}, {Lynam}, {Mina}, {Mora}, {Pichay}, {Ravi}, {Rees}, {Rich}, {Risin}, {Sandford}, {Savino}, {Softich}, {Theissen}, {Vidal}, {Wu}, \& {Zeng}}]{Zheng_2025}
{Zheng}, W., {Dessart}, L., {Filippenko}, A.~V., {et~al.} 2025, \apj, 988, 61, \dodoi{10.3847/1538-4357/ade0bf}

\bibitem[{{Zimmerman} {et~al.}(2024){Zimmerman}, {Irani}, {Chen}, {Gal-Yam}, {Schulze}, {Perley}, {Sollerman}, {Filippenko}, {Shenar}, {Yaron}, {Shahaf}, {Bruch}, {Ofek}, {De Cia}, {Brink}, {Yang}, {Vasylyev}, {Ben Ami}, {Aubert}, {Badash}, {Bloom}, {Brown}, {De}, {Dimitriadis}, {Fransson}, {Fremling}, {Hinds}, {Horesh}, {Johansson}, {Kasliwal}, {Kulkarni}, {Kushnir}, {Martin}, {Matuzewski}, {McGurk}, {Miller}, {Morag}, {Neil}, {Nugent}, {Post}, {Prusinski}, {Qin}, {Raichoor}, {Riddle}, {Rowe}, {Rusholme}, {Sfaradi}, {Sjoberg}, {Soumagnac}, {Stein}, {Strotjohann}, {Terwel}, {Wasserman}, {Wise}, {Wold}, {Yan}, \& {Zhang}}]{Zimmerman_2023}
{Zimmerman}, E.~A., {Irani}, I., {Chen}, P., {et~al.} 2024, \nat, 627, 759, \dodoi{10.1038/s41586-024-07116-6}

\bibitem[{{Zs{\'\i}ros} {et~al.}(2024){Zs{\'\i}ros}, {Szalai}, {De Looze}, {Sarangi}, {Shahbandeh}, {Fox}, {Temim}, {Milisavljevic}, {Van Dyk}, {Smith}, {Filippenko}, {Brink}, {Zheng}, {Dessart}, {Jencson}, {Johansson}, {Pierel}, {Rest}, {Tinyanont}, {Niculescu-Duvaz}, {Barlow}, {Wesson}, {Andrews}, {Clayton}, {De}, {Dwek}, {Engesser}, {Foley}, {Gezari}, {Gomez}, {Gonzaga}, {Kasliwal}, {Lau}, {Marston}, {O'Steen}, {Siebert}, {Skrutskie}, {Strolger}, {Wang}, {Williams}, {Williams}, \& {Xiao}}]{Zsiros_2024}
{Zs{\'\i}ros}, S., {Szalai}, T., {De Looze}, I., {et~al.} 2024, \mnras, 529, 155, \dodoi{10.1093/mnras/stae507}

\end{thebibliography}
\bibliographystyle{aasjournal}

\end{document}